\documentclass[prd,superscriptaddress,twocolumn,amssymb,amsmath,amsfonts,aps,nofootinbib]{revtex4-1}
\newcommand{\gx}{\textsc{GlueX}}
\usepackage[pdftex]{graphicx}
\usepackage[colorlinks=true,allcolors=blue,bookmarks=true]{hyperref}
\pdfoutput=1
\usepackage{color}
\definecolor{orange}{RGB}{153,86,0}

\begin{document}

\begin{flushleft}
\hspace{-260pt} {\large C12-12-002}
\end{flushleft}

\title{A study of decays to strange final states with GlueX in Hall~D \\ using components of the BaBar DIRC \\
 {\small  (A proposal to the 42$^\mathrm{nd}$ Jefferson Lab Program Advisory Committee)} }

\author{M.~Dugger}
\author{B.~Ritchie}
\author{I.~Senderovich}
\affiliation{Arizona State University}
\author{E.~Anassontzis}
\author{P.~Ioannou}
\author{C.~Kourkoumeli}
\author{G.~Vasileiadis}
\author{G.~Voulgaris}
\affiliation{University of Athens}
\author{N.~Jarvis}
\author{W.~Levine}
\author{P.~Mattione}
\author{W.~McGinley}
\author{C.~A.~Meyer}
\author{R.~Schumacher}
\author{M.~Staib}
\affiliation{Carnegie Mellon University}
% ...
\collaboration{The \textsc{GlueX} Collaboration}

\date{June 1, 2014}

\begin{abstract}

We propose to enhance the kaon identification capabilities of the \gx{} detector by constructing
an FDIRC (Focusing Detection of Internally Reflected Cherenkov) detector utilizing the decommissioned
BaBar DIRC components.  The \gx{} FDIRC would significantly enhance the \gx{} physics program
by allowing one to search for and study hybrid mesons decaying into kaon final states.  Such 
systematic studies of kaon final states are essential for inferring the quark flavor content of 
hybrid and conventional mesons.  The \gx{} FDIRC would reuse one-third of the synthetic fused
silica bars that were utilized in the BaBar DIRC.  A new focussing photon camera, read out
with large area photodetectors, would be developed.  We propose operating the enhanced
\gx{} detector in Hall~D for a total of 220 days at an average intensity of $5\times10^{7}~\gamma$/s,
a program that was conditionally approved by PAC39.

\end{abstract}

\maketitle


\section{Preamble}

In 2012, the \gx{} Collaboration submitted a proposal to PAC39 titled ``A study of meson and
baryon decays to strange final states with \gx{} in Hall~D"~\cite{pac39}.  The proposal requested
220 days of data taking with the \gx{} detector operating at an intensity of $5\times 10^{7}~\gamma$/s
on target.  A necessary component of the broad physics program put forth in this proposal was
an upgrade to the \gx{} particle identification (PID) capability.  At the time of the proposal, a design
for an upgraded PID system had not been finalized.  The PAC granted conditional approval
stating in its summary document:
\begin{quote}
GlueX is the flagship experiment in Hall~D; the theoretical motivation for the proposed extension
of running is very sound.  However, the success of the experiment depends crucially on the final
design of the kaon identification system.  The PAC39 therefore recommends C2 conditional
approval, contingent upon the final design of the particle ID system.
\end{quote}

In 2013, the \gx{} Collaboration returned to PAC40 with a proposal~\cite{pac40} demonstrating that some
states with hidden and open strangeness were accessible with the baseline \gx{} design,
provided that the statistical precision of the data sample was sufficient.  PAC40 approved
an additional 200 days of running at $5\times 10^{7}~\gamma$/s on target with an ``A" 
scientific rating for this program.  The PAC40 report noted:
\begin{quote}
The PAC was impressed by the level of sophistication of the GlueX software and analysis which is essential for the achievement of a significant kaon and hyperon program even in the absence of dedicated hardware. Still the complete mapping of the spectrum of conventional and exotic hadrons will ultimately require the implementation of dedicated particle ID in the forward direction, extending the kaon identification capability to 10 GeV/$c$. The PAC therefore encourages the collaboration to move forward with the design of such system and aim at an early installation, if at all possible.
\end{quote}
The 10 GeV/$c$ momentum cutoff cited by the PAC was motivated by preliminary designs
for a dual-radiator RICH discussed in our previous proposals rather than any specific physics
requirement.  Concurrently with our preparation for PAC40, it became apparent that there was an opportunity
to utilize components of the BaBar DIRC detector\footnote{At the time of our original proposal
the BaBar DIRC was to be used in Super$B$, a next-generation $B$-factory experiment.  The
cancellation of Super$B$ in late 2012 made the DIRC components available for reuse.}.  In fall
of 2013, a team of five members of the \gx{}~Collaboration visited SLAC to understand the
condition of the DIRC components and details about their utilization.  The conclusion of
this visit was that there are no insurmountable technical challenges in utilizing the BaBar DIRC;
the opportunity presents a unique and cost-effective solution to upgrade the PID capability
of \gx{}, providing a significant enhancement in our kaon identification capability.

In December 2013, members of the collaboration submitted a detailed conceptual design
for utilizing the DIRC to SLAC for consideration.  As this was precisely the request of
PAC39, a final design of the particle ID system, we present this design in the document
that follows.  The physics case remains largely the same from that presented in our PAC39
proposal~\cite{pac39} and is repeated in part here for completeness.  


\section{Introduction and Motivation}

The GlueX experiment, currently under construction and scheduled to start operating in Hall~D at Jefferson Lab in fall 2014, will provide the data necessary to construct quantitative tests of non-perturbative QCD by studying the spectrum of light-quark mesons and baryons. The primary goal of the GlueX experiment is to search for and study the spectrum of so-called hybrid mesons that are formed by exciting the gluonic field that couples the quarks. QCD-based calculations predict the existence of hybrid meson states, including several that have exotic quantum numbers that cannot be formed from a simple quark/anti-quark pair. To achieve its goal, GlueX must systematically study all possible decay modes of conventional and hybrid mesons, including those with kaons. The addition of a Cherenkov-based particle identification system utilizing the BaBar DIRC (Detection of Internally Reflected Cherenkov) components will dramatically increase the number of potential hybrid decay modes that GlueX can access and will reduce the experimental backgrounds from misidentified particles in each mode. This enhanced capability will be crucial in order for the GlueX experiment to realize its full discovery potential.

In this section we motivate the \gx{} experiment and discuss the importance of
kaon identification in the context of the \gx{} physics program.  The subsequent
section discusses the baseline \gx{} design and run plan.  Both of these sections
are largely reproduced from Refs.~\cite{pac39,pac40}, documents that were
developed jointly by the \gx{}~Collaboration.

\subsection{The GlueX experiment}
\label{Sec:Experiment}

A long-standing goal of hadron physics has been to understand how the quark 
and gluonic degrees of freedom that are present in the fundamental QCD Lagrangian 
manifest themselves in the spectrum of hadrons.   Of particular interest is how the 
gluon-gluon interactions might give rise to physical states with gluonic excitations.  
One class of such states is the hybrid meson, 
which can be naively thought of as a quark anti-quark pair coupled to a valence 
gluon ($q\bar{q}g$).  Recent lattice QCD calculations~\cite{Dudek:2011bn} predict a rich spectrum 
of hybrid mesons.  A subset of these hybrids has an exotic experimental signature:  
angular momentum ($J$), parity ($P$), and charge conjugation ($C$) that cannot be 
created from just a quark-antiquark pair.  The primary goal of the \gx~experiment in 
Hall~D is to search for and study these mesons.

Our understanding of how gluonic excitations manifest themselves within QCD 
is maturing thanks to recent results from lattice QCD. This numerical approach 
to QCD considers the theory on a finite, discrete grid of points in a manner that 
would become exact if the lattice spacing were taken to zero and the spatial extent 
of the calculation, {\it i.e.,} the ``box size," 
was made large. In practice, rather fine spacings and large boxes are used so that 
the systematic effect of this approximation should be small. The main limitation of
these calculations at present is the poor scaling of the numerical algorithms 
with decreasing quark mass.  In practice most contemporary calculations use a 
range of artificially heavy light quarks and attempt to observe a trend as the light 
quark mass is reduced toward the physical value. Trial calculations at the physical 
quark mass have begun, and regular usage is anticipated within a few years.

The spectrum of eigenstates of QCD can be extracted from correlation functions 
of the type $\langle 0 | \mathcal{O}_f(t) \mathcal{O}_i^\dag(0) | 0 \rangle$, where 
the $\mathcal{O}^\dag$ are composite QCD operators capable of interpolating a 
meson or baryon state from the vacuum. The time-evolution of the Euclidean 
correlator indicates the mass spectrum ($e^{-m_\mathfrak{n} t}$), and information 
about quark-gluon substructure can be inferred from matrix-elements 
$\langle \mathfrak{n} | \mathcal{O}^\dag |0 \rangle$. In a series of recent 
papers~\cite{Dudek:2009qf,Dudek:2010wm,Dudek:2011tt,Edwards:2011jj}, 
the Hadron Spectrum Collaboration has explored the spectrum of mesons and 
baryons using a large basis of composite QCD interpolating fields, extracting a 
spectrum of states of determined $J^{P(C)}$, including states of high internal excitation.

\begin{figure*}
\begin{center}
\includegraphics[width=\linewidth]{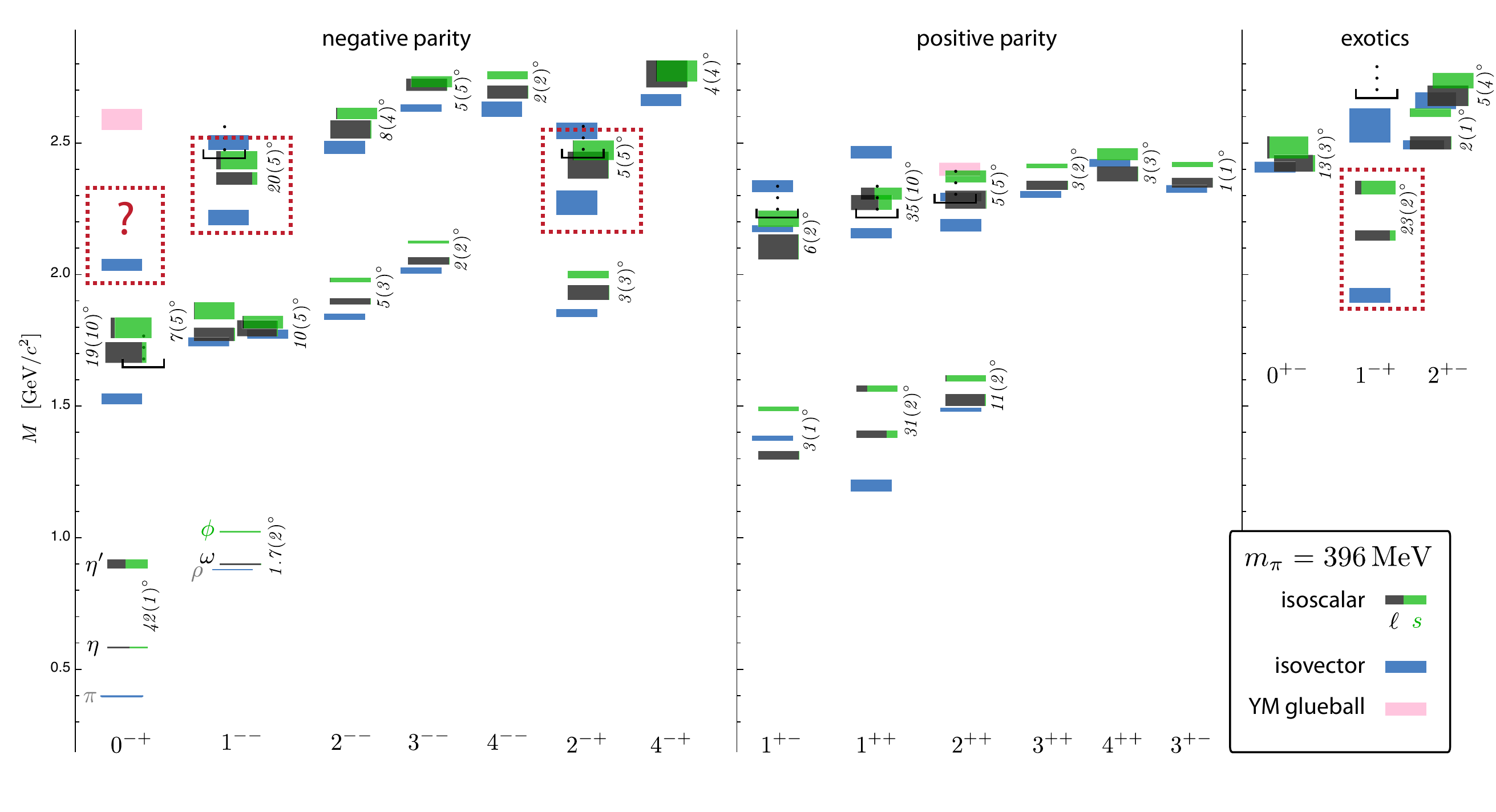}
\caption{\label{fig:lqcd_meson}A compilation of recent lattice QCD computations for both the
isoscalar and isovector light mesons from Ref.~\cite{Dudek:2011bn}, including 
$\ell\bar{\ell}$ $\left(|\ell\bar{\ell}\rangle\equiv (|u\bar{u}\rangle+|d\bar{d}\rangle)/\sqrt{2}\right)$ and
$s\bar{s}$ mixing angles (indicated in degrees).  The dynamical computation is
carried out with two flavors of quarks, light ($\ell$) and strange ($s$).  The $s$ quark
mass parameter is tuned to match physical $s\bar{s}$ masses, while the light quark mass parameters are
heavier, giving a pion mass of 396~MeV.  The black brackets with upward ellipses represent regions of the spectrum
where present techniques make it difficult to extract additional states.  
The dotted boxes indicate states that are interpreted as the lightest
hybrid multiplet -- the extraction of clear $0^{-+}$ states in this region is difficult in practice.}
\end{center}
\end{figure*}

As shown in Fig.~\ref{fig:lqcd_meson}, these calculations show a clear and detailed spectrum of exotic 
$J^{PC}$ mesons, with a lightest $1^{-+}$ state lying a few hundred MeV below a $0^{+-}$ 
and two $2^{+-}$ states. Through analysis of the matrix elements 
$\langle \mathfrak{n} | \mathcal{O}^\dag |0 \rangle$ for a range of different quark-gluon 
constructions, $\mathcal{O}$, we can infer \cite{Dudek:2011bn} that although the bulk of the 
non-exotic $J^{PC}$ spectrum has the expected systematics of a $q\bar{q}$ bound 
state system, some states are only interpolated strongly by operators featuring non-trivial 
gluonic constructions. One may interpret these states as non-exotic hybrid mesons, and by combining them with
the spectrum of exotics, it is possible to isolate the lightest hybrid supermultiplet of $(0,1,2)^{-+}$ and $1^{--}$ states at a mass 
roughly 1.3 GeV heavier than the $\rho$ meson. The form of the operator that has the strongest overlap 
onto these states has an $S$-wave $q\bar{q}$ pair in a color octet configuration and 
an exotic gluonic field in a color octet with $J_g^{P_gC_g}=1^{+-}$, a \emph{chromomagnetic} 
configuration. The heavier $(0,2)^{+-}$ states, along with some positive parity 
non-exotic states, appear to correspond to a $P$-wave coupling of the $q\bar{q}$ pair 
to the same chromomagnetic gluonic excitation.

A similar calculation for isoscalar states uses both $u\bar{u} + d\bar{d}$ and $s\bar{s}$ 
constructions and is able to extract both the spectrum of states and also their hidden 
flavor mixing.  (See Fig.~\ref{fig:lqcd_meson}.)  The basic experimental pattern of significant 
mixing in the $0^{-+}$ and $1^{++}$ 
channels and small mixing elsewhere is reproduced, and for the first time, we are able to say 
something about the degree of mixing for exotic-$J^{PC}$ states.  In order to
probe this mixing experimentally, it is essential to be able to reconstruct decays to both strange and non-strange
final state hadrons.

\subsection{The importance of kaon identification}
\label{Sec:KaonMotivation}

The primary goal of the \gx~experiment is to conduct a definitive mapping of states in the light 
meson sector, with an emphasis on searching for exotic mesons.  Ideally, we would like to produce 
the experimental analogue of the lattice QCD spectrum pictured in Fig.~\ref{fig:lqcd_meson}, enabling 
a direct test of our understanding of gluonic excitations in QCD.  In order to achieve this, one must be 
able to reconstruct strange final states, as observing decay patterns of mesons has been one of the 
primary mechanisms of inferring quark flavor content.  An example of this can be seen by examining 
the two lightest isoscalar $2^{++}$ mesons in the lattice QCD calculation in Fig.~\ref{fig:lqcd_meson}.  
The two states have nearly pure flavors, with only a small ($11^\circ$) mixing in the $\ell\bar{\ell}$ and 
$s\bar{s}$ basis.  A natural experimental assignment for these two states are the $f_2(1270)$ and the 
$f_2'(1525)$.  An experimental study of the branching ratios shows that 
$\mathcal{B}(f_2(1270)\to K K)/\mathcal{B}(f_2(1270)\to \pi\pi)\approx 0.05$ and 
$\mathcal{B}(f_2'(1525)\to \pi\pi)/\mathcal{B}(f_2'(1525)\to K K) \approx 0.009$~\cite{Beringer:1900zz}, 
which support the prediction of an $f_2(1270)$ ($f_2'(1525)$) with a dominant $\ell\bar{\ell}$ ($s\bar{s}$) 
component.  By studying both strange and non-strange decay modes of mesons, \gx~hopes to provide
similarly valuable experimental data to aid in the interpretation of the hybrid spectrum.

\subsubsection{Exotic $s\bar{s}$ states}

While most experimental efforts to date have focused on the lightest isovector exotic meson, the 
$J^{PC}=1^{-+}$ $\pi_1(1600)$, lattice QCD clearly predicts a rich spectrum of both isovector and
isoscalar exotics, the latter of which may have mixed $\ell\bar{\ell}$ and $s\bar{s}$ flavor content.
A compilation of the ``ground state" exotic hybrids is listed in Table~\ref{tab:exotic_modes}, along with
theoretical estimates for masses, widths, and key decay modes.  It is expected that initial searches 
with the baseline \gx~hardware will target primarily the $\pi_1$ state.  Searches for the $\eta_1$, $h_0$, and $b_2$ may be 
statistically challenging, depending on the masses of these states and the production cross sections.
With increased statistics and kaon identification, the search scope can be broadened to include these
heavier exotic states in addition to the $s\bar{s}$ states:  $\eta_1'$, $h_0'$, and $h_2'$.  The $\eta_1'$ and
$h_2'$ are particularly interesting because some models predict these states to be relatively narrow, and
that they should decay through well-established kaon resonances.

Observations of various $\pi_1$ states have been reported in the literature for over fifteen
years, with some analyses based on millions of events~\cite{Meyer:2010ku}.  
However, it is safe to say that there exists
a fair amount of skepticism regarding the assertion that unambiguous experimental evidence
exists for exotic hybrid mesons.  If the scope of exotic searches with \gx~is narrowed to only include
the lightest isovector $\pi_1$ state, the ability for \gx~to comprehensively address the question of 
the existence of gluonic excitations in QCD is greatly diminished.  On the other hand, clear identification of 
all exotic members of the lightest hybrid multiplet, the three exotic $\pi_1^{\pm,0}$ states and the exotic 
$\eta_1$ and $\eta_1'$, which can only be done by systematically studying a large number of
strange and non-strange decay modes, would provide unambiguous experimental confirmation of 
exotic mesons. A study of decays to kaon final states could demonstrate that the $\eta_1$ candidate 
is dominantly $\ell\bar{\ell}$ while the $\eta_1'$ candidate is $s\bar{s}$, as predicted by initial lattice 
QCD calculations.  Such a discovery would represent a substantial improvement in the experimental 
understanding of exotics.  In addition, further identification of members of the 
$0^{+-}$ and $2^{+-}$ nonets as well as measuring the mass splittings with the $1^{+-}$ states will 
validate the lattice QCD inspired phenomenological picture of these states as $P$-wave 
couplings of a gluonic field with a color-octet $q\bar{q}$ system.

\begin{table*}\centering
\caption{\label{tab:exotic_modes}
A compilation of exotic quantum number hybrid approximate masses, widths, and decay predictions.
Masses are estimated from dynamical LQCD calculations with $M_\pi = 396~\mathrm{MeV}/c^2$~\cite{Dudek:2011bn}.  
The PSS (Page, Swanson and Szczepaniak) and IKP (Isgur, Kokoski and Paton) model widths are from Ref.~\cite{Page:1998gz},
with the IKP calculation based on the model in Ref.~\cite{Isgur:1985vy}.  The total widths have a mass 
dependence, and Ref.~\cite{Page:1998gz} uses somewhat different mass values than suggested by the most recent
lattice calculations~\cite{Dudek:2011bn}.
Those final states marked with a dagger ($\dagger$) are ideal for experimental exploration 
because there are relatively few stable particles in the final state or moderately narrow 
intermediate resonances that may reduce combinatoric background.  
(We consider $\eta$, $\eta^\prime$, and $\omega$ to be stable final state particles.)}
\begin{tabular}{ccccccc}\hline\hline
 &  Approximate & $J^{PC}$ & \multicolumn{2}{c}{Total Width (MeV)} & 
Relevant Decays & Final States \\ 
         & Mass (MeV) &  & PSS & IKP & &  \\ \hline
$\pi_{1}$   & 1900 & $1^{-+}$ &  $80-170$ & $120$ & 
$b_{1}\pi^\dagger$, $\rho\pi^\dagger$, $f_{1}\pi^\dagger$, $a_{1}\eta$, $\eta^\prime\pi^\dagger$ & $\omega\pi\pi^\dagger$, $3\pi^\dagger$, $5\pi$, $\eta 3\pi^\dagger$, $\eta^\prime\pi^\dagger$  \\
$\eta_{1}$  & 2100 & $1^{-+}$ &  $60-160$ & $110$ &
$a_{1}\pi$, $f_{1}\eta^\dagger$, $\pi(1300)\pi$ & $4\pi$, $\eta 4\pi$, $\eta\eta\pi\pi^\dagger$ \\ 
$\eta^{\prime}_{1}$ & 2300 & $1^{-+}$ &  $100-220$ & $170$ &
$K_{1}(1400)K^\dagger$, $K_{1}(1270)K^\dagger$, $K^{*}K^\dagger$ & $KK\pi\pi^\dagger$, $KK\pi^\dagger$, $KK\omega^\dagger$ \\ \hline
$b_{0}$     & 2400 & $0^{+-}$ & $250-430$ & $670$ &
$\pi(1300)\pi$, $h_{1}\pi$ & $4\pi$ \\
$h_{0}$     &  2400 & $0^{+-}$ & $60-260$  & $90$  &
$b_{1}\pi^\dagger$, $h_{1}\eta$, $K(1460)K$ & $\omega\pi\pi^\dagger$, $\eta3\pi$, $KK\pi\pi$ \\
$h^{\prime}_{0}$    & 2500& $0^{+-}$ & $260-490$ & $430$ &
$K(1460)K$, $K_{1}(1270)K^\dagger$, $h_{1}\eta$ & $KK\pi\pi^\dagger$, $\eta3\pi$ \\ \hline
$b_{2}$     & 2500 & $2^{+-}$ &    $10$ & $250$ &
$a_{2}\pi^\dagger$, $a_{1}\pi$, $h_{1}\pi$ & $4\pi$, $\eta\pi\pi^\dagger$ \\
$h_{2}$     & 2500 & $2^{+-}$ &    $10$ & $170$ &
$b_{1}\pi^\dagger$, $\rho\pi^\dagger$ & $\omega\pi\pi^\dagger$, $3\pi^\dagger$ \\
$h^{\prime}_{2}$  & 2600 & $2^{+-}$ &    $10-20$ &  $80$ &
$K_{1}(1400)K^\dagger$, $K_{1}(1270)K^\dagger$, $K^{*}_{2}K^\dagger$ & $KK\pi\pi^\dagger$, $KK\pi^\dagger$\\
\hline\hline
\end{tabular}
\end{table*}

\subsubsection{Non-exotic $s\bar{s}$ mesons}
\label{sec:normalss}

As discussed above, one expects the lowest-mass hybrid multiplet to contain 
$(0,1,2)^{-+}$ states and a $1^{--}$ state that all have about the same mass and correspond to an 
$S$-wave $q\bar{q}$ pair coupling to the gluonic field in a $P$-wave.  For each $J^{PC}$ we 
expect an isovector triplet and a pair of isoscalar states in the spectrum.  Of the four sets of $J^{PC}$
values for the lightest hybrids, only the $1^{-+}$ is exotic. The other hybrid states will appear as 
supernumerary states in the spectrum of conventional mesons. The ability to clearly identify these 
states depends on having a thorough and complete understanding of the meson spectrum.  Like 
searching for exotics, a complete mapping of the spectrum of non-exotic mesons requires the 
ability to systematically study many strange and non-strange final states.  Other experiments, such 
as BESIII or COMPASS, are carefully studying this with very high statistics data samples
and have outstanding capability to cleanly study any possible final state.  While the production 
mechanism of \gx~is complementary to that of charmonium decay or pion beam production and is 
thought to enhance hybrid production, it is essential that the detector capability and statistical 
precision of the data set be competitive with other contemporary experiments in order to maximize the collective
experimental knowledge of the meson spectrum. 

Given the numerous discoveries of unexpected, apparently non-$q\bar{q}$ states in the charmonium spectrum,
a state that has attracted a lot of attention in the $s\bar{s}$ spectrum is the $Y(2175)$, which is
assumed to be an $s\bar{s}$ vector meson ($1^{--}$).  The $Y(2175)$ (also denoted as $\phi(2170)$) has been observed to decay to 
$\pi\pi\phi$ and has been produced in both $J/\psi$ decays~\cite{Ablikim:2007ab} and $e^+e^-$ 
collisions~\cite{Aubert:2006bu,Shen:2009zze}. The state is a proposed analogue of the $Y(4260)$ in 
charmonium, a state that is also about 1.2 GeV heavier than the ground state triplet ($J/\psi$)
and has a similar decay mode:  $Y(4260)\to\pi\pi J/\psi$~\cite{Aubert:2005rm,Coan:2006rv,He:2006kg,Yuan:2007sj}.  
The $Y(4260)$ has no obvious interpretation
in the charmonium spectrum and has been speculated to be a hybrid 
meson~\cite{Close:2005iz,Zhu:2005hp,Kou:2005gt,Luo:2005zg}, which, by loose analogy,
leads to the implication that the $Y(2175)$ might also be a hybrid candidate.  It should be noted
that the spectrum of $1^{--}$ $s\bar{s}$ mesons is not as well-defined experimentally as the $c\bar{c}$ system; 
therefore, it is not clear that the $Y(2175)$ is a supernumerary state.  However, \gx~is ideally suited
to study this system.  We know that vector mesons are copiously produced in photoproduction; therefore,
with the ability to identify kaons, a precision study of the $1^{--}$ $s\bar{s}$ spectrum can be conducted
with \gx.  Some have predicted~\cite{Ding:2007pc} that the potential hybrid nature of the $Y(2175)$ can be explored by studying ratios of branching fractions into various kaonic final states.  
In addition, should \gx~be able to conclude that the $Y(2175)$ is in fact a supernumerary 
vector meson, then a search can be made for the exotic $1^{-+}$ $s\bar{s}$ member of the multiplet 
($\eta_1'$),  evidence of which would provide a definitive interpretation of the $Y(2175)$ and likely have
implications on how one interprets charmonium data.

\section{The baseline \gx{} program}

\subsection{Detector design and construction}
\label{sec:baseline}

\begin{figure*}
\begin{center}
\includegraphics[width=0.8\linewidth]{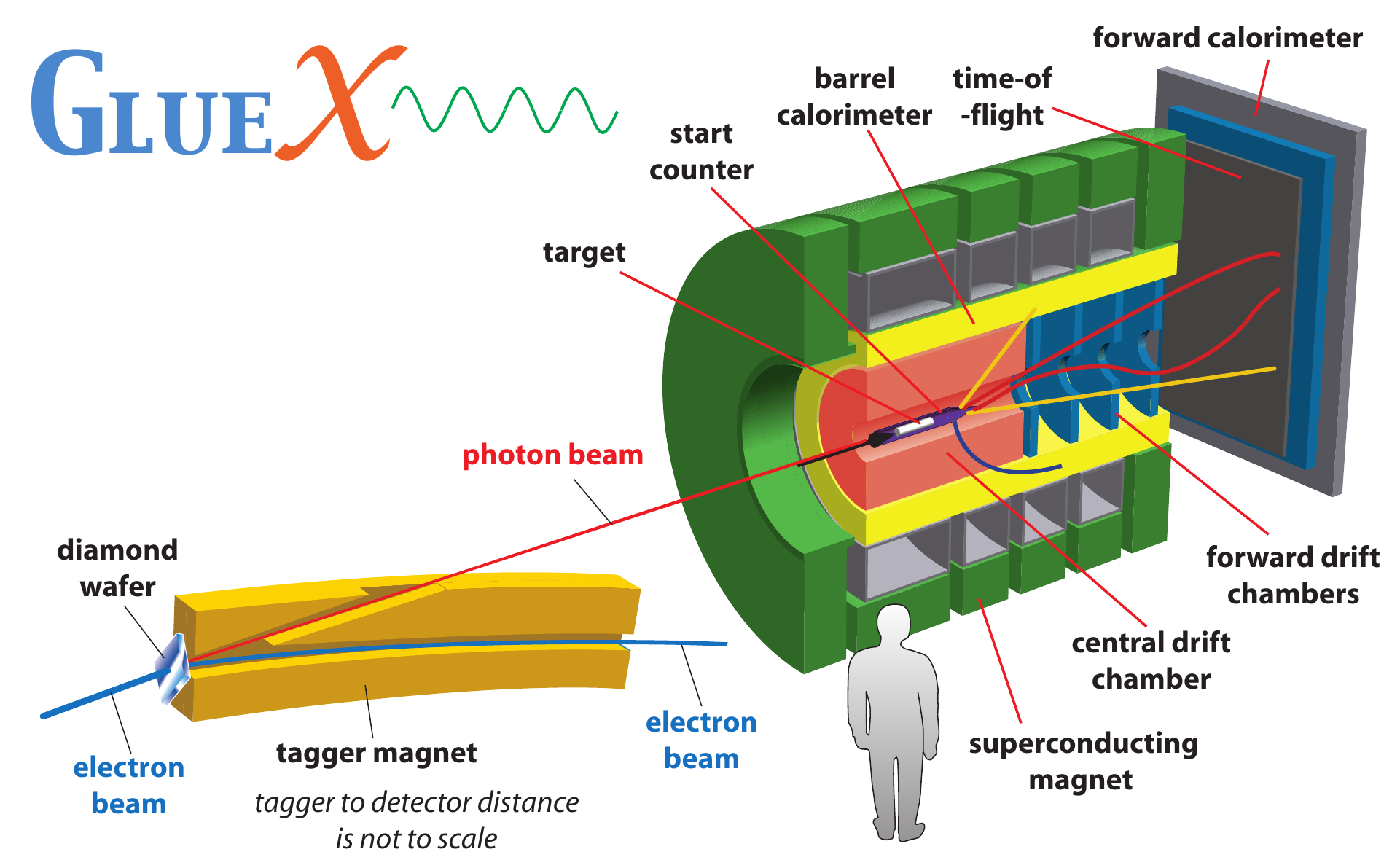}
\caption{\label{fig:detector}A schematic of the \gx~detector and beam.}
\end{center}
\end{figure*}

A schematic view of the \gx~detector is shown in Fig.~\ref{fig:detector}.  
The civil construction of Hall D is complete and the collaboration gained control of both
Hall~D and the Hall~D tagger hall in 2012.  All major detector subsystems have
been installed in Hall~D, and commissioning with beam is expected to begin
in the Fall of 2014.  The collaboration consists of over a 
hundred members, including representation from the theory community.

The \gx~photon beam originates from coherent bremsstrahlung radiation produced by the 12~GeV electron beam
impinging on a $20~\mu$m diamond wafer.  Orientation of the diamond and downstream collimation
produce a photon beam peaked in energy around 9~GeV with about 40\% linear polarization.
A coarse tagger tags a broad range of electron energy, while precision tagging in the coherent peak 
is performed by a tagger microscope.  A downstream pair spectrometer is utilized to measure photon 
conversions and determine the beam flux.  Polarization will be measured independently by measuring
the angular distribution of pair production in the field of an atomic electron (triplet photoproduction).

At the heart of the \gx~detector is the $2.2$~T superconducting solenoid, which provides
 the essential magnetic field for tracking.  The solenoidal geometry also has the benefit of reducing 
electromagnetic backgrounds in the detectors since low energy $e^+e^-$ pairs spiral within a small 
radius of the beamline.  Charged particle tracking is performed by two systems:  a central 
straw-tube drift chamber (CDC) and four six-plane forward drift chamber (FDC) packages.  
The CDC is composed of 28 layers of 1.5-m-long 
straw tubes.  The chamber provides $r-\phi$ measurements for charged tracks.  Sixteen of the 28 layers 
have a $6^\circ$ stereo angle to supply $z$ measurements.  Each FDC package is composed of six planes 
of anode wires.  The cathode strips on either side of the anode cross at $\pm75^\circ$ angles, providing
a two-dimensional intersection point on each plane. 

Like tracking, the \gx~calorimetry system consists of two detectors:  a barrel calorimeter with a 
cylindrical geometry (BCAL) and a forward lead-glass calorimeter with a planar geometry (FCAL).  
The primary goal of these systems is to detect photons that can be used to reconstruct $\pi^0$'s and 
$\eta$'s, which are produced in the decays of heavier states.  The BCAL is a relatively high-resolution 
sampling calorimeter, based on 1~mm double-clad Kuraray scintillating fibers embedded in a lead matrix.
It is composed of 48 four-meter-long modules; each module having a radial thickness of 14.7 radiation lengths.
Modules are read out on each end by silicon SiPMs, which are not adversely 
affected by the high magnetic field in the proximity of the \gx~solenoid flux return.
The forward calorimeter is composed of 2800 lead glass modules, stacked in a circular array.  Each
bar is coupled to a conventional phototube.  The fractional energy resolution of the combined calorimetry
system $\delta(E)/E$ is approximately $5\%$-$6\%/\sqrt{E~[\mathrm{GeV}]}$. 

The particle ID capabilities of \gx{} are derived from several subsystems.  A dedicated 
forward time-of-flight wall (TOF), which is constructed from two planes of 2.5-cm-thick scintillator 
bars, provides about $70$~ps timing resolution on forward-going tracks within about $10^\circ$ of 
the beam axis.  This information is complemented by time-of-flight data from the BCAL and specific
 ionization ($dE/dx$) measured with the CDC, both of which are particularly important for identifying 
the recoil proton in $\gamma p\to Xp$ reactions.  Finally, identification of the beam bunch, which 
is critical for timing measurements, is performed by a thin start counter that surrounds the target.

As of June 2014, the CDC, FDC, BCAL, FCAL, and TOF have all be assembled in the hall.
The beam-line instrumentation and the target and start counter package remain to be installed.
Both calorimeters are fully cabled and have successfully recorded cosmic ray tracks.  
Studies using the tracking chambers in conjunction with the BCAL to observe cosmic
ray events are planned.  Commissioning with beam is expected to take place in Fall 2014.

\subsection{Proposed run plan}

The GlueX physics program was presented initially to the Jefferson Lab Program Advisory Committee (PAC) in 2006~\cite{pac30}.
The first beam time allocations were made for the commissioning phases of GlueX after the presentation to the PAC in
2010~\cite{pac36}.  This allocation covered phases I-III of the run plan highlighted in Table~\ref{tab:params}.
In 2012, the collaboration presented a proposal to the PAC~\cite{pac39} 
for running at design intensity with enhanced particle
identification capability (noted as Phase IV+ in Table~\ref{tab:params}).  
The PAC conditionally approved
this proposal pending a final design of the particle identification hardware.  In 2013 the collaboration returned
to the PAC to present a proposal for running at design intensity with limited PID capability~\cite{pac40}, this was approved
by the PAC and 200 days of beam for Phase IV was granted.  

 In this document, we present our conceptual design for developing a FDIRC for
\gx{} using the BaBar components and are therefore seeking approval of Phase IV+, 
first proposed in 2012~\cite{pac39}.  Our goal is to pursue construction of this design on a 
time scale that allows us to merge both 
Phase IV and Phase IV+ into a single run.

\begin{table*}
\begin{center}
\caption{\label{tab:params}A table of relevant parameters for the various phases of \gx~running.}
\begin{tabular}{l|cccc|c}\hline\hline
 &\multicolumn{4}{c}{Approved} & Conditionally Approved \\
 & Phase I & Phase II & Phase III & Phase IV & Phase IV+ \\ \hline 
 Duration (PAC days) & 30 & 30 & 60 & 200 & 220\footnote{Twenty days are allocated for FDIRC commissioning.} \\
 Minimum electron energy (GeV) & 10 & 11 & 12 & 12 & 12 \\
 Average photon flux  ($\gamma$/s) & $10^6$ & $10^7$ & $10^7$ & $5 \times 10^7$  & $5 \times 10^7$ \\
 Level-one (hardware) trigger rate (kHz) & 2 & 20 & 20 & 200 & 200 \\
 Raw Data Volume (TB)\footnote{Phase IV(+)  assume a level-three software trigger is implemented.} & 60 & 600 & 1200 & 2300 & 2300 \\
\hline\hline
 \end{tabular}
 \end{center}
 \end{table*}


\section{An FDIRC for \gx{}:  Conceptual Design}

In the following section we discuss the conceptual design of a DIRC particle identification detector that is built from the BaBar DIRC components.

\subsection{Mechanical design and optics}

  
The world's first DIRC detector was developed and utilized by the BaBar experiment. It provided excellent particle identification performance up to about 4 GeV/c~\cite{NIM}.
The radiator of the BaBar DIRC consisted of a barrel made up of twelve boxes each containing twelve synthetic fused silica (henceforth referred to as quartz\footnote{In this document, we will refer to synthetic fused silica as quartz for the sake of brevity; however, it is worth noting that quartz is birefringent and, thus, not suitable for use in the DIRC.}) bars.  Quartz was chosen because of the following properties of the material: it has a large index of refraction ($n$) and a small chromatic dispersion; it has a long attenuation length; it is highly resistant to ionizing radiation; and it is possible to polish its surface.  Each box is hermetically sealed and nitrogen gas constantly flows through the box to prevent any contamination which would compromise the preservation of the Cherenkov angle by total internal reflection.

\begin{figure}[]
	\begin{center}
	\includegraphics[height=5.cm]{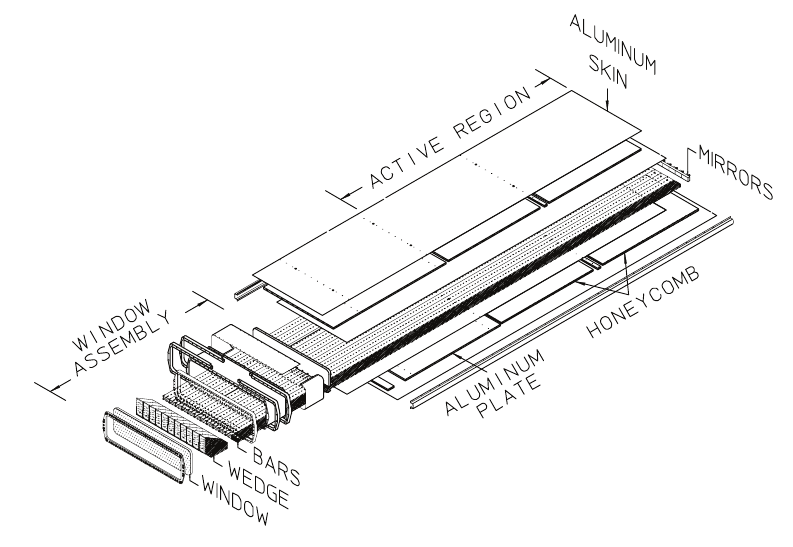}
	\caption{Schematic diagram of one BaBar box showing the 12 quartz bars (4.9m long), mirror ends, wedges and window ends~\cite{NIM}.
} 
	\label{BOXE}
	\end{center}
\end{figure}

Figure~\ref{BOXE} shows the assembly of one DIRC box.  Each bar is 17.25~mm thick, 35~mm wide and 4.9~m long and was produced by glueing four smaller bars end-to-end.  One end of each box is  coupled to a volume instrumented with photodetectors (the photon camera), while the other end has a mirror that reflects light back to the readout side.  The readout side also has a quartz wedge glued to it.  Neighboring bars are optically isolated by a 0.15~mm gap created using aluminum shims.

\begin{figure}[]
	\begin{center}
		\includegraphics[width=0.42\textwidth]{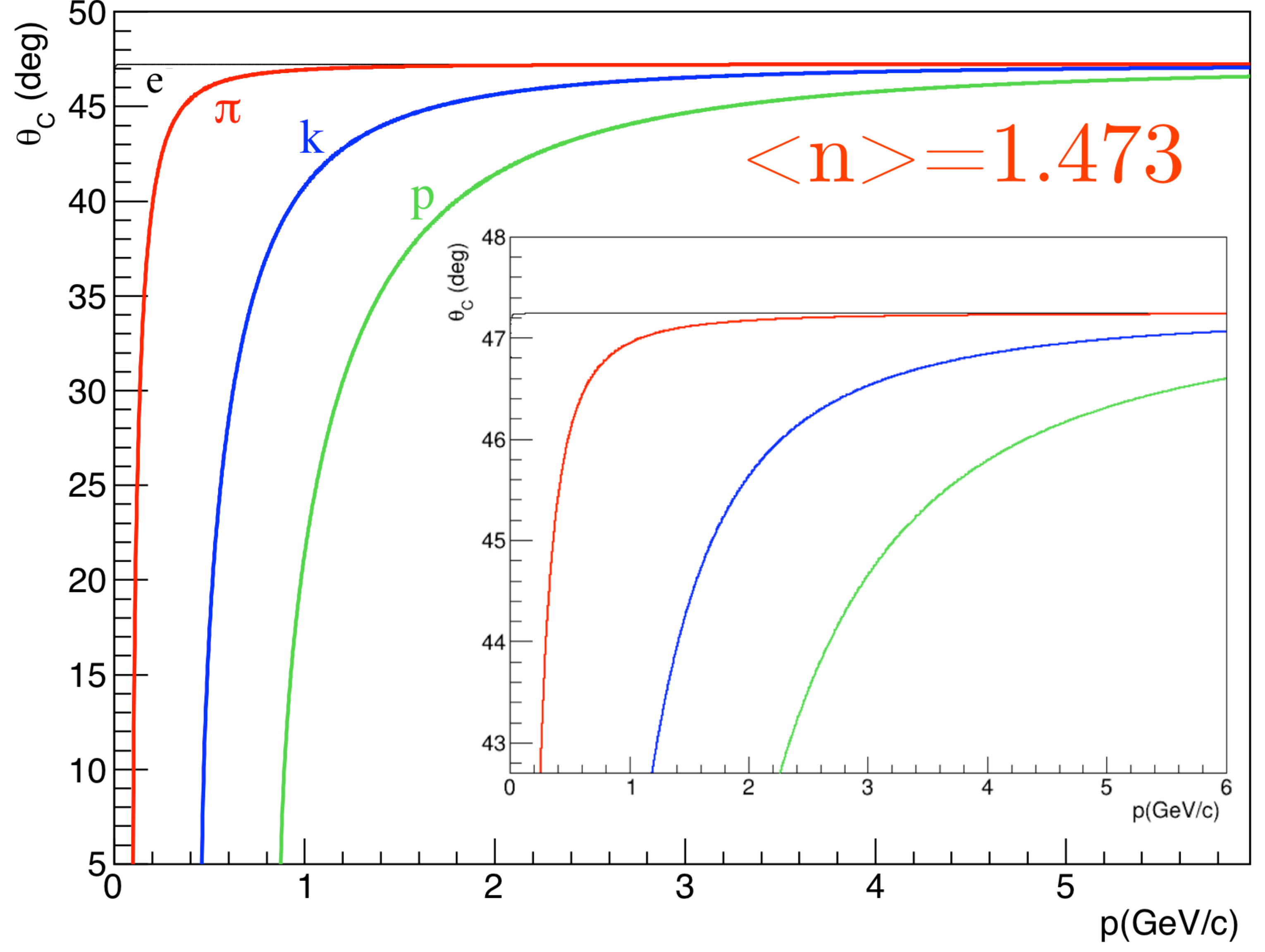}
		\includegraphics[width=0.5\textwidth]{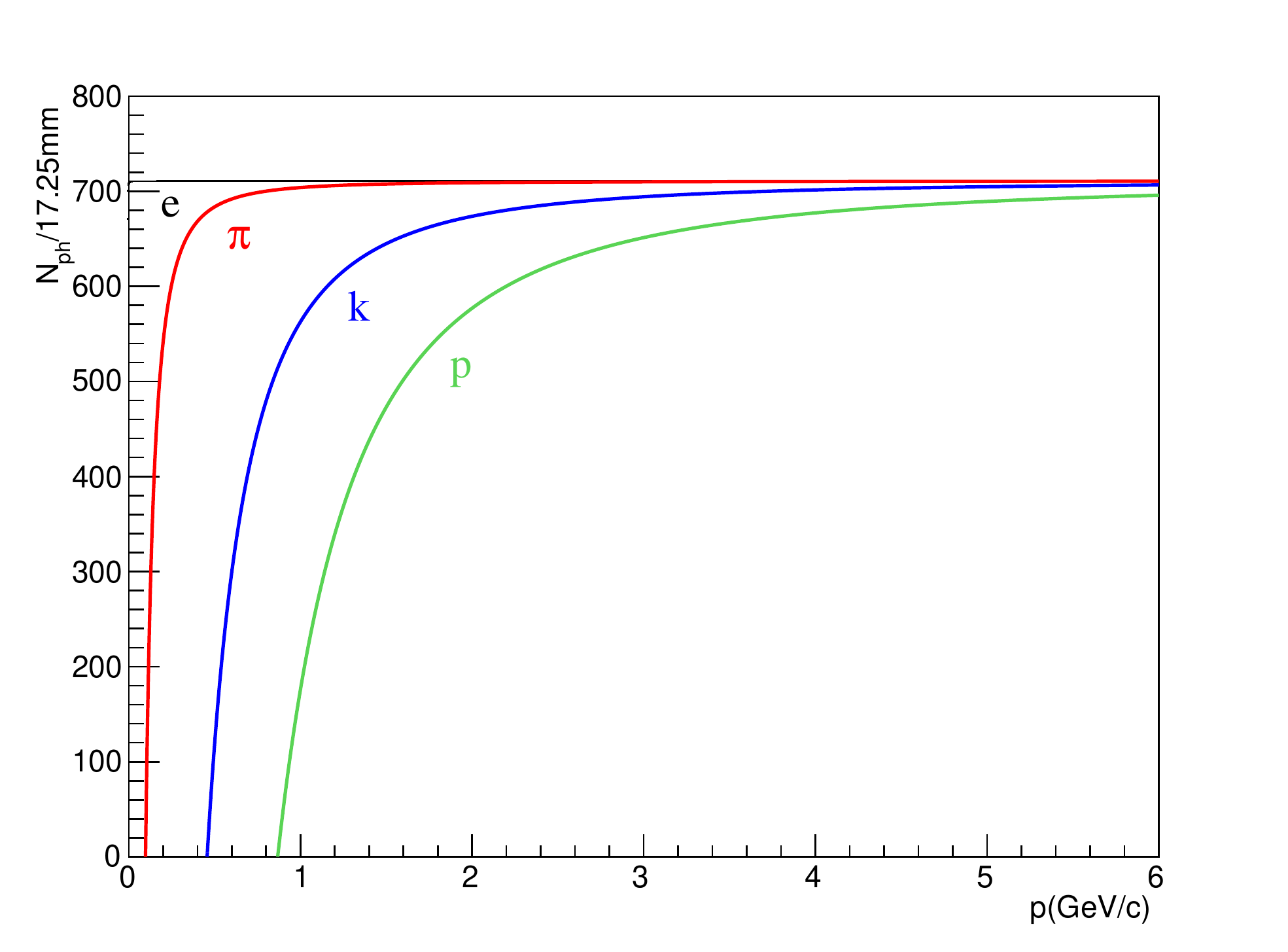}		
	\caption{(top) Cherenkov angle computed for four different charged particles (e, pion, kaon and proton), as a function of the momentum, for a fixed $< n>=1.473$ quartz index of refraction. (bottom) Number of Cherenkov photons produced in 17.25~mm of quartz material and within the photon wavelength range 300-600~nm, for different particles, as a function of their momentum. } 
	\label{figthresholds}
	\end{center}
\end{figure}

The quartz bars are used both as radiators and as light guides for the Cherenkov light trapped in the bars by total internal reflection. 
The number of photons produced per unit path length ($x$) of a particle with charge $q$ per unit photon wavelength ($\lambda$) can be estimated using the following expression:
 \begin{equation}
\frac{d^{2}N}{dxd\lambda}=\frac{2\pi \alpha q^{2}}{\lambda^{2}}\Bigg[1-\frac{1}{\beta^{2}n^{2}(\lambda)}\Bigg],
\label{CherenkoNumber}
\end{equation}
where $\alpha $ is the electromagnetic coupling constant and $\beta$ is the velocity of the incoming particle divided by the speed of light.  The index of refraction of the material $n$ is a function of the wavelength of the emitted photon.  The large index of refraction of the quartz material leads to a large number of Cherenkov photons produced within the wavelength acceptance of the DIRC (300-600~nm) (see Fig.~\ref{figthresholds}).

\begin{figure}[]
	\begin{center}
	\includegraphics[height=3.cm]{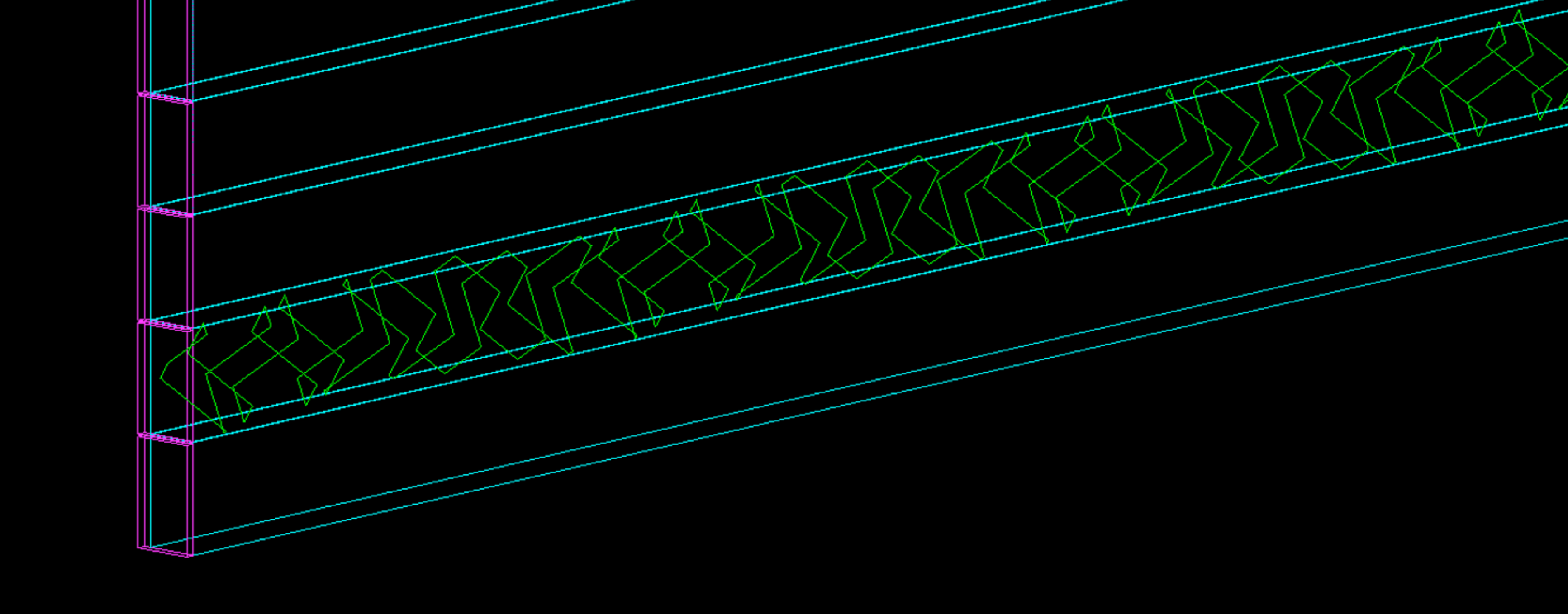}
	\caption{A single photon bouncing within a bar. If the photon angle is bigger than the critical angle, the light is internally reflected and the Cherenkov angle is preserved as the photon travels through the bar. } 
	\label{Sgphoton}
	\end{center}
\end{figure}

The Cherenkov light produced by a particle of velocity $\beta$ is emitted at an angle with respect to the direction of the particle's velocity, referred to as the Cherenkov angle ($\theta_{C}$), given by
 \begin{equation}
\cos \theta_{C}=\frac{1}{\beta n(\lambda)} .
\label{CherenkovAngle}
\end{equation}
Figure~\ref{figthresholds} shows the Cherenkov angle for different particle types.  One can see that for an average quartz index of refraction $\langle n \rangle =1.473$, the maximal Cherenkov angle is about $47^{\circ}$.  The critical angle for trapping light via total internal reflection at the quartz-nitrogen boundary, given by the ratio of the indices of refraction, is $\theta_{\rm critical} \approx 42.7^{\circ}$; thus, $\theta_C > \theta_{\rm critical}$ over most of the momentum range of interest for all particle types.
An example of the path followed by a single photon trapped within a bar is shown in Fig.~\ref{Sgphoton}.  The Cherenkov angle is preserved as the photon travels through the bar to the photodetectors.

\subsection{Focusing DIRC design}

The initial photon camera of the BaBar DIRC detector was very large and filled with 6000 liters of purified water. 
Recently, a new design with focusing mirrors has been developed that permits detection of  the Cherenkov light produced in the quartz radiator using a much more compact design~\cite{NIM2, NIM3, NIM4, NIM5, NIM6}. 
The focusing DIRC (FDIRC) was designed at SLAC with the constraint that the BaBar boxes cannot be altered~\cite{NIM7}. The new photon camera system is about 25 times smaller than the camera used at BaBar yet has approximately the same Cherenkov angle resolution.  The focusing design has the following advantages: the background rate is lower; the chromatic effect can be corrected for; the thickness of the bars can be corrected for; and the total number of photo-multipliers required is greatly reduced.  

\begin{figure}[]
	\begin{center}
			\includegraphics[width=0.45\textwidth]{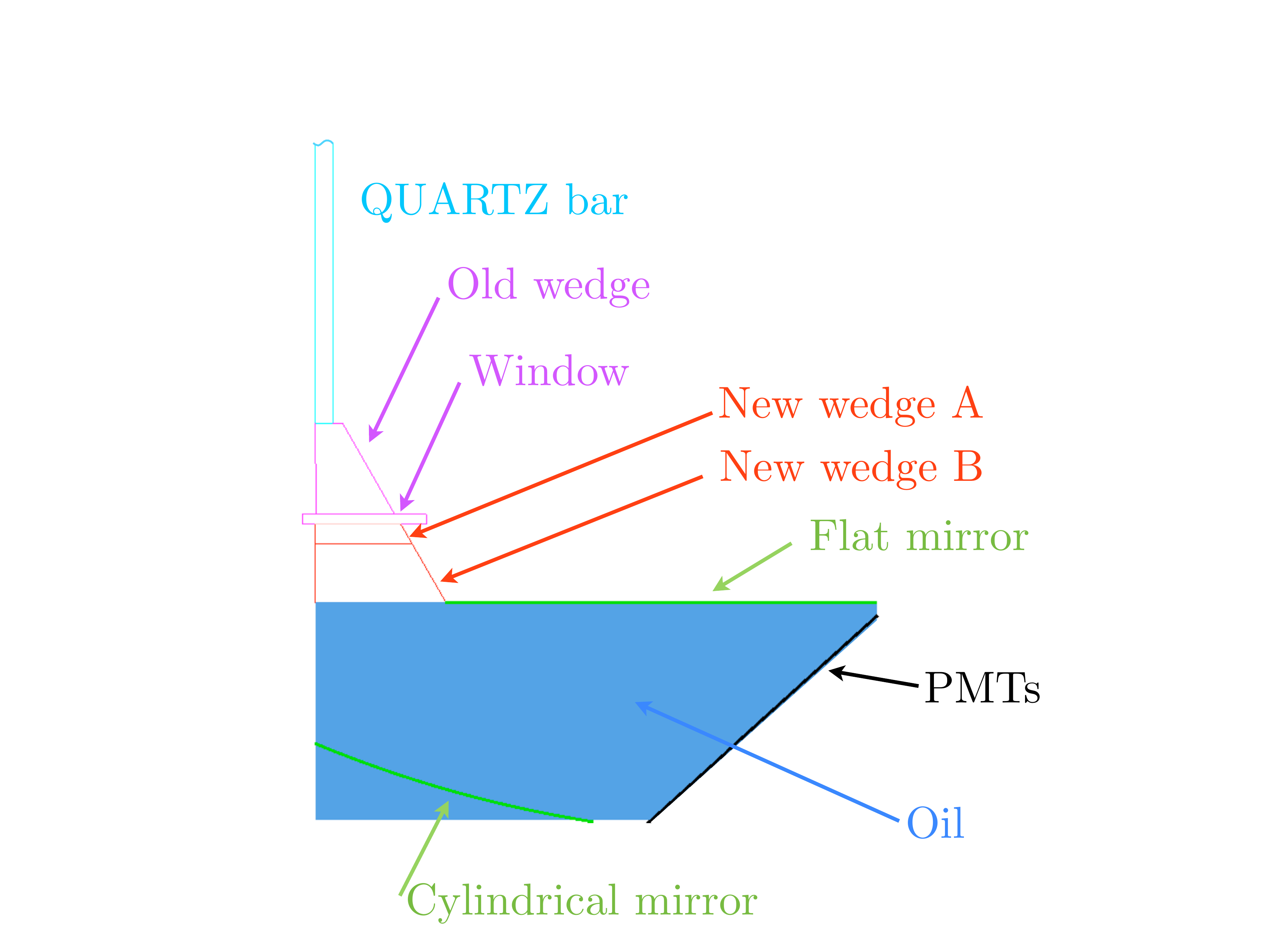}\quad 
      \includegraphics[width=0.45\textwidth]{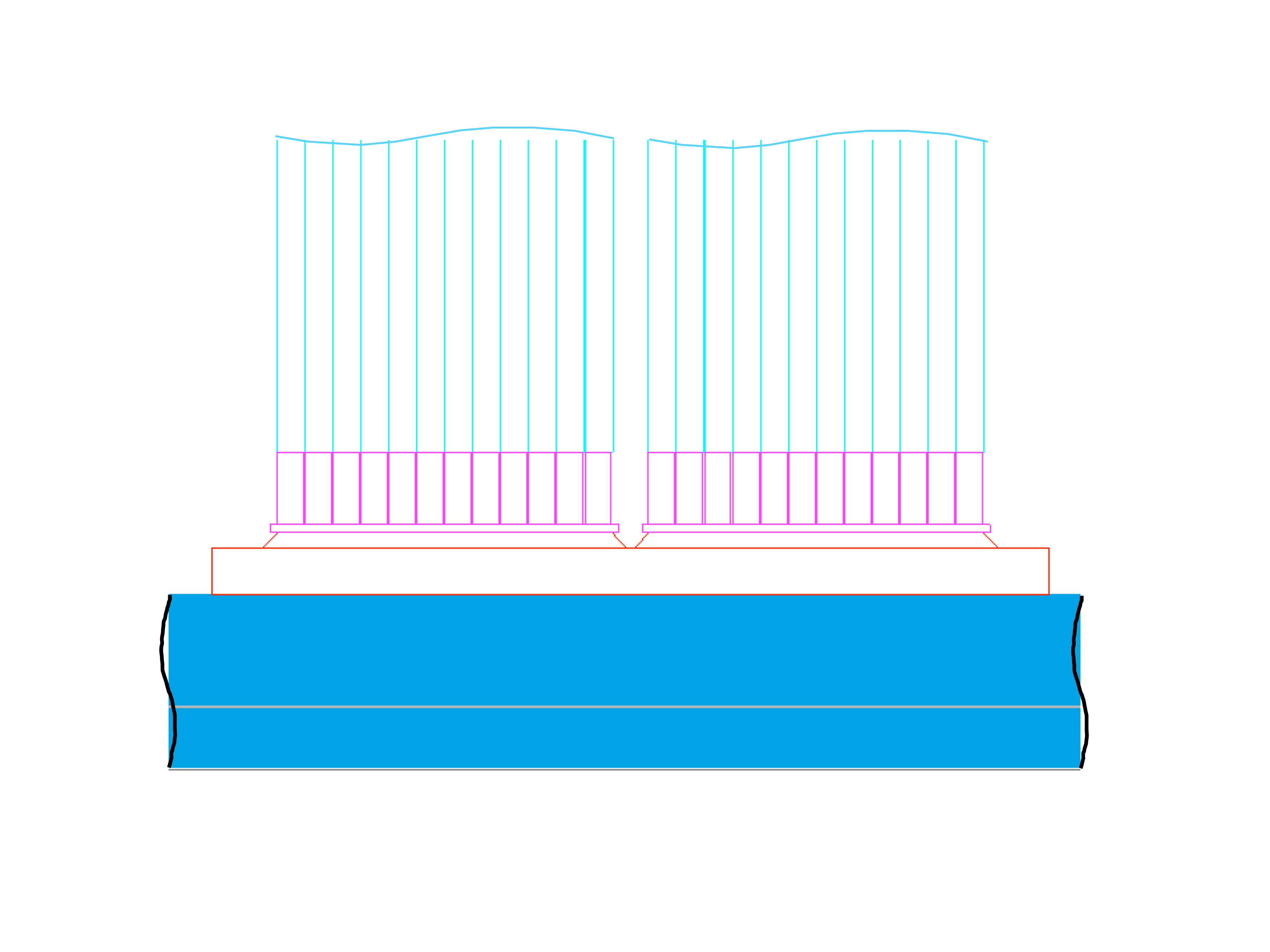}  \\
	\caption{Side and rear views of the FDIRC. The bars, old wedges and windows are part of the original BaBar boxes. The new focusing camera consists of the new wedges ($A$ and $B$) and the FOB containing the cylindrical and flat mirrors.} 
	\label{figFIRC}
	\end{center}
\end{figure}

Figure~\ref{figFIRC} shows the focusing scheme of the FDIRC prototype developed at SLAC \cite{NIM7}, adapted for use in the \gx~detector.  The photon camera consists of two new quartz wedges and a Focusing Oil Box (FOB). The bars, window and wedges are glued end-to-end and are all made of quartz. The FOB consists of cylindrical and flat mirrors to focus the light onto the PMT plane immersed in an oil bath.  The geometry of the FOB has not yet been optimized; it is shown here as a simplified rectangular volume. 

The cylindrical mirror removes the effect of the bar thickness on the Cherenkov angle resolution since parallel rays are focused on the same point on the detector plane. 
The flat mirror then reflects the light almost perpendicularly to the detector plane.  The total PMT surface to be covered is 2668~mm x 312~mm.
 
The addition of the new wedges ensures that the photons are reflected by the cylindrical mirrors.  
The first wedge (458~mm $\times$ 20~mm) is required to account for the flange geometry and support structure.  The second wedge (1051~mm $\times$ 58~mm) covers the full length of two neighboring boxes and eliminates side reflections that lead to ambiguities in the reconstruction and reduction in the FDIRC performance.
The expansion volume of the FOB is filled with a specific oil (CARGILLE 50350 or BICRON BC-599-14) whose index of refraction closely matches that of quartz; thus,  large refraction between the different media is avoided.

\subsection{\gx~FDIRC}

\begin{figure}[]
	\begin{center}
		\includegraphics[width=0.45\textwidth]{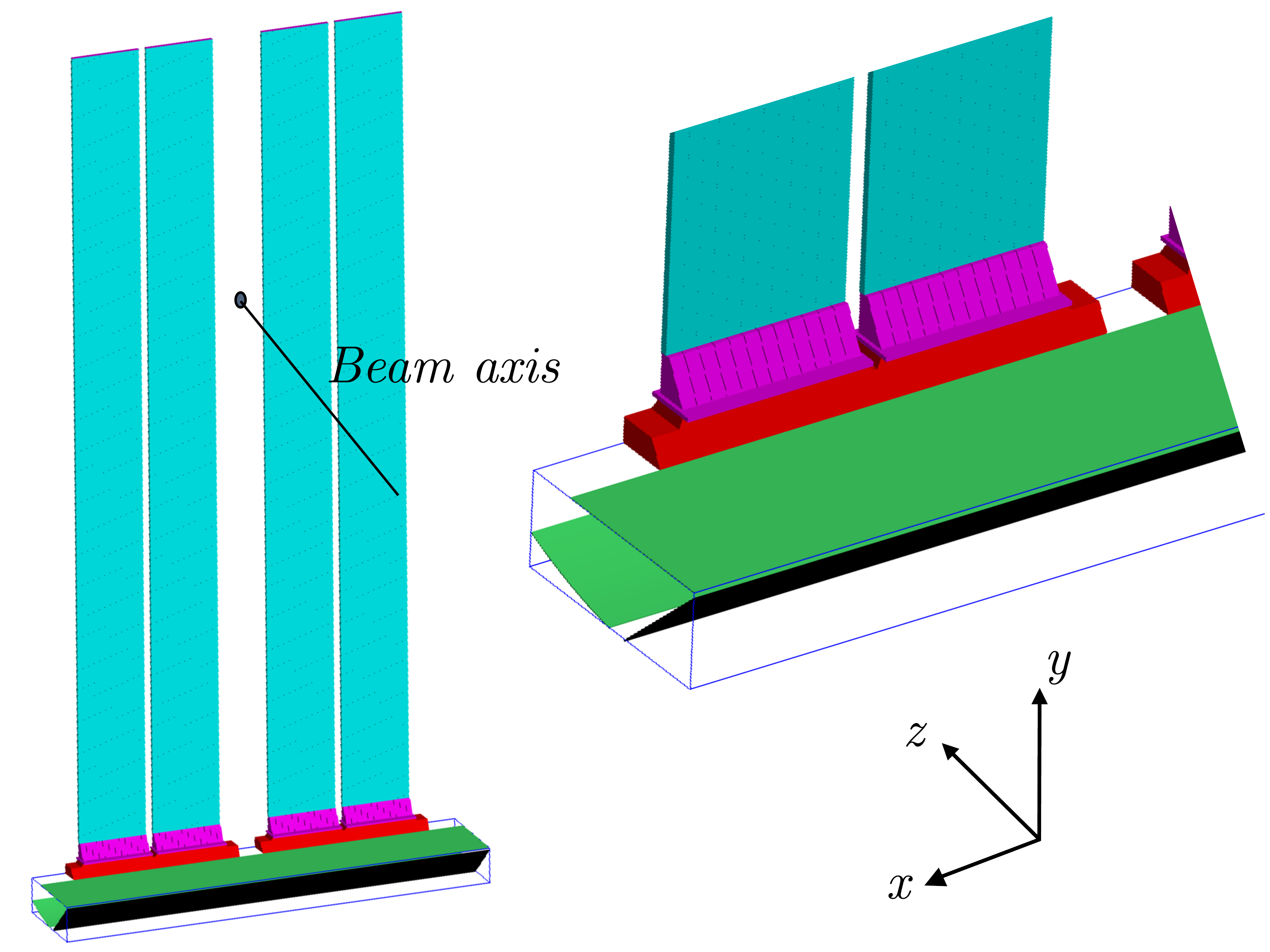}
	\caption{Schematic diagrams of the \gx~FDIRC detector. Four BaBar boxes are required to cover the full acceptance.  The bars are oriented vertically and placed symmetrically around the beam line.  One FOB covers the full length of the four boxes.} 
	\label{Design3D}
	\end{center}
\end{figure}

Figure~\ref{Design3D} shows a schematic diagram of the proposed \gx{} FDIRC detector.  The acceptance in the forward region of \gx{} is limited by the solenoid at $\approx 11^{\circ}$; thus, to fully cover the acceptance requires four BaBar boxes, each containing 12 quartz bars. The bar boxes will be oriented vertically in the \gx{} hall and placed symmetrically around the beam line.  A single FOB will be used for all of the bars and placed below the bar boxes.
The FDIRC detector will fit into the reserved space between the downstream end of the \gx{} detector solenoid and the time-of-flight wall. 

\begin{figure}[]
           \includegraphics[width=0.4\textwidth]{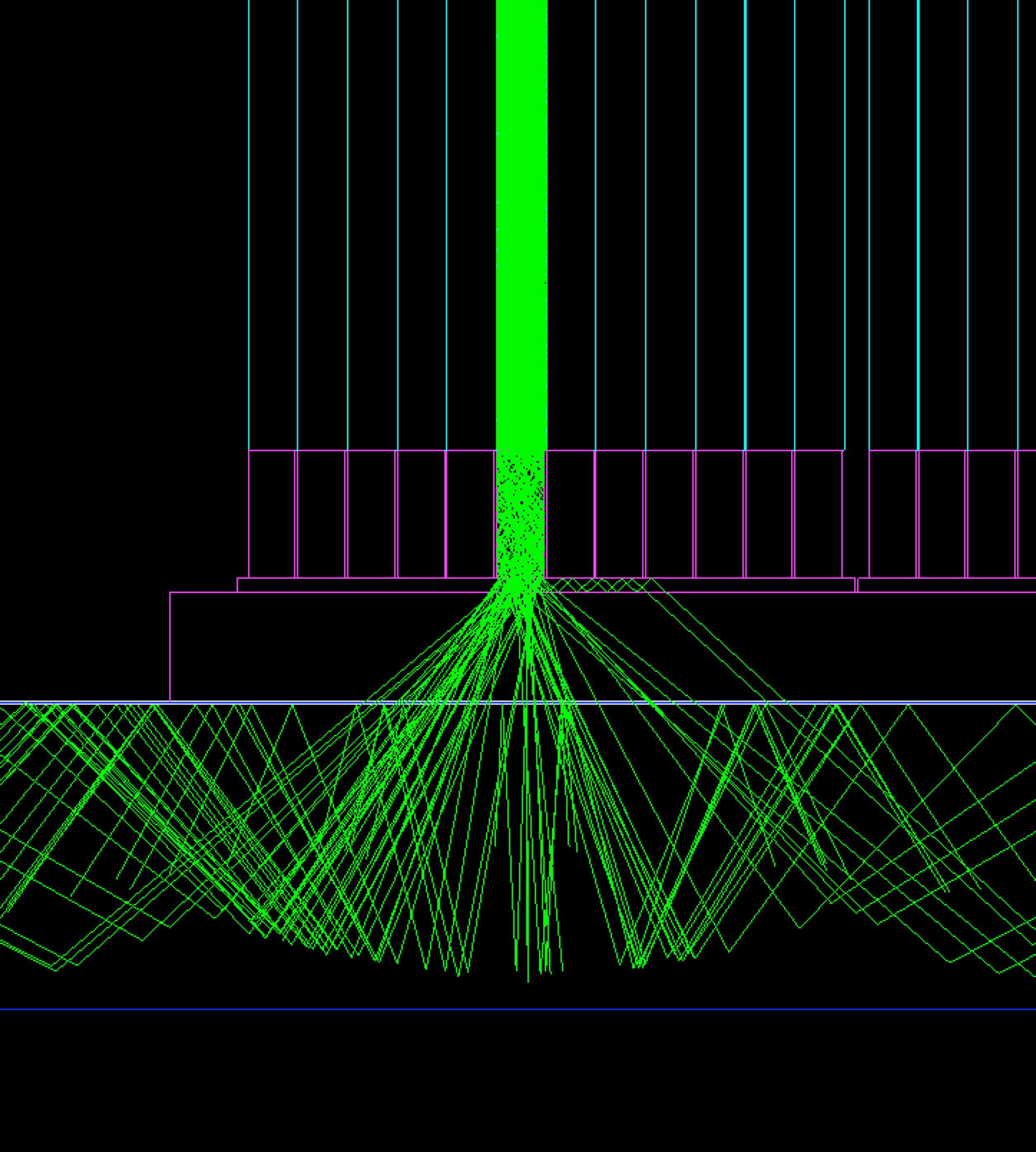}\quad 
            \includegraphics[width=0.4\textwidth]{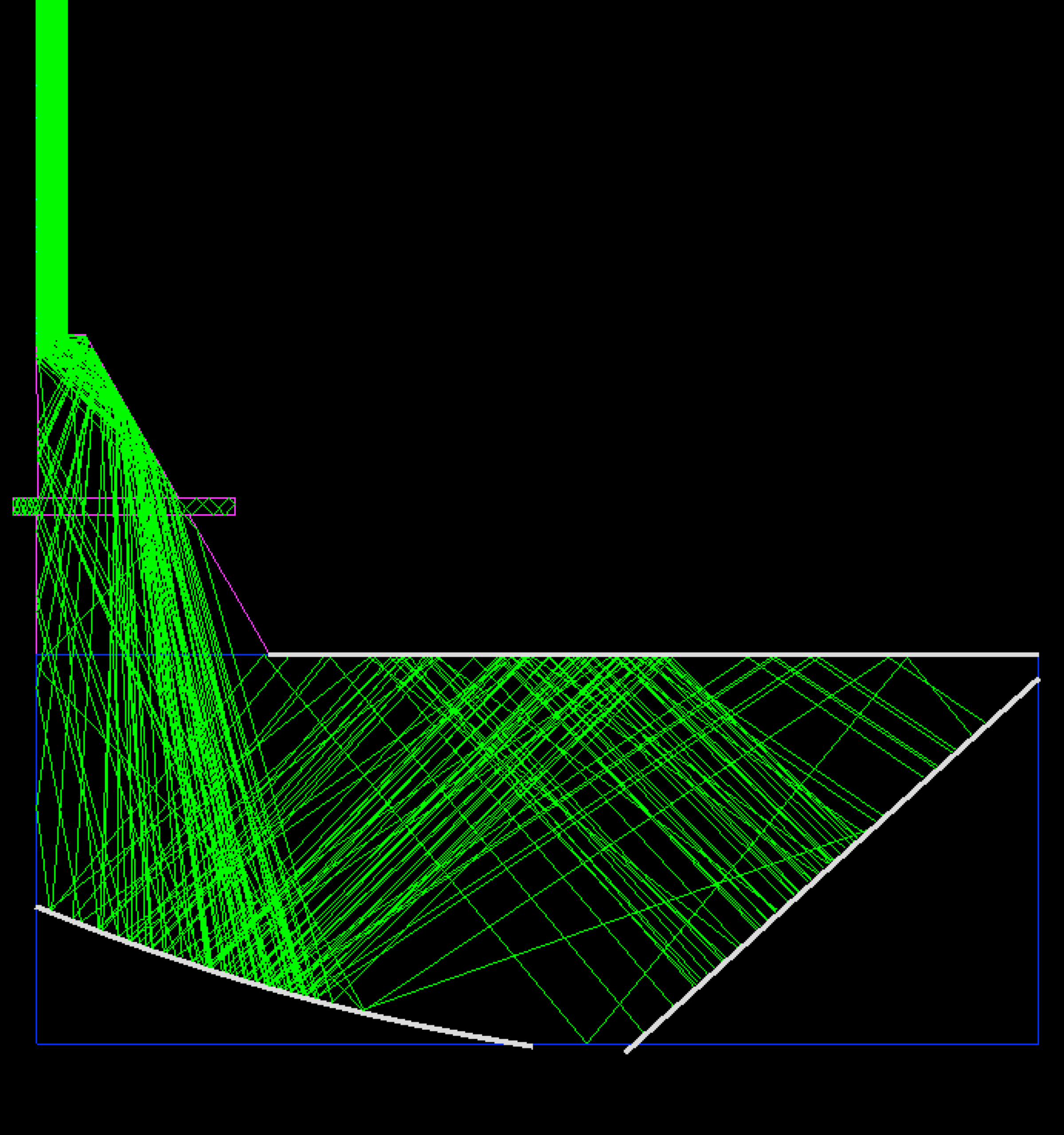}  \\
	\caption{View of the \gx~FDIRC detector from the rear (top panel) and side (bottom panel). The propagation of the Cherenkov light through the different elements of the detector is visible.  The rear view shows the collection of the light within a bar, while the side view shows the focusing scheme of the light on the PMTs surface.} 
	\label{1eventleftright}
\end{figure}

\begin{figure}[]
	\begin{center}
		\includegraphics[width=0.5\textwidth]{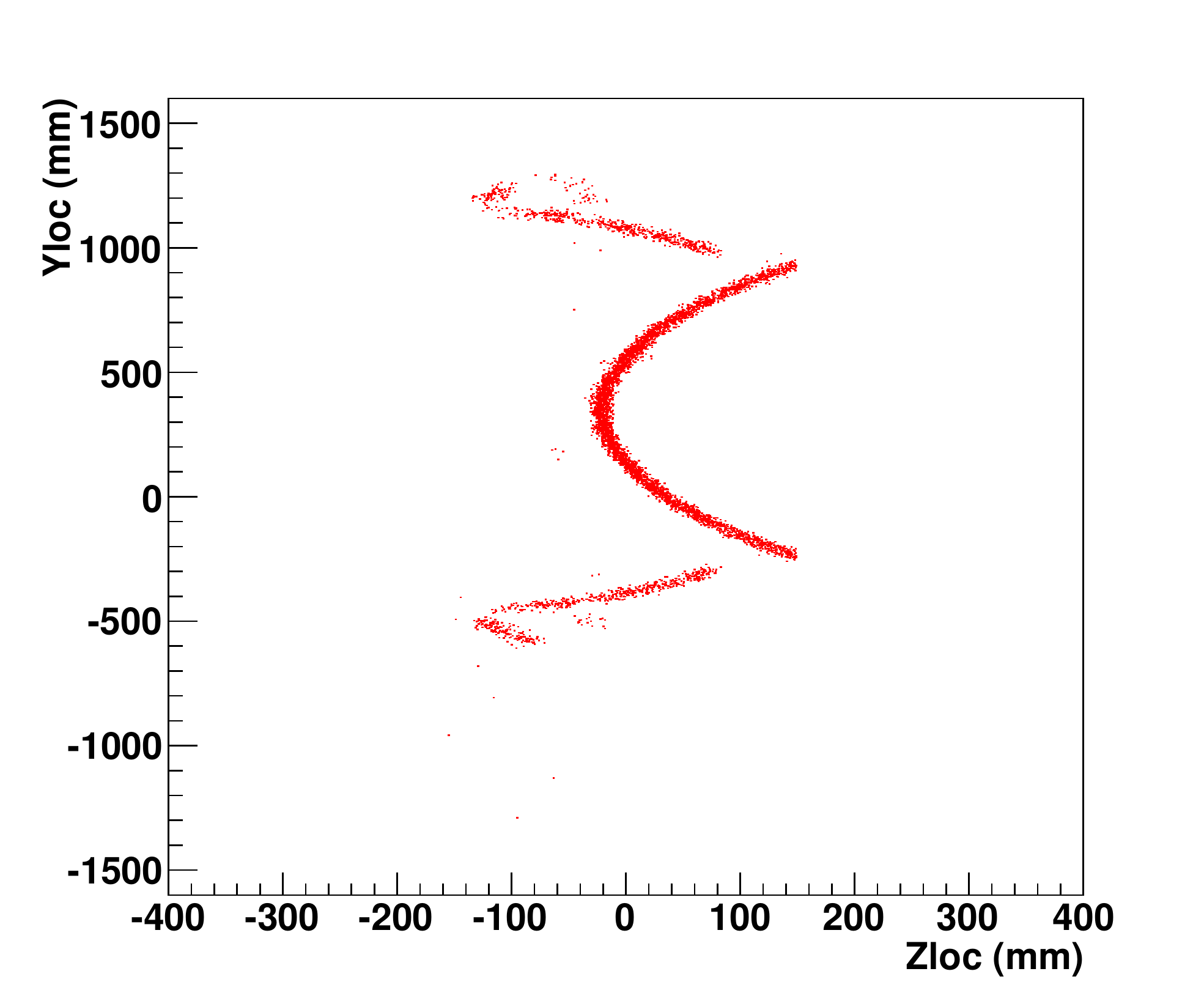}
		\includegraphics[width=0.5\textwidth]{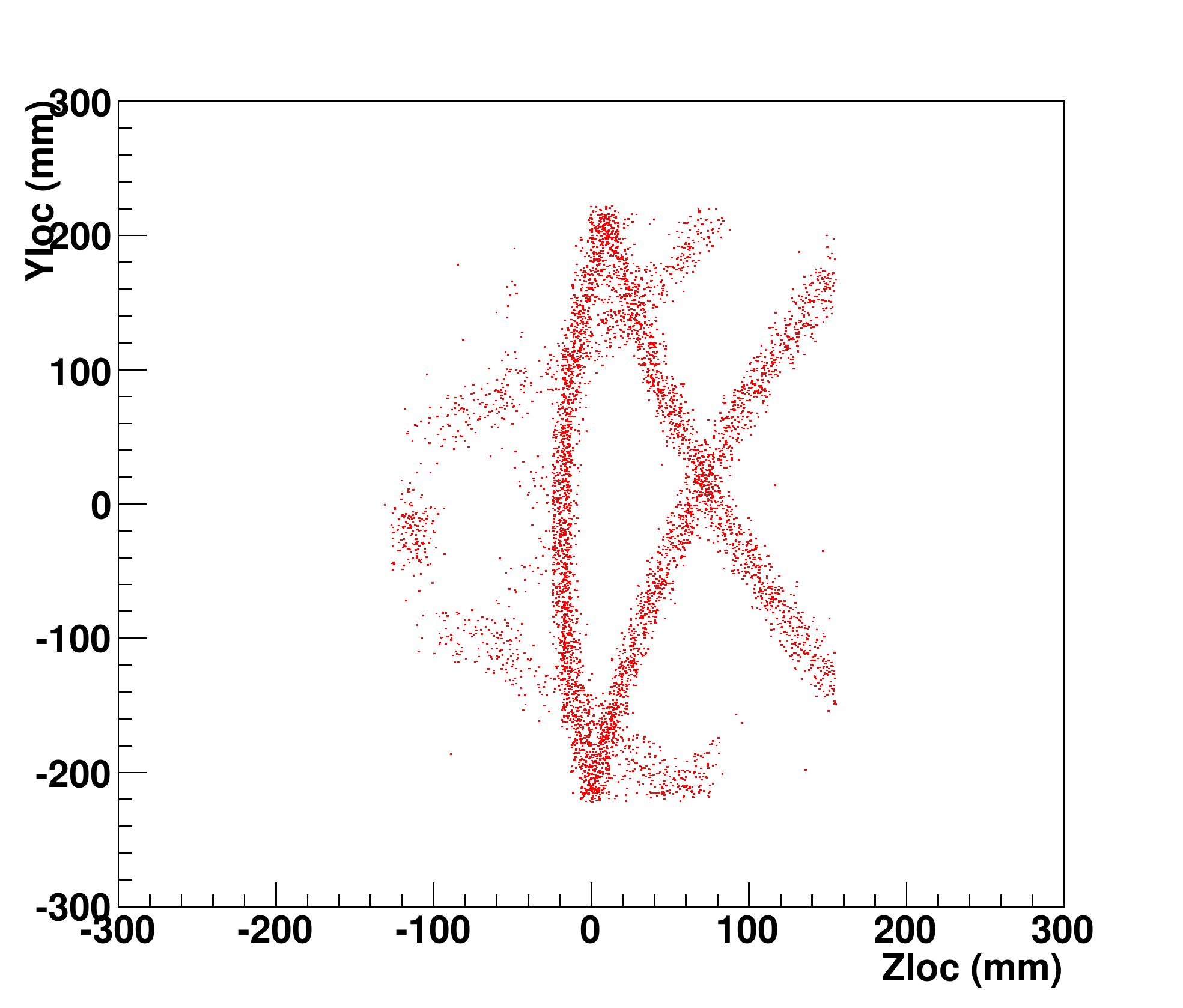}
	\caption{Occupancy on the photodetector plane for a charged particle hitting a bar perpendicularly (efficiency not accounted for in this image) for the \gx~(top panel) and SLAC (bottom panel) designs.} 
	\label{figYlocZloc}
	\end{center}
\end{figure}

GEANT4 simulations of the \gx~FDIRC are under development.  An example of Cherenkov light propagation in the \gx~FDIRC is shown in Fig.~\ref{1eventleftright}.  Figure~\ref{figYlocZloc} shows the occupancy in the PMT plane for many identical charged particles thrown perpendicularly to a bar for both the \gx~and SLAC designs. 
In the SLAC design, photons entering the focusing block at large angles reflect from the sides giving rise to the crossed pattern.  
The alignment of the boxes in \gx~permits using one common FOB with a single readout system and avoids the side reflections seen with the focusing block design.   Removing side reflections is highly desirable in the reconstruction as it avoids introducing ambiguities in the pattern recognition. 
A more detailed comparison of the patterns observed in the two designs is shown in the Appendix Fig. ~\ref{YZcomparison}.

\begin{figure}[!h]
           \includegraphics[width=0.45\textwidth]{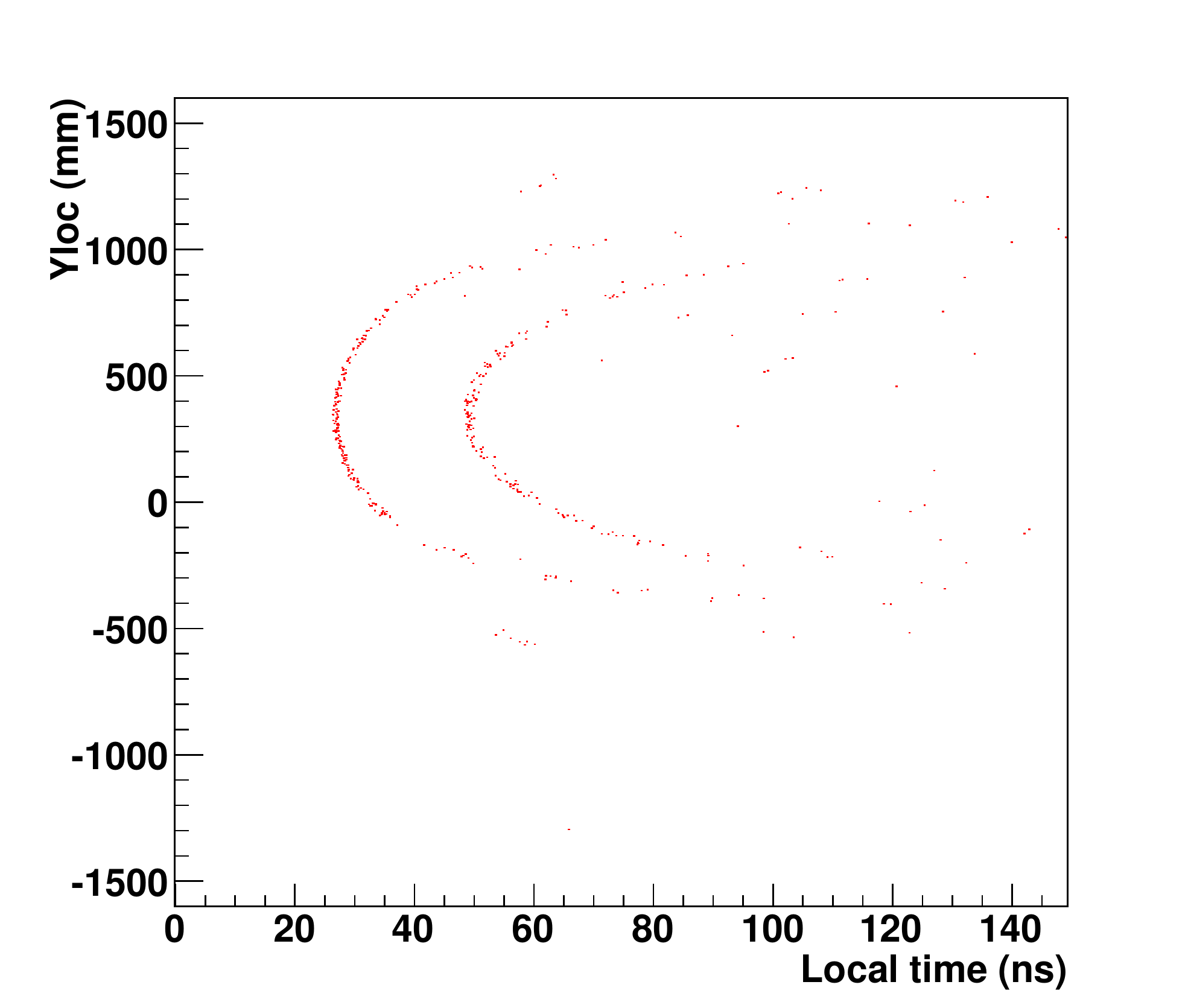}\quad 
            \includegraphics[width=0.45\textwidth]{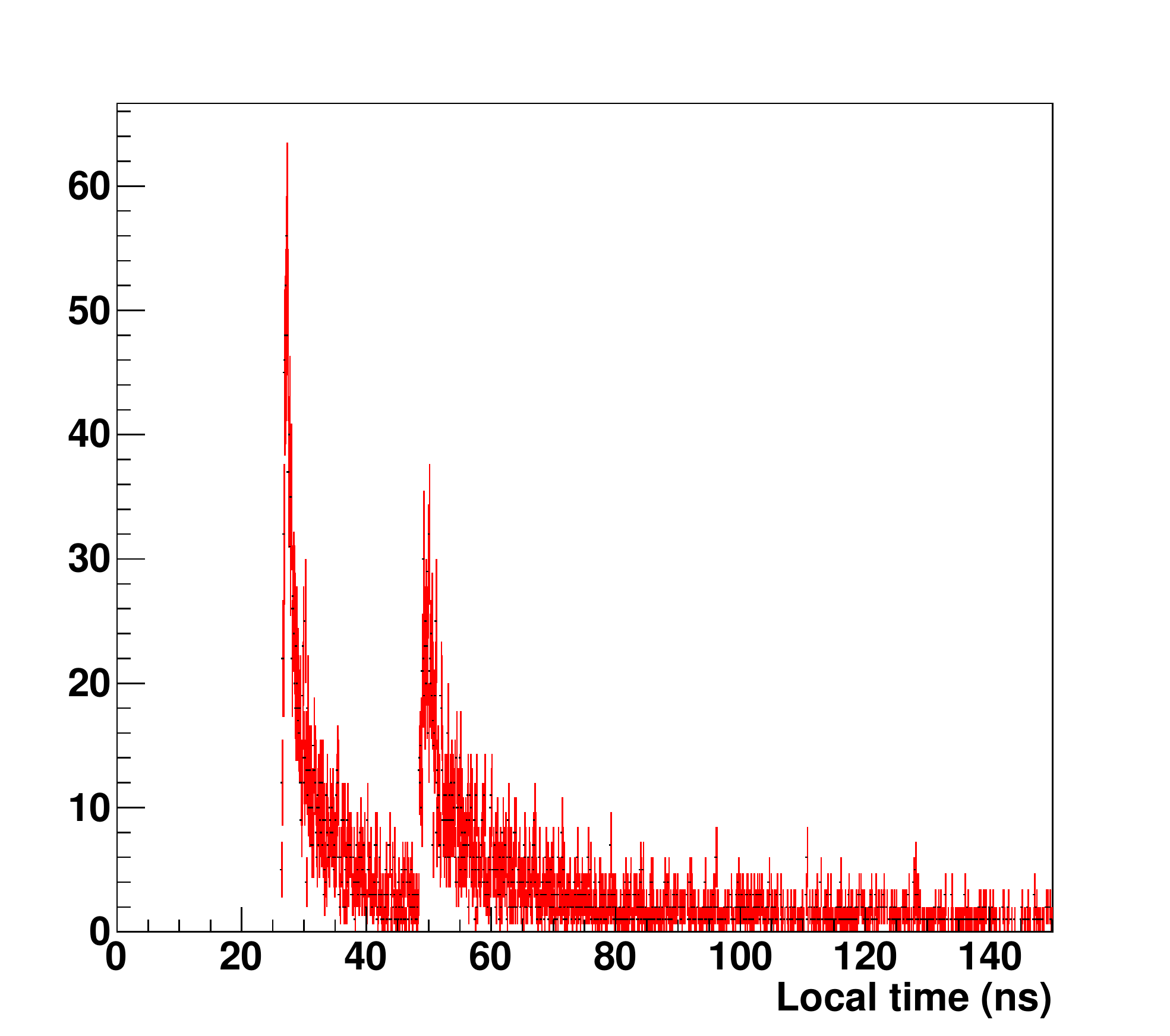}  \\
	\caption{(top) Y position in the local photodetector coordinates {\em vs} the local arrival time of the photon. (bottom) one-dimensional projection of the local time (using 50 identical pions).
The local arrival time is defined as the time between the photon production to its detection.} 
	\label{YlocTiming}
\end{figure}

\begin{figure}[!h]
           \includegraphics[width=0.45\textwidth]{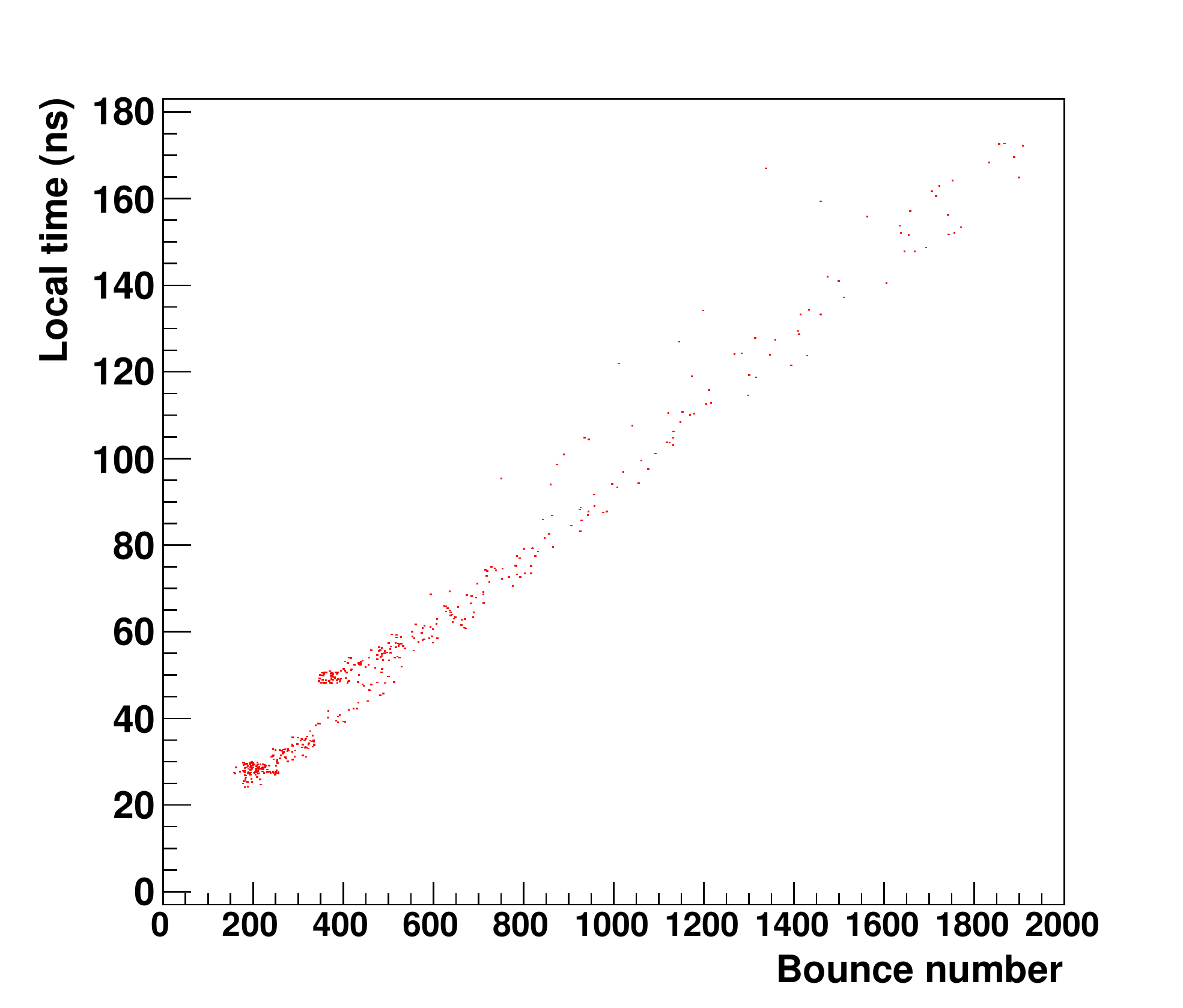}\quad 
            \includegraphics[width=0.45\textwidth]{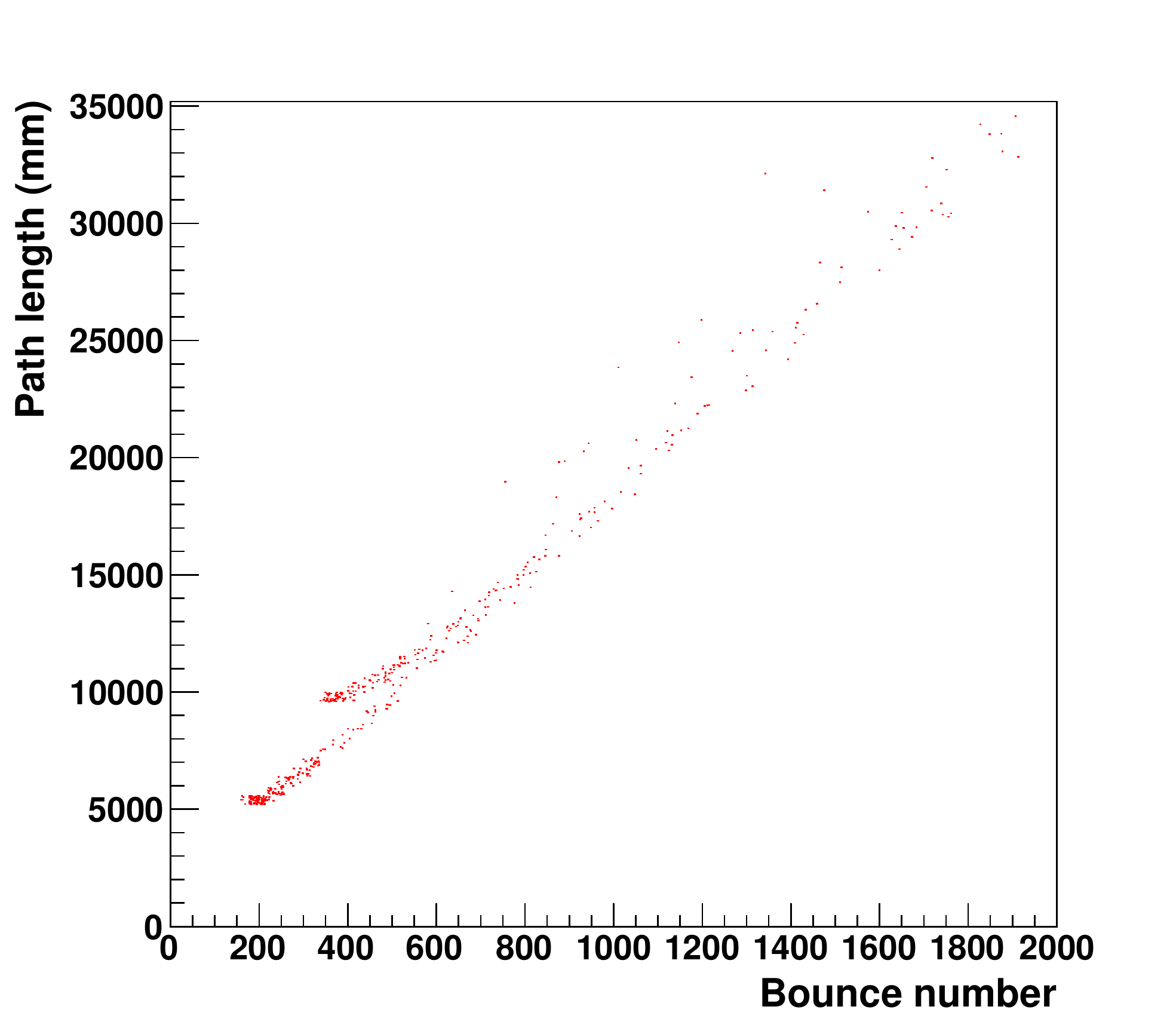}  \\
	\caption{(top) local arrival time of the photons {\em vs} the number of bounces required to reach the detector. (bottom) path length of the photons {\em vs} the number of bounces required to reach the detector.  The photon timing is defined as in Fig.~\ref{YlocTiming}.} 
	\label{TimingandBounce}
\end{figure}

Figure \ref{YlocTiming} shows the photon position in the photodetector plane as a function of the arrival time of the photon, while Fig.~\ref{TimingandBounce} shows the photon arrival time and path length as a function of the number of bounces that the photon makes before being detected.  The two bunches separated in time correspond to forward and backward emitted photons. The forward photons go directly from the creation point to the readout side while the backward photons are reflected by the mirrors and then traverse the entire length of the bar.  In the current configuration, the forward photons begin to arrive about 27~ns after production corresponding to a path length of about 5~m making 200 bounces. The backward photons travel about twice the distance making about 400 bounces. The large number of bounces made by the photons requires that the bars have excellent surface quality in order to preserve the Cherenkov angle.
The position of the bars in the vertical direction has not yet been optimized.  There is some freedom in their placing along this axis; thus, we are studying how the placement of the bars along the vertical axis affects the chromatic correction and other aspects of the FDIRC performance.  

The \gx{} application has two key differences from the focussing block design developed at SLAC
for SuperB.  First the variation in entry angle of charged particles into the FDIRC is relatively small given
its downstream location.  Second, all bar boxes can be arranged in a common plane,
as opposed to the barrel shape of both BaBar and SuperB.
It is these two properties that motivated us to explore the focussing oil box design in an attempt to find
a simpler, more cost effective solution, that reduces ambiguities in the reconstruction.
We recognize that our design, as sketched above, presents some mechanical challenges in 
construction.  For example, coupling two $A$ wedges from two different bar boxes to a 
common $B$ wedge will be very challenging.  Our goal at present is to develop the optical properties of the
system that are optimal for \gx{}.  We may achieve similar optical performance by using a mirror
submerged in the oil box instead of a $B$ wedge in air.  As we work towards a final technical 
design we plan to examine and optimize these details considering cost, performance, and technical risk.
If we cannot achieve our goals with a focussing oil box solution, we may always implement
the focussing block design developed and tested at SLAC for SuperB.

\subsection{Readout}



For satisfactory Cherenkov ring reconstruction, the DIRC detector needs a 2-dimensional photoreadout with a resolution on the order of a few millimeters.
Although the yield of Cherenkov photons is proportional to $1/\lambda^{2}$, due to the materials used in the BaBar DIRC bars especially the EPOTEK 301-2 glue~\cite{NIM2}, only photons with wavelength longer than 300~nm can exit the DIRC bar boxes.
Therefore the readout only needs to cover the range above 300~nm.
In addition, due to the fringe field from the open solenoid used by the \gx{} spectrometer, the readout has to be able to tolerate a magnetic field of about 100 Gauss.

Several readout options have been evaluated including multianode photomultiplier (MaPMT), Silicon photo-multiplier (SiPM) and a newly developed large area pico-second photodetector (LAPPD) using the renovated micro-channel plate (MCP) technology~\cite{lappd_proposal}.
At the end, we chose to focus on two types of photodetectors: the MaPMT and the LAPPD, with the LAPPD as our primary readout choice.

\subsubsection{Multianode Photomultiplier}

Multianode photomultipliers have been recently tested for various Cherenkov detectors, including SuperB's focusing DIRC detector~\cite{mapmt_superb}, Jefferson Lab CLAS12's RICH detector~\cite{mapmt_clas12} and Jefferson Lab SoLID's light gas Cherenkov counter~\cite{mapmt_solid}.
Most of these works focus on the H8500 MaPMT assembly~\cite{h8500_spec} manufactured by Hamamatsu Corp. and it appears to be a solid solution for the DIRC readout.

\begin{figure}
  \centering
  \includegraphics[width=0.45\textwidth]{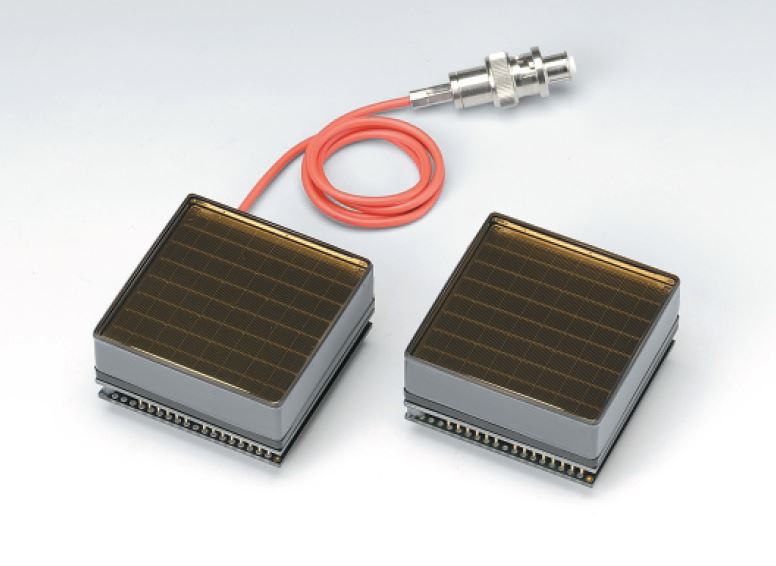}\\
  \caption{The H8500 multianode photomultiplier manufactured by Hamamatsu.}\label{fig:mapmt}
\end{figure}

The H8500 flat panel MaPMT assembly has an active area of 49$\times$49~mm$^{2}$.
It has an 8$\times$8 anode readout array, and each anode covers an area of 5.8$\times$5.8~mm$^{2}$.
The packing factor of H8500 is a very tight 89\% and this makes it very suitable for large area photon detection.
The H8500 uses bialkali photocathode and borosilicate glass window, and is sensitive to photons of wavelength between 300$\sim$650~nm and the maximum quantum-efficiency in this range is close to 30\%.
The H8500 does have a variation using UV glass which extends the sensitive range down to 180~nm.
But this won't be necessary for our DIRC design due to the wavelength cut-off that was mentioned.

If finer resolution is desired for more accurate Cherenkov angle measurement, the H9500 MaPMT~\cite{h9500_spec} from Hamamatsu can be used instead.
H9500 has the same dynode structure and geometry as H8500 and it has a 16$\times$16 anode readout array with 256 3$\times$3~mm$^{2}$ pixels.

\begin{figure}
  \centering
  \includegraphics[width=.45\textwidth]{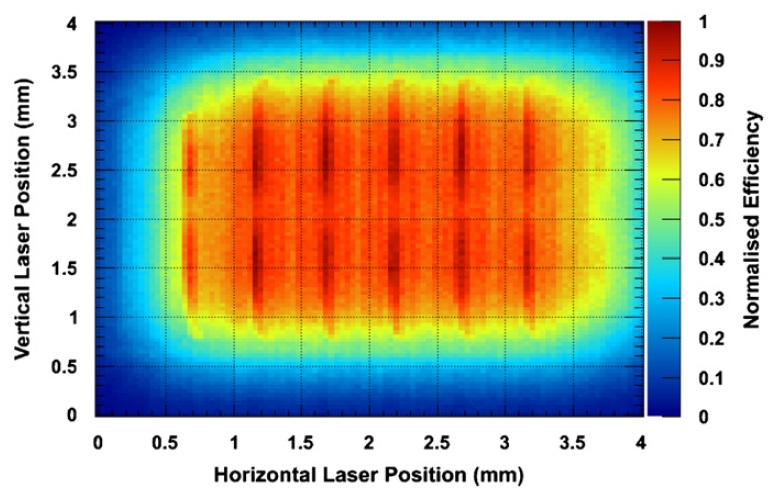}\\
  \caption{The relative uniformity of one H9500 pixel using laser scan~\cite{mapmt_clas12}. }\label{fig:h9500_uniformity}
\end{figure}

\begin{figure}
  \centering
  \includegraphics[width=.45\textwidth]{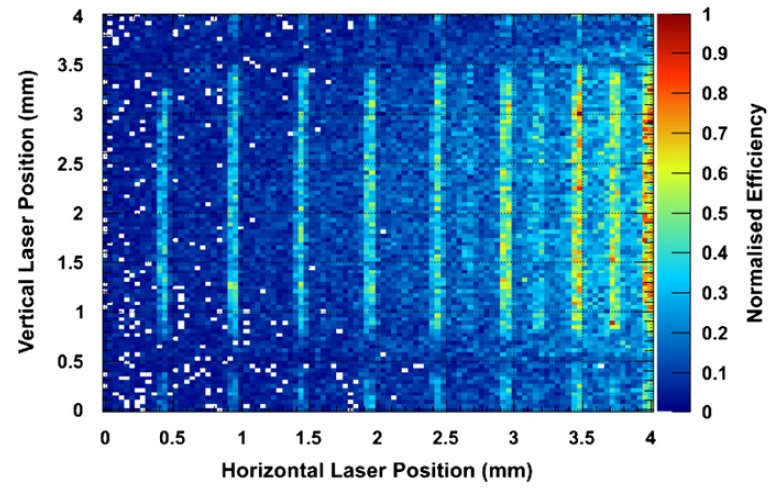}\\
  \caption{The normalized crosstalk map of one H9500 pixel using single photoelectron laser scan~\cite{mapmt_clas12}. }\label{fig:h9500_crosstalk}
\end{figure}

The photodetection uniformity and the crosstalk between adjacent anode pixels of H8500 and H9500 were reported in literature and some results can be found in Ref~\cite{mapmt_superb,mapmt_clas12}.
A relative variation up to ~25\% has been observed in the uniformity test as shown in Figure~\ref{fig:h9500_uniformity}.
The crosstalk pattern in Figure~\ref{fig:h9500_crosstalk} shows a clear dependency upon the dynode mesh construction of the MaPMT.
Although it is postulated that these behaviours will not be problematic for single photon detection of a RICH detector, further investigation will be needed to optimize the readout design and reconstruction algorithm.

In addition, the performance of H8500 in magnetic field was also studied at Jefferson Lab~\cite{mapmt_solid}.
The test demonstrated that the H8500 MaPMT can operate without much degradation in a longitudinal field up to 300~Gauss.
Although the drop of performance in transverse magnetic field is significantly more pronounced, up to 100~Gauss, such transverse field is also the easiest to shield in practice.
Therefore, we conclude that MaPMTs will be able to operate in Hall-D's fringe field without shielding.

\begin{figure}
  \centering
  \includegraphics[width=.48\textwidth]{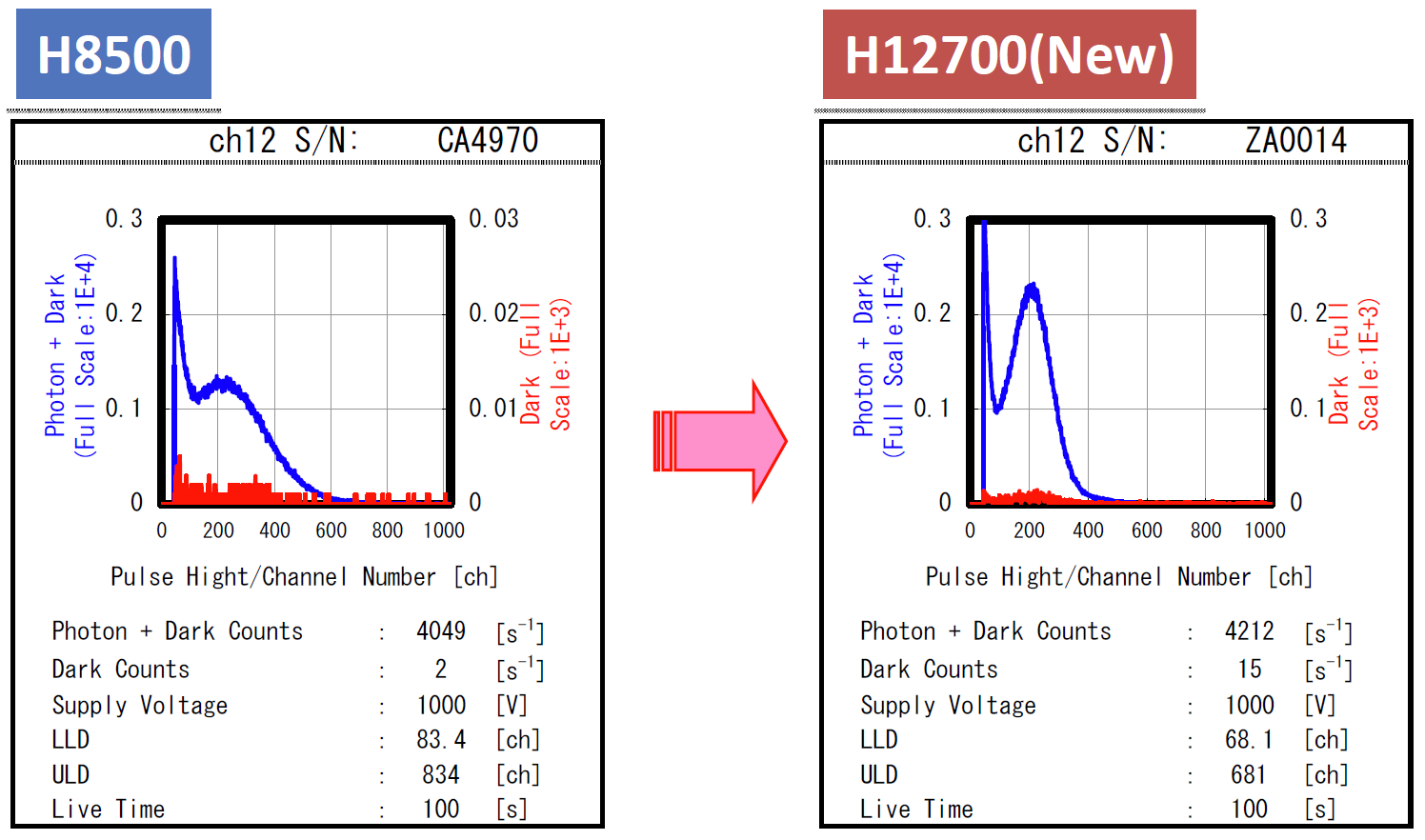}\\
  \caption{Improvement of H12700's single photon detection capability~\cite{adarvan}.}\label{fig:h12700}
\end{figure}

Recently, an upgraded version of the H8500 MaPMT has been revealed by Hamamatsu.
The new H12700 MaPMT has the same geometry and output pin layout as the H8500.
With a newly optimized dynode structure and voltage scheme, the collection efficiency has been greatly improved~\cite{adarvan} and it now has a better separation of single photon signals from background, as shown in Figure~\ref{fig:h12700}.

\begin{figure}
  \centering
  \includegraphics[width=.48\textwidth]{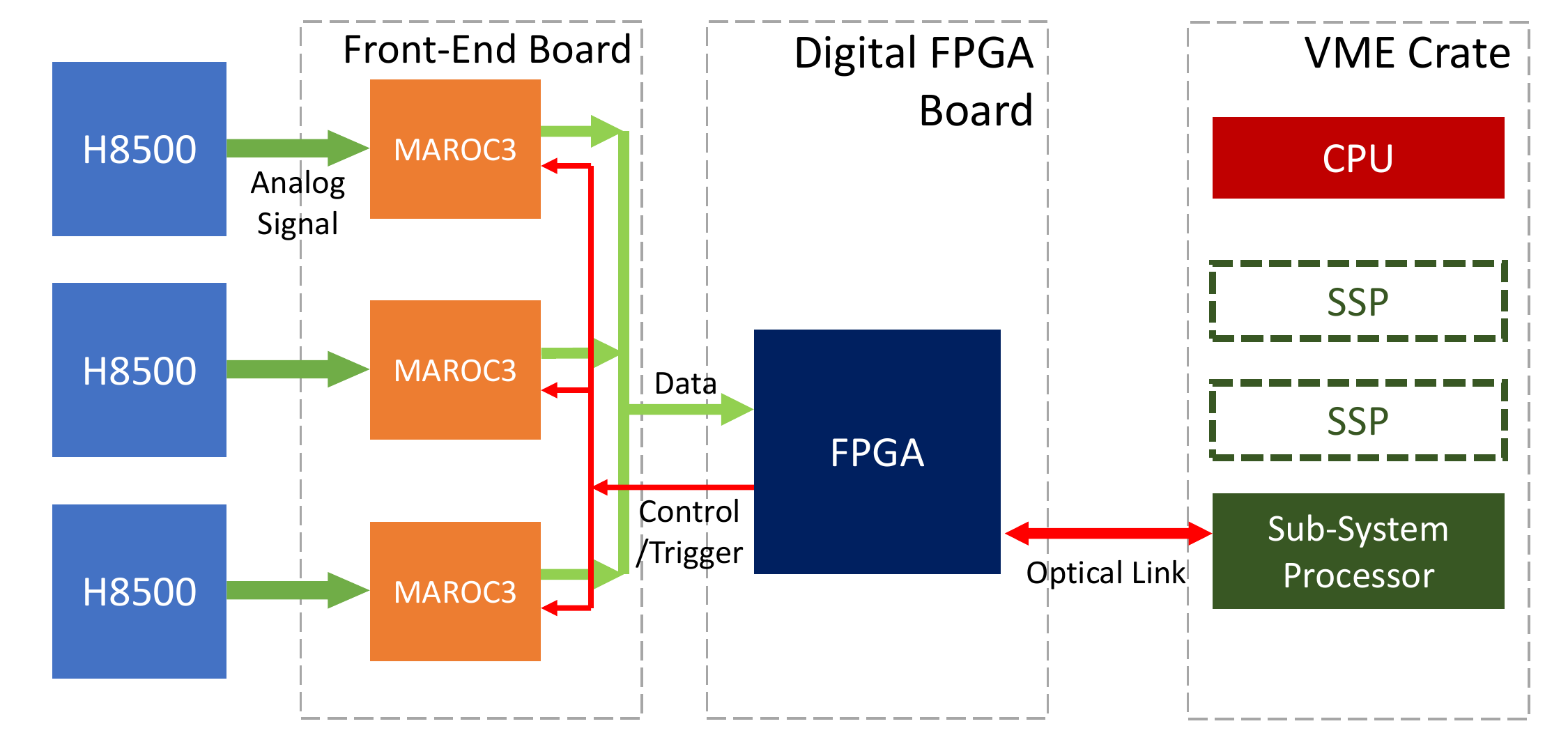}\\
  \caption{Readout scheme of Jefferson Lab CLAS12's RICH detector.}\label{fig:mapmt_readout}
\end{figure}

As for the readout electronics, we will use the design of Jefferson Lab CLAS12's RICH detector~\cite{rich_clas12} as a reference.
The core of the design is to use the MAROC3 chip~\cite{maroc3} specifically designed for the readout of 64-channel MaPMTs.
As shown in Figure~\ref{fig:mapmt_readout}, the MAROC3 chips digitize the analog signals from MaPMTs and pass the resulting binary data stream to a digital FPGA board.
The FPGA on board not only processes the data but also controls and provides triggers to the MAROC3 chips.
The processed data from the FPGA will then be transmitted to a Jefferson Lab developed Sub-System Processor (SSP)~\cite{ssp} hosted in a VME crate through high speed optical links.
The frontend board is currently under development by a group at INFN, and the digital FPGA board will be developed by Jefferson Lab's electronics group.
These two groups together have demonstrated the feasibility of using MAROC3 chips for the RICH readout in a recent DOE project review.
It's also worth mentioning that by using the Jefferson Lab SSP, such a readout system can be seamlessly integrated into the Hall D DAQ system.

\subsubsection{Large Area Picosecond Photodetector}

\begin{figure}
  \centering
  \includegraphics[width=0.45\textwidth]{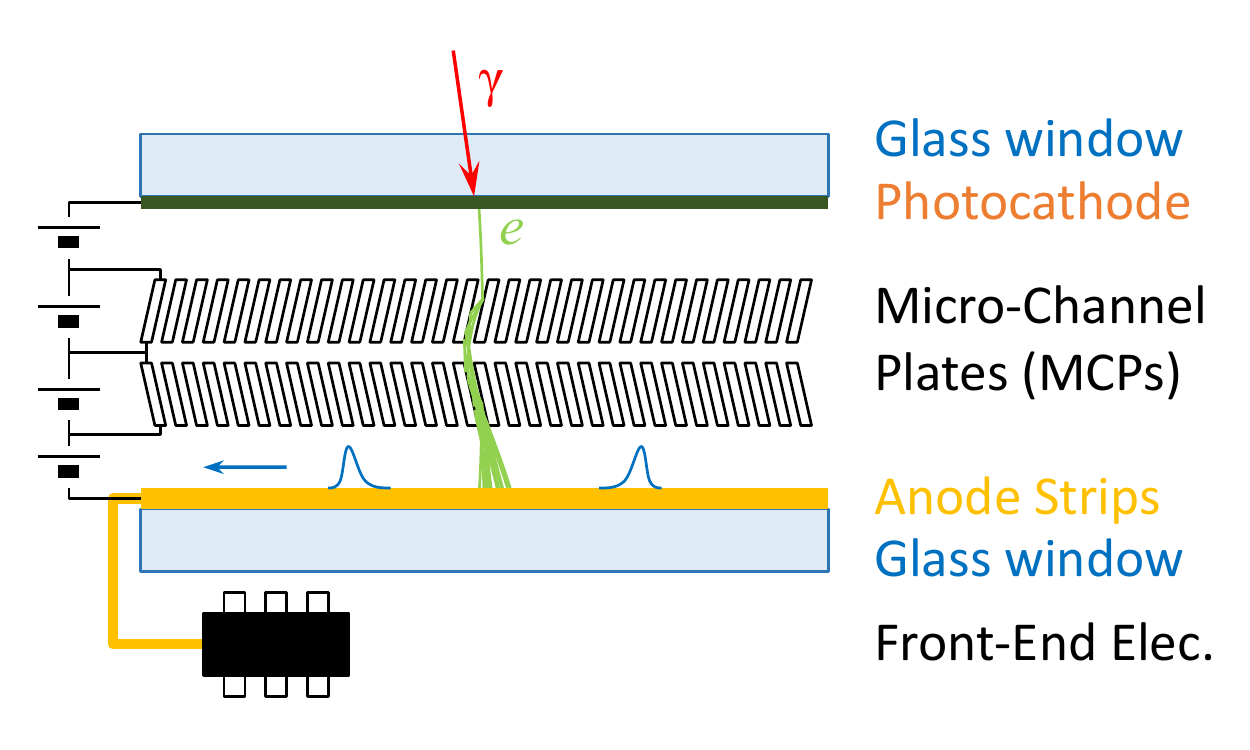}\\
  \caption{Schematic of LAPPD's MCP-PMT.}\label{fig:lappd_scheme}
\end{figure}

Since 2009, a new development of a large-area fast photo-detector using micro-channel plate (MCP-PMT)~\cite{lappd_tofrich} is being carried out by the large-area picoseconds photo-detector (LAPPD) collaboration~\cite{lappd_proposal} and they provide a very attractive, low cost, high performance readout solution for RICH detectors.
The goal of this R\&D program is to develop a family of large-area robust photo-detectors that can be tailored for a wide variety of applications where large-area economical photon detection is needed.
The approach is to apply microchannel plate (MCP) technology to produce large-area photo-detectors with excellent space and time resolution.
The schematic of such a detector is shown in Figure~\ref{fig:lappd_scheme}.
In addition to having excellent resolution, the new devices should be relatively economical to produce in quantity.
Such a detector can be used in many applications, such as precision time-of-flight measurements, readout of Cherenkov counters, and positron-emission tomography (PET) for medical imaging.

\begin{figure}
  \centering
  \includegraphics[width=0.35\textwidth]{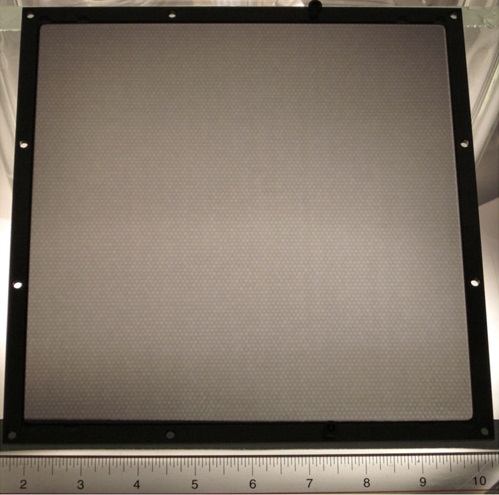}\\
  \caption{Photograph of a 20$\times$20~cm$^{2}$ MCP made using ALD treatment of a borosilicate glass micro-capillary array.
  20 mm pores, L/D$\sim$60:1, pore bias 8$^{\circ}$.
  The multifiber hexagonal boundaries are visible in this backlit image.}\label{fig:lappd_mcp}
\end{figure}

\begin{figure}
  \centering
  \includegraphics[width=0.45\textwidth]{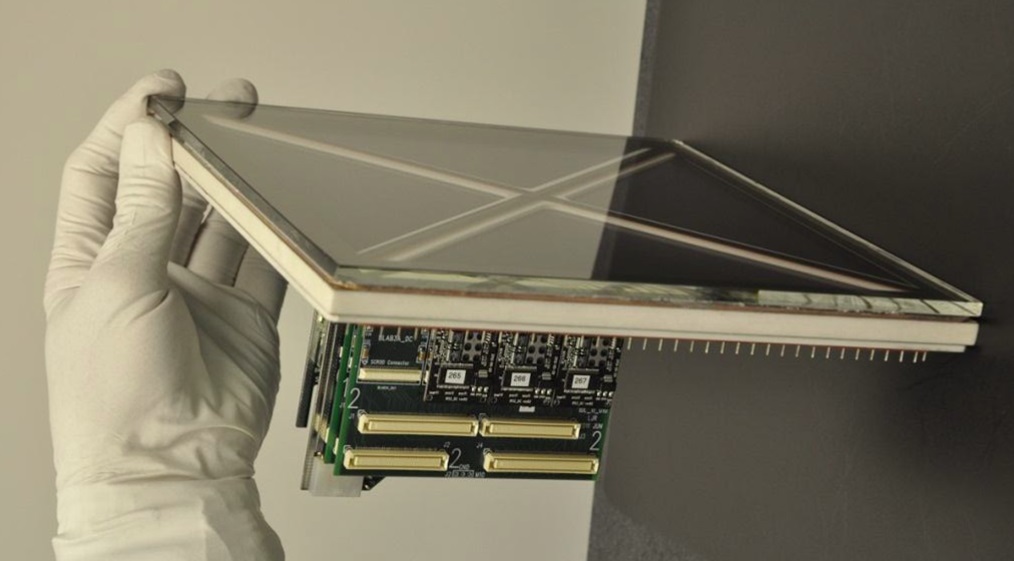}\\
  \caption{A 20$\times$20~cm$^{2}$ ceramic body MCP-PMT prototype.}\label{fig:lappd}
\end{figure}

As the project is in its fourth year, excellent progress has been made.
In particular, chemical vapor deposition (CVD) technique is being studied to form a photocathode on a large area glass window, and the resulting quantum efficiency is now over 25\% for 350 nm wavelength.
The collaboration applied atomic layer deposition (ALD) on capillary glass channel substrates (see Fig.~\ref{fig:lappd_mcp}) to produce MCPs~\cite{lappd_mcp} and has achieved better performance at much lower cost than standard commercial MCPs.
The anode readout will use strip transmission lines~\cite{lappd_anode} sampled by front-end waveform sampling chips.
The LAPPD collaboration has assembled several prototypes using ceramic bodies (see Fig.~\ref{fig:lappd}) and small samples are expected to be available to early adopters in 2014.

\begin{figure}
  \centering
  \includegraphics[width=0.45\textwidth]{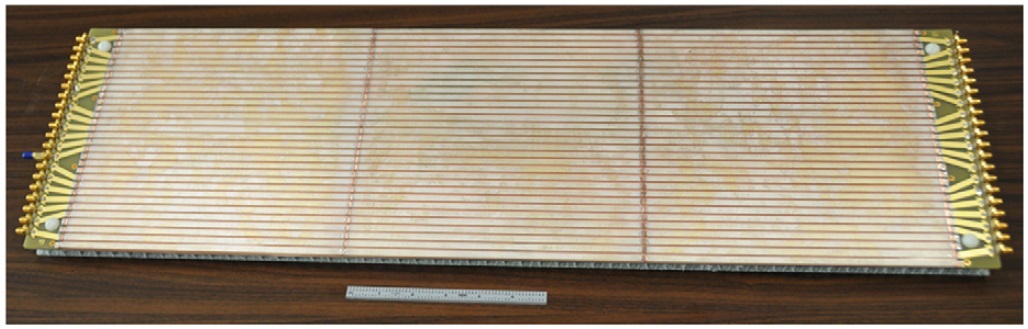}\\
  \caption{The 3-tile anode.
  The connections between anode strips on neighboring tiles have been made by soldering small strips of copper to the silver silk-screened strips on the glass.}\label{fig:lappd_anode}
\end{figure}

When produced in large quantities, the manufacturing cost of LAPPD MCP-PMTs is expected to be
less expensive than existing pixelated photo detectors such as Silicon Photomultipliers and Multi-anode Photomultiplier Tubes while still being able to provide comparable spatial resolution ($<5$~mm).
Since LAPPD uses stripe-line readout, this design significantly reduces the total channel count particularly for applications that need to cover a very large area such as a RICH detector.
Under a low rate condition, the readout can be chained as shown in Figure~\ref{fig:lappd_anode} to further reduce the number of channels.

\begin{figure}
  \centering
  \includegraphics[width=0.45\textwidth]{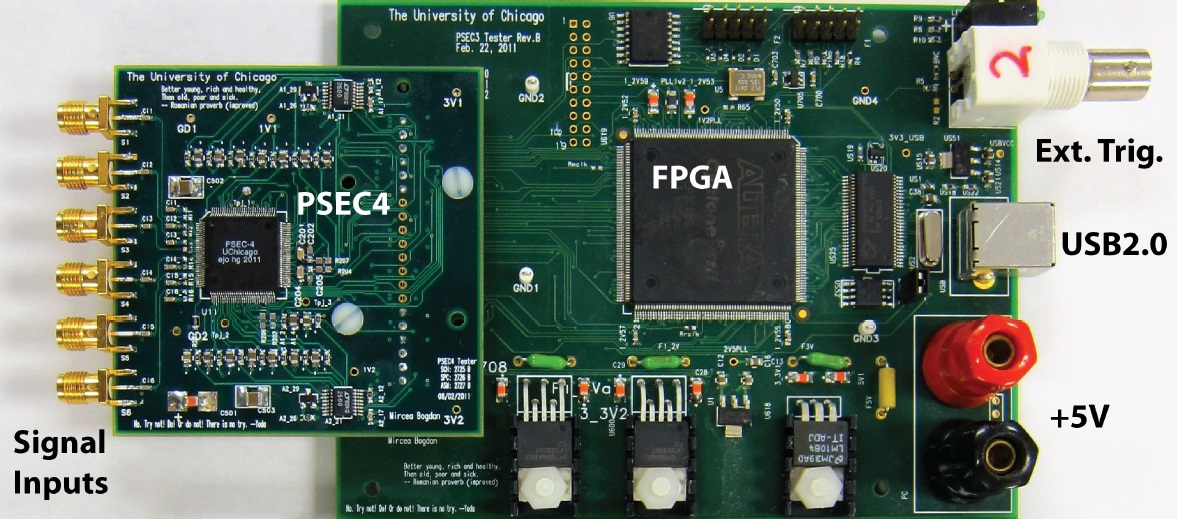}\\
  \caption{The PSEC4 evaluation board~\cite{lappd_psec4}.
  The board uses a Cyclone III Altera FPGA (EP3C25Q240) and a USB 2.0 PC interface.}\label{fig:lappd_psec4}
\end{figure}

For the readout electronics, we can adopt the PSEC4~\cite{lappd_psec4} ACIS developed by the LAPPD collaboration.
It is a waveform sampling chip with a rate up to 15~GSample/s.
A PSEC4 evaluation board in shown in Figure~\ref{fig:lappd_psec4}.
Alternatively, as the total number of channels is expected to be a few hundred, we are also considering using Jefferson Lab's F1TDC~\cite{f1tdc} modules to readout signals from both sides of individual strips for a more versatile and sharable setup.
Each F1TDC module has 16 TDCs with a resolution of 60~ps.
In this scenario, flash ADCs will be connected to some of the channels for monitoring.

\begin{figure}
  \centering
  \includegraphics[width=0.45\textwidth]{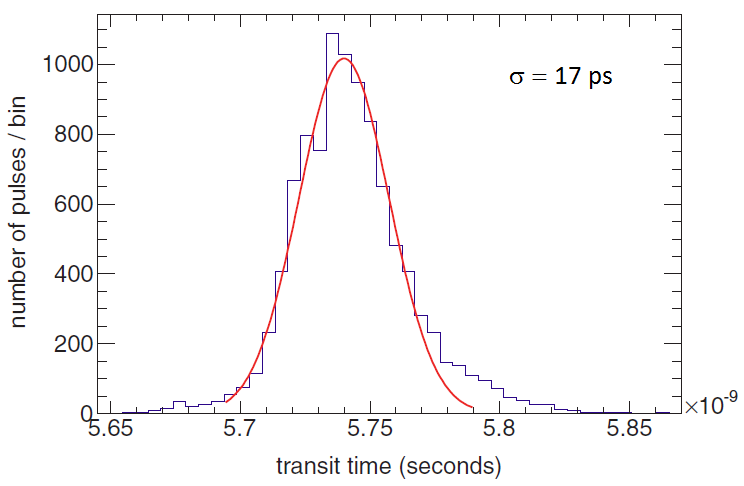}\\
  \caption{The measured time distribution of signals from a focused fs laser source~\cite{lappd_test}.
  The signal was read from one side of a single strip line, fitted with a Gaussian.}\label{fig:lappd_tof}
\end{figure}

Another huge advantage that the LAPPD's MCP-PMT can bring is the exceptional time resolution.
The observed timing resolution for single photon hits has been measured to be better than 20~ps from a demountable prototype using metal photocathode~\cite{lappd_test}, as shown in Figure~\ref{fig:lappd_tof}.
With such an excellent time resolution, the time-of-flight measured by the MCP-PMTs can be used for complementary particle identification.

The LAPPD's MCP-PMTs are a very attractive readout solution and therefore we make them our primary choice.
A lot of properties of these detectors, including the radiation hardness and magnetic tolerance, have yet to be thoroughly tested.  We will work together with the LAPPD collaboration to perform corresponding tests as early as possible. 

\subsection{Integration and installation into the \gx{} detector}

\begin{figure}
\begin{center}
\includegraphics[width=\linewidth]{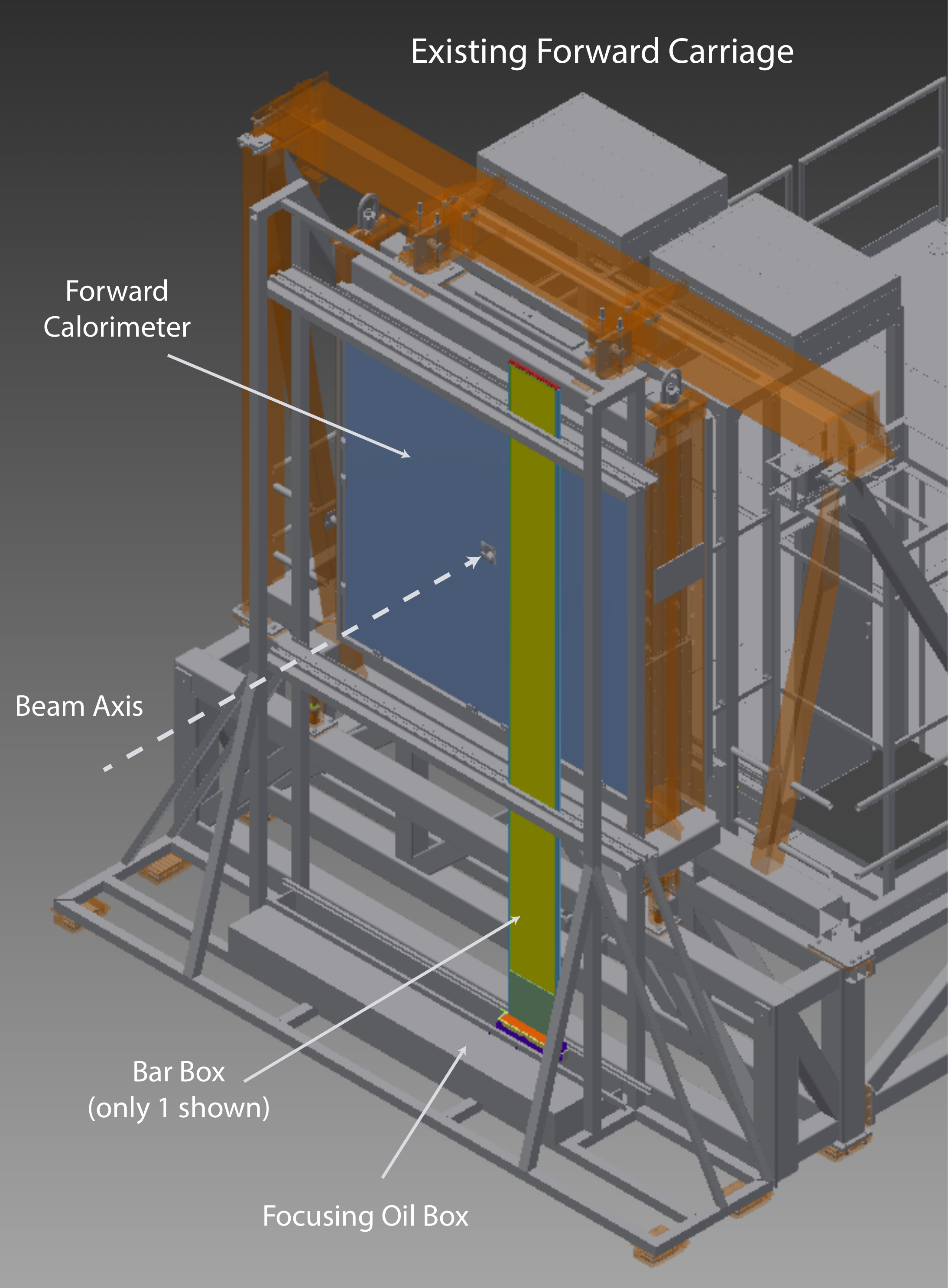}
\caption{\label{fig:carriage}A preliminary mechanical design showing the integration of the
FOB and single bar box into the \gx{} forward carriage.}
\end{center}
\end{figure}

As noted earlier, the existing \gx{} design has space reserved for a particle identification device 
between the downstream end of the solenoid and the forward carriage that supports the time-of-flight
and the forward calorimeter.  Given the fixed length of the DIRC bar boxes, the height of the \gx{}
beamline off of the floor, and considerations about accessibility and hydrostatic pressure in the FOB, 
the most desirable orientation of the boxes is with the long axis oriented vertically with the
existing window down.  A sketch of the proposed installation is shown in Figure~\ref{fig:carriage}.
Since such an orientation was never envisioned in the design of the boxes, 
potential mechanical problems must be carefully evaluated to ensure no damage will result.
It should be noted that nothing prohibits arranging the boxes horizontally if, at a later point, it is determined
that a vertical orientation presents an unacceptable risk of damage.  A horizontal arrangement
would only complicate the support structure needed for the detector and consume large amounts of
space in the existing Hall.  The performance characteristics of the detector would remain unchanged.

\begin{figure}
\begin{center}
\includegraphics[width=0.75\linewidth]{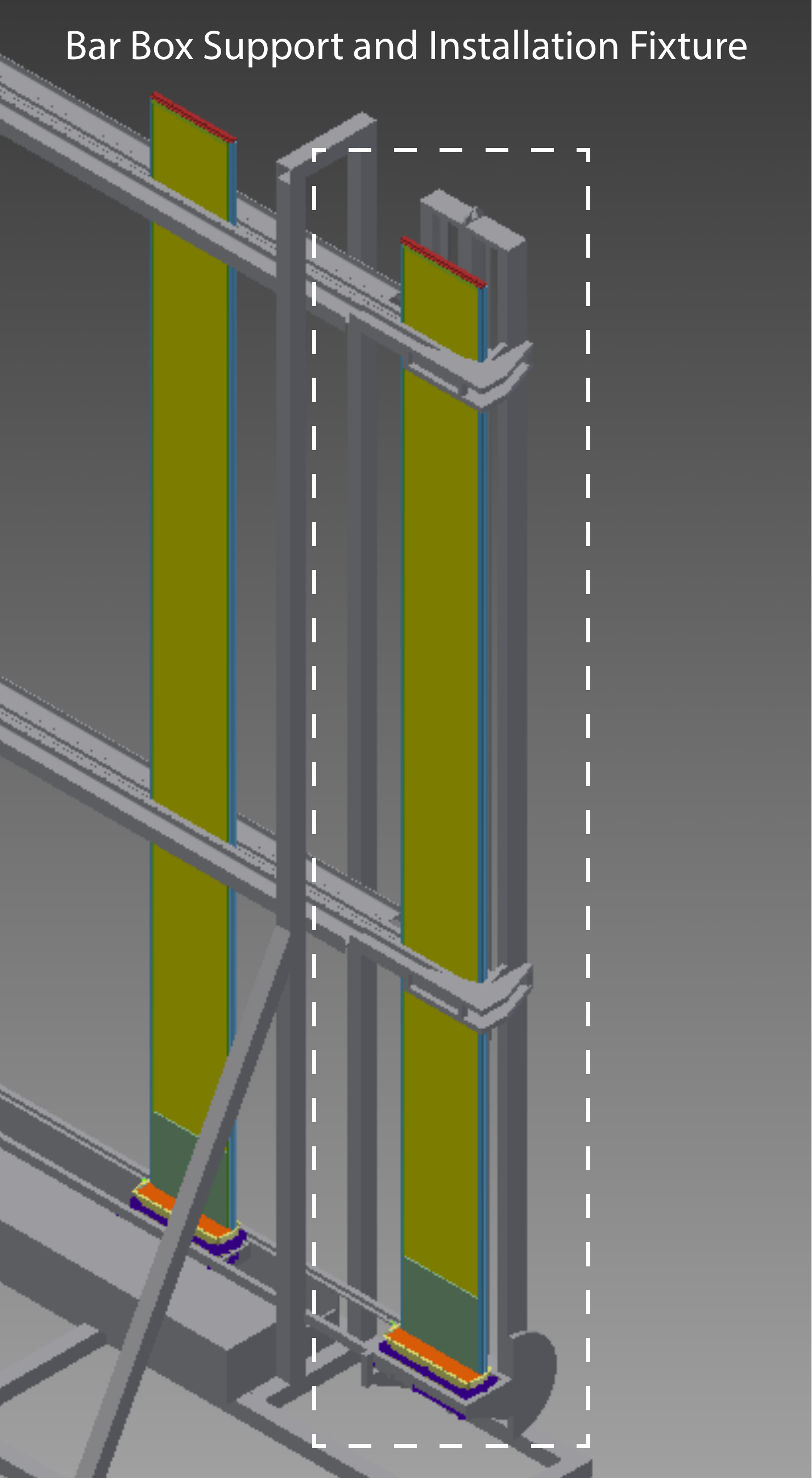}
\caption{\label{fig:install}A conceptual design for the installation jig that will allow the
box to be oriented vertically and installed into the support fixture.}
\end{center}
\end{figure}

In the vertical orientation the boxes themselves would remain rigid and dimensionally stable because
the gravitational torques about the support points are less than they are in the horizontal orientation.
One concern is that the vertical orientation causes the existing window to be loaded with the weight
of the bars plus the spring load of the mirror.  Because of the geometry of the existing wedge, the bulk
of the load would be dispersed on the window near the supporting flange.  A simple analysis
of the worst case scenario, the load concentrated at a single point in the center of the window, shows
a deflection of the window at center of about 0.0006", which is unlikely to result in breakage.  The actual
deflection is certain to be less.  A second, finite element analysis using the exact geometry 
and a 45 psi load, which accounts for the weight of the bars and the spring pressure, 
indicated a maximum bending
stress around the edges of the window of 250 psi, much lower than the 7600 psi rupture 
strength of fused silica.
This leads us to conclude that breakage or stress caused by deformation of the window
is not a concern.  If deemed necessary, we
may procure a sample of the window material and conduct a destructive test to evaluate the deformation
and breakage threshold.

An additional concern is maintaining the rigidity of the bar box in the transition from the horizontal to
vertical orientation.  In order to do this, the box will first be fixed to a rigid installation jig in the horizontal
orientation.  The jig will then be oriented vertically and fixed to the support structure on the forward
carriage (see Fig.~\ref{fig:install}).  The box will then be rolled, using the integrated rollers, off the jig and into the support structure,
where it will be locked in place.  We propose to test this installation technique utilizing the existing
prototype bar box that was constructed using ordinary glass.


\section{Expected FDIRC performance}

In this section we detail the expected performance of the proposed \gx{} FDIRC design.  We begin
by examining the discrimination power for single tracks.  We conclude by folding this information
into a simulation of inclusive photoproduction to estimate the background rejection power that the 
FDIRC provides.

\subsection{Single track particle identification}
\label{Sec:singleTrack}

\subsubsection{\gx{} tracking resolution}

Charged particle reconstruction in the \gx~detector is provided by the forward and central drift chambers (CDC and FDC) as described in Sec.~\ref{sec:baseline}.  In this section we study the reconstructed track resolutions in the forward angle region of \gx{} ($\theta < 11^{0}$) which is relevant for the FDIRC detector.  The reconstructed track variables which impact the particle identification performance of the FDIRC detector are the angle of incidence and position of the track crossing point, as well as the magnitude of the momentum as the particle enters the quartz bar.

To determine the resolution with which we can expect to measure these variables, we simulate single charged pions originating in the target of the \gx{} detector, produced uniformly in azimuth for a range of polar angles and momenta.  A complete GEANT model is used to simulate the \gx{} detector response, and a Kalman Filter tracking algorithm is used to reconstruct the tracks in the drift chambers.  The reconstructed track helix is then extrapolated through the magnetic field to the plane of the FDIRC detector.  

\begin{figure*}
	\begin{center}
 		\includegraphics[width=0.4\textwidth]{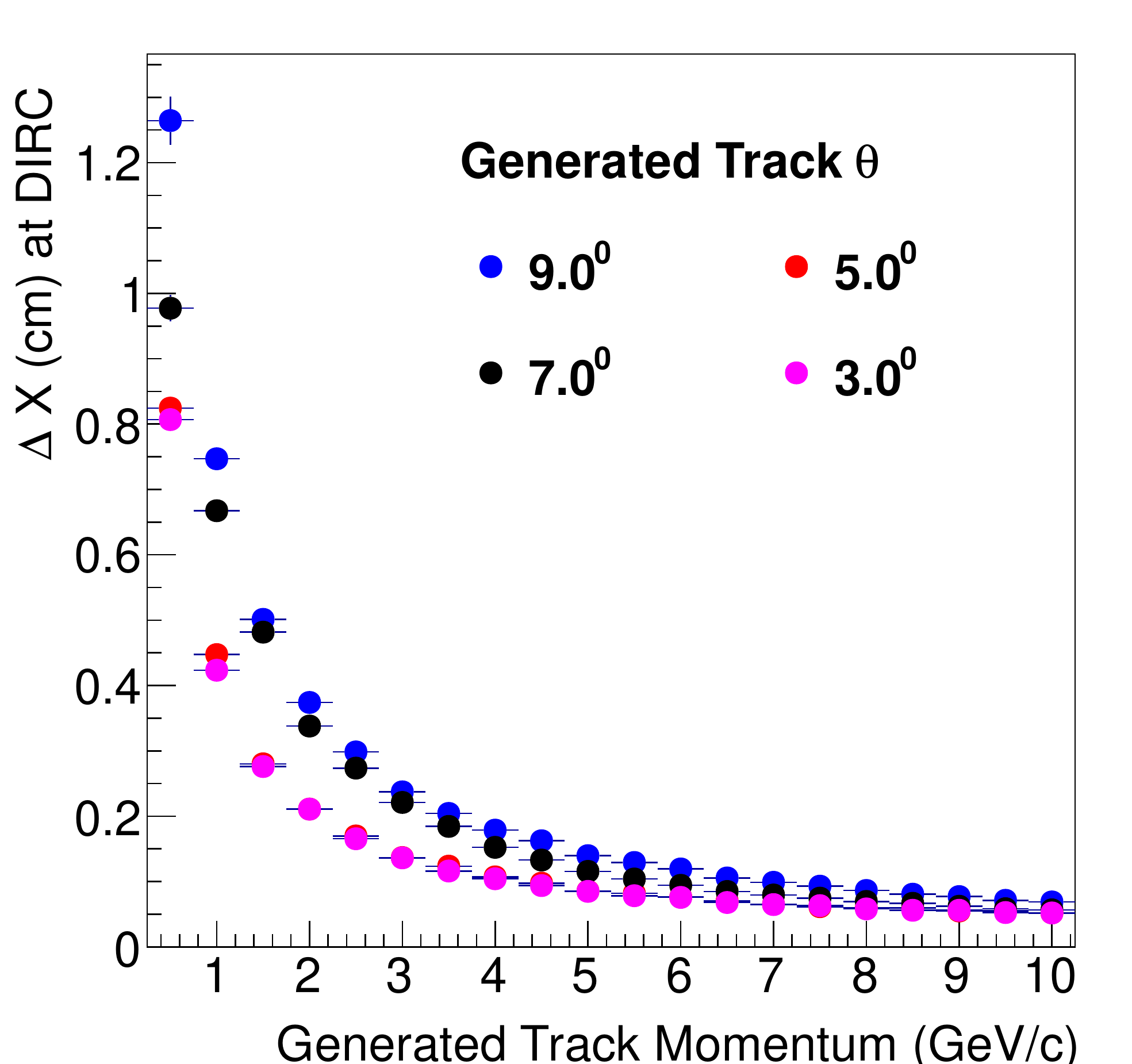} 
		\includegraphics[width=0.4\textwidth]{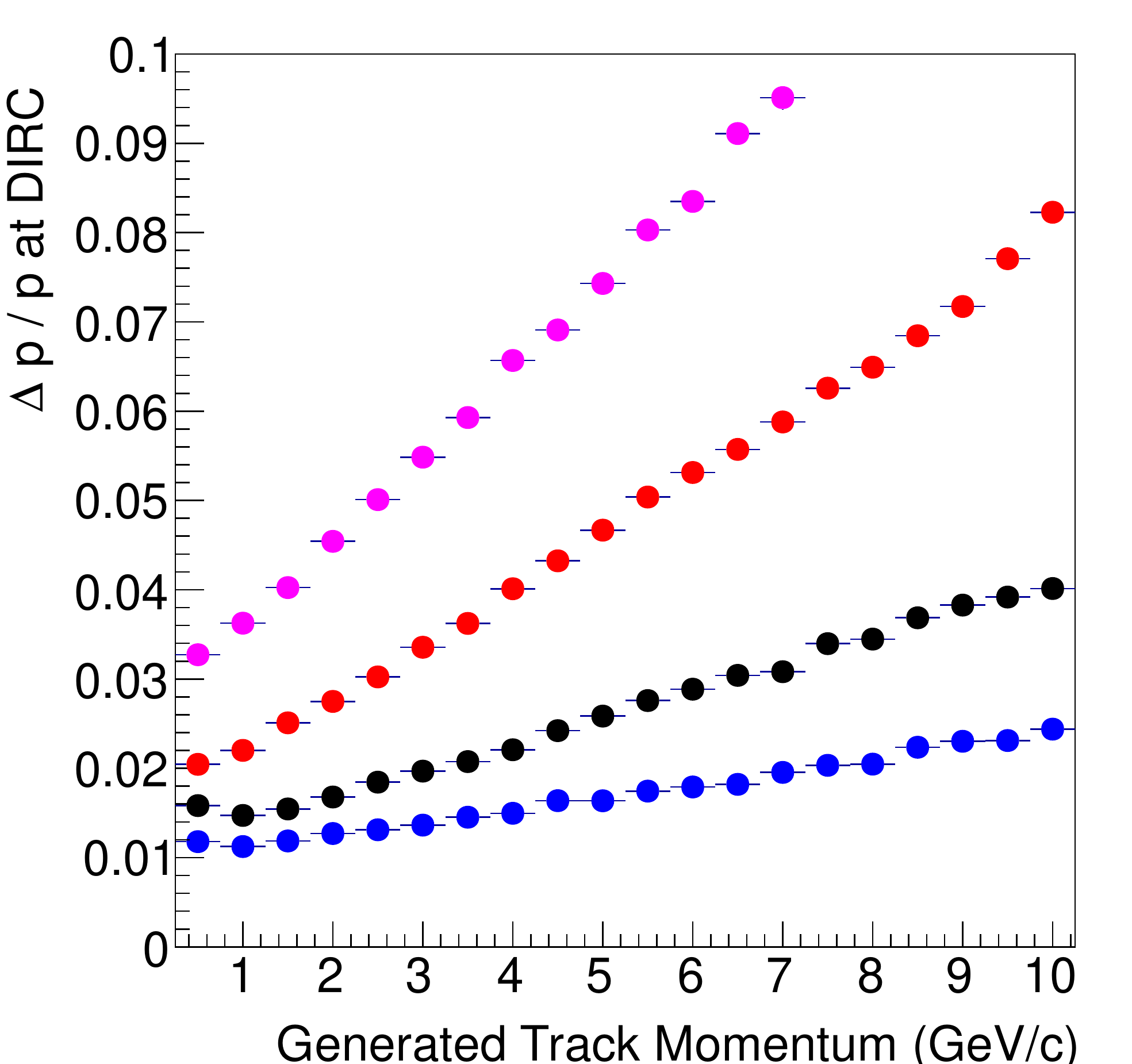}
		\caption{Track (left) position and (right) momentum resolution at the FDIRC. }
  		\label{Fig:TrackPosMomRes}
	\end{center}
\end{figure*}

\begin{figure*}
	\begin{center}
 		\includegraphics[width=0.4\textwidth]{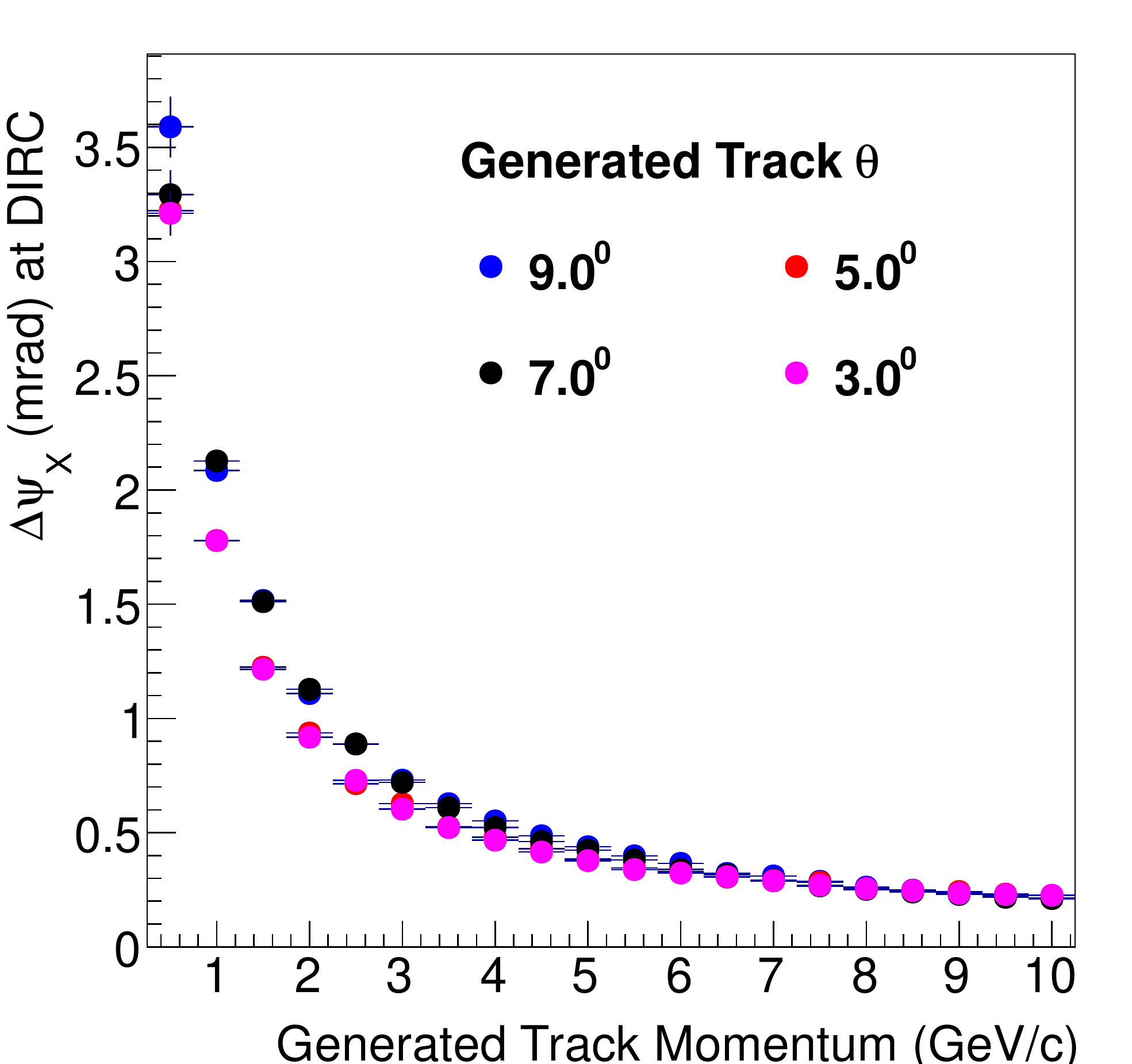} 
		\includegraphics[width=0.4\textwidth]{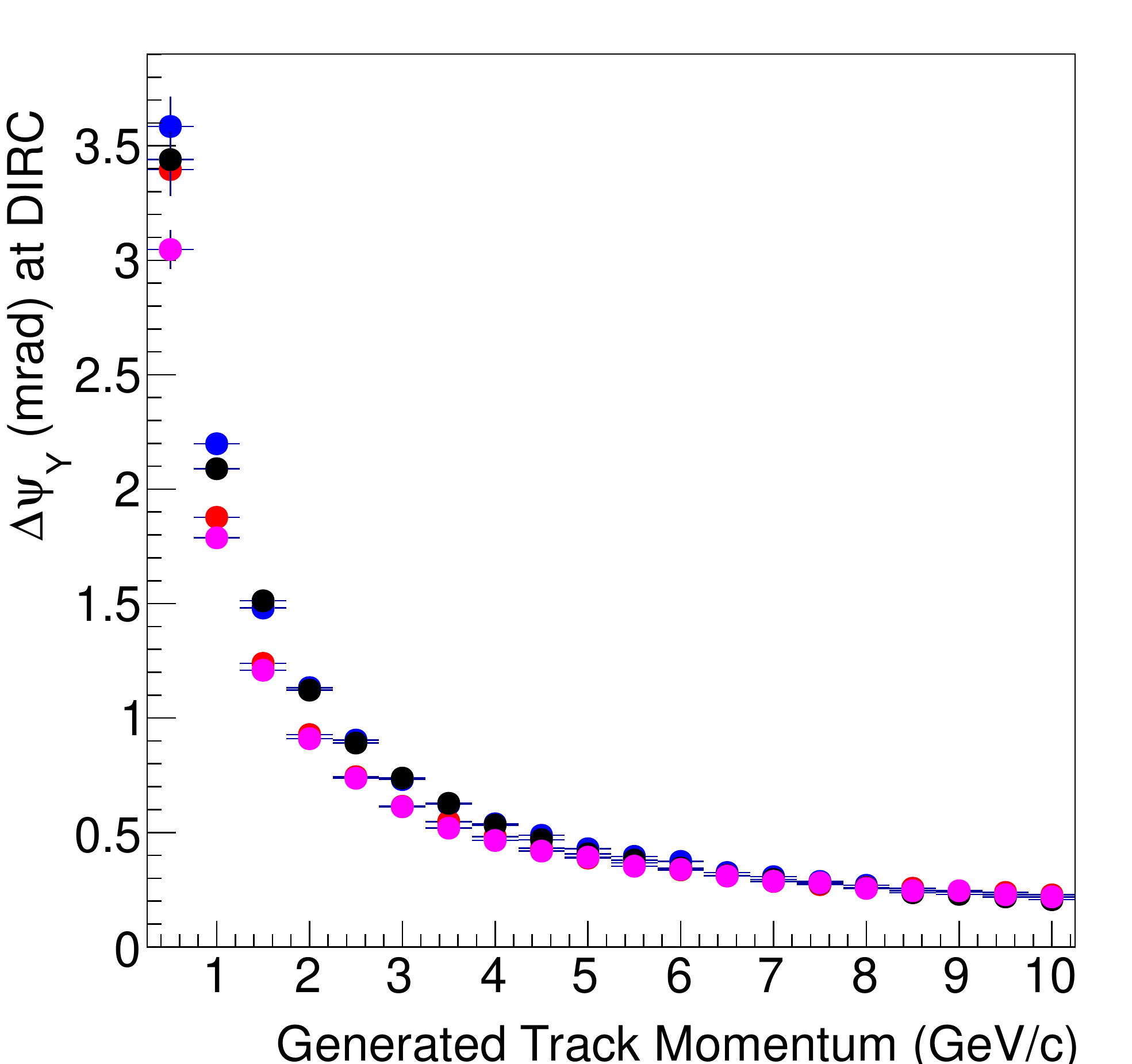}
		\caption{Track incident angle resolution at the FDIRC. }
  		\label{Fig:TrackAngleRes}
	\end{center}
\end{figure*}

The position and momentum resolution of the track as it enters the FDIRC detector are shown in Fig.~\ref{Fig:TrackPosMomRes}.  The incident angle resolution is shown in Fig.~\ref{Fig:TrackAngleRes} for two variables, $\psi_X$ and $\psi_Y$, which denote the angles with respect to the planes perpendicular and parallel to the bar's long axis, respectively.  The resolution is momentum dependent.  The \gx{} detector has adequate particle identification for particles with momenta below about 2~GeV/$c$ using time-of-flight information; thus, for the FDIRC we are primarily concerned with particles above 2~GeV/$c$.  In this regime, the resolution on the input quantities to the FDIRC shown in Figs.~\ref{Fig:TrackPosMomRes} and \ref{Fig:TrackAngleRes} are better than is required (as we will show in the following section).  

\subsubsection{Cherenkov resolution and separation}

The resolution on the Cherenkov angle of a single photon ($\sigma_{ \theta_{\gamma}}$) has many different contributions; these are listed in Table~\ref{tab:RESOLUTIONs}. 
The Cherenkov angle resolution for a track ($\sigma_{\theta_{C}}$) is given by
\begin{equation}
\sigma_{\theta_{C}}= \sqrt{ \frac{ \sigma_{ \theta_{\gamma}}^{2} }{N_{\gamma}} + \sigma_{ \theta_{track}}^{2} },
\label{EqCherenkovResolution}
\end{equation}
where $N_{\gamma}$ is the number of Cherenkov photons detected.
Based on the SLAC FDIRC prototype results, we expect the mean number of detected photons to be 25. The \gx{} tracking system provides an angular resolution better than 1.5~mrad in the momentum range of interest; thus, the total Cherenkov angle resolution is expected to be better than 2.5~mrad (2.7~mrad) using a 5mm (6mm) detector pixel resolution.  

\begin{table}[!h]
\caption{Cherenkov angle error contributions for the FDIRC detector for a single Cherenkov photon. Table extracted from\cite{NIM7}. }
\label{tab:RESOLUTIONs}
\centering
\begin{tabular}{lc}
  \hline\hline
  Source of uncertainty & Contribution [mrad] \\
  \hline
Chromatic error & 5.5 \\
Pixel contribution 5mm (6mm) & 5.8 (7) \\
Optical aberration & 4.5 \\
Transport along the bar & 2-3 \\
Bar thickness (after focusing) & $\approx$ 1 \\
Old wedge bottom inclined surface & 3.5 \\
\hline
Final error w/o chromatic correction & 10 (11) \\
\hline\hline
\end{tabular}
\end{table}
 
For a particle with $\beta \approx 1$ and momentum ($p$) well above threshold entering the quartz bar, the number of $\sigma$ separation ($N_{\sigma}$) between pions and kaons can be approximated as
 \begin{equation}
N_{\sigma} \approx \frac{|m_{\pi}^{2} -m_{k}^{2}|}{2p^{2}\sigma\left[ \theta_{C}\right]  \sqrt{n^{2}-1}},
\label{CherenkovAngleResolution}
\end{equation}
where $m_{\pi}$ ($m_{k}$) is the pion (kaon) mass. 
A 2.5~mrad Cherenkov angle resolution provides $K/\pi$ separation of at least 3$\sigma$ up to around 4~GeV/$c$. 


\subsubsection{EM background}

\begin{figure}[]
  \includegraphics[width=0.45\textwidth]{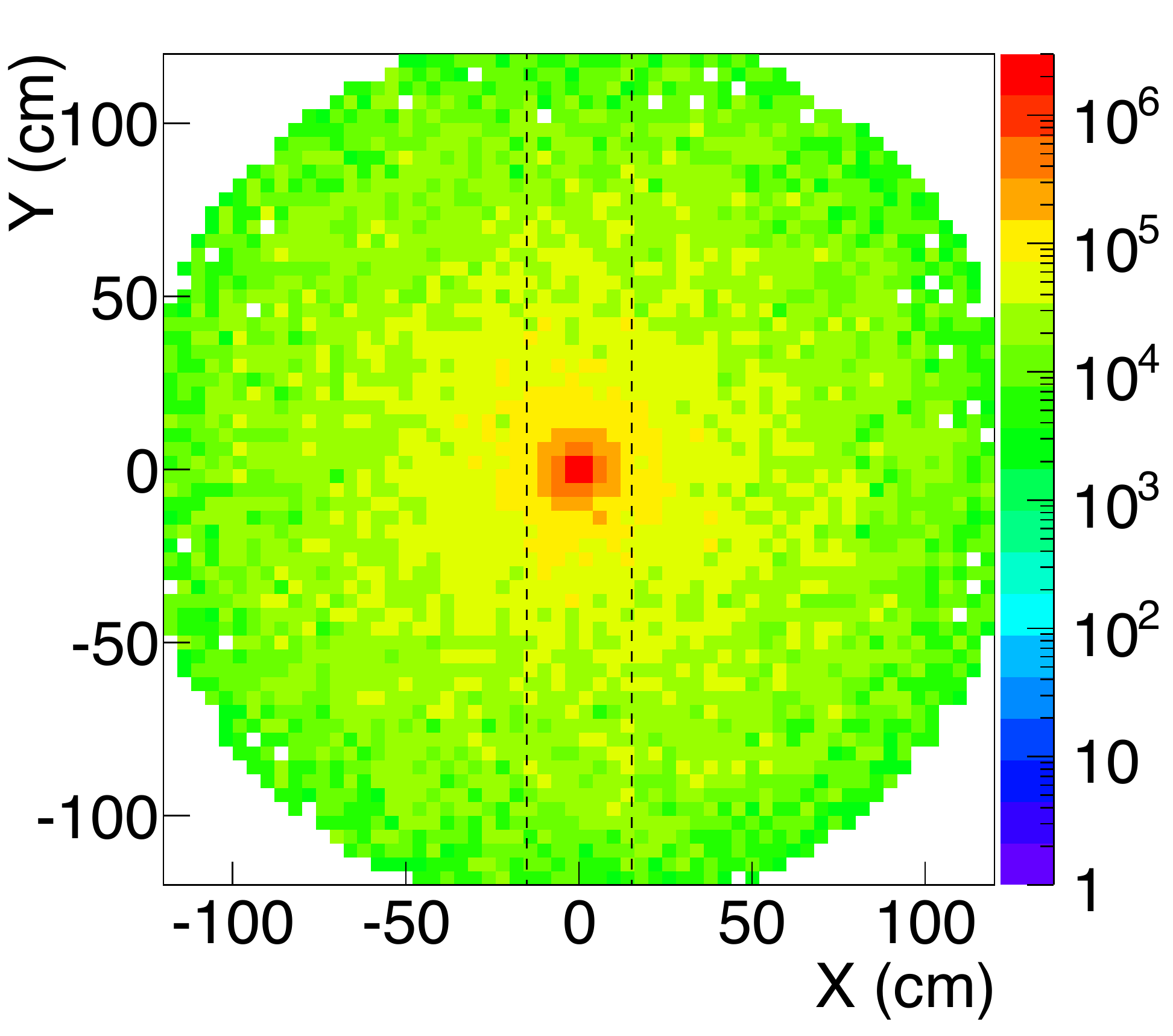}
\caption{EM background expected in a plane near the FDIRC (arbitrary normalization).  Each bin has
a size of 4~cm$\times$4~cm.  The dashed lines show location of the inner edge of the first FDIRC bar.}
\label{fig:em_bkgd}
\end{figure}

Interactions of the photon beam inside the \gx~detector produce background in the FDIRC from secondary electrons, positrons and photons. 
This EM background rate is important for two reasons: it will result in noise making determination of Cherenkov angles more difficult and it will cause more electronic channels to be read out increasing the event data size.  The EM background from the beam was simulated using GEANT and the full \gx~MC.   The rate at which particles produced by EM interactions enter the FDIRC is highly position dependent.  As can be seen in Fig.~\ref{fig:em_bkgd}, the rate falls off exponentially with distance from the beam line.  The closest FDIRC bar will be placed 15~cm from the beam.  Integrating over the region covered by the FDIRC, we expect roughly 8 photoelectrons per time segment readout by the data acquisition; thus, the EM background rate will not cause any significant increase in event size or reduction in performance.  

\subsection{Strangeness reactions of interest}

As described in section~\ref{Sec:KaonMotivation}, to fully explore the spectrum of hybrid mesons, a systematic study of many hadronic final states is necessary, including those with kaons.  The hybrid mesons with exotic quantum-number states that decay to kaons include the $\eta_{1}^{\prime}$, $h_{0}^{\prime}$, the $h_{2}^{\prime}$, which are all expected to couple to the $KK\pi\pi$ final state, while both the $\eta_{1}^{\prime}$ and the $h_{2}^{\prime}$ are expected to couple to the $KK\pi$ final state.  To study the \gx{} sensitivity to these two final states, we have modeled two decay chains. For the $KK\pi$ state, we assume one of the kaons is a $K_{S}$, which leads to a secondary vertex and the $K^{+}\pi^{-}\pi^{+}\pi^{-}$ final state:
\begin{eqnarray}
\label{eq:eta1}
\eta_{1}^{\prime}(2300) & \rightarrow & K^{*}K_{S} \nonumber \\
                                  & \rightarrow & (K^{+}\pi^{-})(\pi^{+}\pi^{-}) \nonumber \\ 
                                  & \rightarrow & K^{+}\pi^{-}\pi^{+}\pi^{-}. 
\end{eqnarray}
For the $KK\pi\pi$ state we assume no secondary vertex:
\begin{eqnarray}
\label{eq:h2}
h_{2}^{\prime}(2600)      & \rightarrow & K_{1}^{+}K^{-} \nonumber \\
                                  & \rightarrow & (K^{*}(892)\pi^{+})K^{-} \nonumber \\
                                  & \rightarrow & K^{+}K^{-}\pi^{-}\pi^{+}. 
\end{eqnarray}

In addition to the exotic hybrid channels, there is an interest in non-exotic $s\bar{s}$ mesons. 
In order to study the sensitivity to conventional $s\bar{s}$ states, we consider 
an excitation of the normal $\phi$ meson, the known $\phi_{3}(1850)$, which decays to $K\bar{K}$ 
\begin{eqnarray}
\label{eq:phi1850}
\phi_{3}(1850) & \rightarrow & K^{+}K^{-} \, .
\end{eqnarray}
The detection efficiency of this state will be typical of $\phi$-like states decaying to the same
final state.
Finally, as noted in Section~\ref{Sec:KaonMotivation}, the $Y(2175)$ state is viewed 
as a potential candidate for a non-exotic hybrid and has been reported in the decay mode
\begin{eqnarray}
\label{eq:Y2175}
Y(2175) & \rightarrow & \phi f_{0}(980) \nonumber \\
             & \rightarrow & (K^{+}K^{-})(\pi^{+}\pi^{-})  \, .
\end{eqnarray}
While this is the same $KK\pi\pi$ state noted in reaction~\ref{eq:h2} above, the intermediate resonances make
the kinematics of the final state particles different from the exotic decay channel noted above.
Therefore, we simulate it explicitly.
The final-state kaons from the reactions~\ref{eq:eta1} - \ref{eq:Y2175}
will populate the \gx{} detector differently, with different overlap of the region
where the time-of-flight system can provide good $K$/$\pi$ separation.

The remainder of this section describes a study of the sensitivity of the baseline \gx{} detector to these reactions of interest involving kaons (Sec.~\ref{Sec:BaselineGlueX}), and the expected increase in sensitivity with the proposed FDIRC detector in \gx{} (Sec.~\ref{Sec:FDIRCglueX}).  The studies were performed using a larger scale \textsc{pythia} simulation of $\gamma p$ collisions processed through a complete \textsc{geant} model of the baseline \gx{} detector and fully reconstructed with the \gx{} analysis software.  Signal samples were obtained from \textsc{pythia} events with the generated reaction topology, and the remainder of the inclusive photoproduction reactions were used as the background sample.  Since many of the cross sections of interest are unknown we use \textsc{pythia} to predict the size of signal topologies of interest.  The reactions listed above are several typical meson photoproduction reactions chosen to demonstrate FDIRC capability.  The benefits of the FDIRC will certainly extend beyond these to other meson and baryon channels, some of which were discussed in our earlier proposals~\cite{pac39,pac40}.

\subsubsection{Performance of the baseline \gx{} detector}
\label{Sec:BaselineGlueX}

The baseline \gx{} detector does not contain any single detector element that is capable of providing discrimination of kaons from pions over the full-momentum range of interest for many key reactions.  However, the hermetic \gx{} detector is capable of exclusively reconstructing all particles in the final state.  In the case where the recoil nucleon is a proton that is detectable by the tracking chamber, this exclusive reconstruction becomes a particularly powerful tool for particle identification because conservation of four-momentum can be checked, via a kinematic fit, for various mass hypotheses for the final state particles.  Many other detector quantities also give an indication of the particle mass, as assumptions about particle mass (pion or kaon) affect interpretation of raw detector information.

An incomplete list of potentially discriminating quantities include:
\begin{itemize}
\item The confidence level (CL) from kinematic fitting that the event is consistent with the desired final
state.
\item The CL(s) from kinematic fitting that the event is consistent with some other final states.
\item The goodness of fit ($\chi^{2}$) of the primary vertex fit.
\item The goodness of fit ($\chi^{2}$) of each individual track fit.
\item The CL from the time-of-flight detector that a track is consistent with the particle mass.
\item The CL from the energy loss ($dE/dx$) that a track is consistent with the particle type.
\item The change in the goodness of fit ($\Delta \chi^{2}$) when a track is removed from the 
primary vertex fit.
\item Isolation tests for tracks and the detected showers in the calorimeter system.
\item The goodness of fit ($\chi^{2}$) of possible secondary vertex fits.
\item Flight-distance significance for particles such as $K_{S}$ and $\Lambda$ that lead to secondary
vertices.
\item The change in goodness of fit ($\Delta \chi^{2}$) when the decay products of a particle that 
produces a secondary vertex are removed from the primary vertex fit. 
\end{itemize}

The exact way that these are utilized depends on the particular analysis, but it is generally better to try to utilize as many of these as possible in a collective manner, rather than simply placing strict criteria on any one of them. This means that we take advantage of correlations between variables in addition to the variables themselves.  One method of assembling multiple correlated measurements into a single discrimination variable, which has been used in this study, is a boosted decision tree (BDT)~\cite{ref:bdt}.  Traditionally, analyses have classified candidates using a set of variables, such as a kinematic fit confidence level, charged-particle time of flight, energy loss ($dE/dx$), {\em etc.}, where cuts are placed on each of the input variables to enhance the signal.  In a BDT analysis, however, cuts on individual variables are not used; instead, a single classifier is formed by combining the information from all of the input variables.

A BDT is a multivariate classifier which is trained on a sample of known signal and background events to select signal events while maximizing a given figure of merit.  The event selection performance is validated using an independent data sample, called a validation sample, that was not used in the training.  If the performance is found to be similar when using the training (where it is maximally biased) and validation (where it is unbiased) samples, then the BDT performance is predictable.  Practically, the output of the BDT is a single number for each event that tends towards one for signal-like events but tends towards negative one for background-like events.  Placing a requirement on the minimum value of this classifier, which incorporates all independent information input to the BDT, allows one to enhance the signal purity of a sample.  For a pedagogical description of BDTs, see Ref.~\cite{ref:roe}.  The BDT algorithms used are contained within ROOT in the TMVA package~\cite{ref:TMVA2007}.

Here we only consider the case where the recoil proton is reconstructed.  A missing recoil nucleon reduces the number of constraints in the kinematic fit, and, consequently, dramatically diminishes the capability of the fit to discriminate pions from kaons. One can build a BDT for the reaction of interest, and look at the efficiency of selecting true signal events as a function of the sample purity.  These studies do not include the efficiency of reconstructing the tracks in the detector, but start at the point where a candidate event containing five charged tracks has been found.  In all cases we set the requirement on the BDT classifier in order to obtain a fixed final sample purity.  For example, a purity of 90\% implies a background at the 10\% level.  Any exotic signal in the spectrum would likely need to be larger than this background to be robust.  Therefore, with increased purity we have increased sensitivity to smaller signals, but also lower efficiency.  In Table~\ref{tab:bdt_eff} we present the signal selection efficiencies (post reconstruction) for our four reactions of interest for the baseline \gx{} detector and including a FDIRC detector in \gx{} (more in Section~\ref{Sec:FDIRCglueX}).  As noted earlier, these assume that the tracks have been reconstructed and do not include that efficiency.  With the baseline \gx{} detector, higher signal purities of 95\% to 99\%, which may be necessary to search for more rare final states, result in the signal efficiency dropping dramatically.   This exposes the limit of what can be done with the baseline \gx{}~hardware.  

\begin{table*}\centering
\caption[]{\label{tab:bdt_eff}Efficiencies for identifying several final states in \gx~excluding reconstruction of the final state tracks.}
\begin{tabular}{c|cccccccc} \hline\hline
~~~~~~~~~~~~  & \multicolumn{2}{c}{~~~$\eta_{1}^{\prime}(2300)\to K^*K_S$~~~} & \multicolumn{2}{c}{~~~$h_{2}^{\prime}(2600)\to K_1^+K^-$~~~} & \multicolumn{2}{c}{~~~$\phi_{3}(1850)\to K^+K^-$~~~} &  \multicolumn{2}{c}{~~~$Y(2175)\to \phi f_0(980)$~~~} \\ \hline
Purity & ~~Baseline & ~~FDIRC & ~~Baseline & ~~FDIRC &  ~~Baseline & ~~FDIRC & ~~Baseline & ~~FDIRC\\
~0.90~ & ~~0.36 & ~~0.48 & ~~0.33 & ~~0.49 & ~~0.67 & ~~0.74 & ~~0.46 & ~~0.65 \\
~0.95~ & ~~0.18 & ~~0.33 & ~~0.16 & ~~0.34 & ~~0.61 & ~~0.68 & ~~0.20 & ~~0.55 \\ 
~0.99~ & ~~0.00 & ~~0.05 & ~~0.00 & ~~0.08 & ~~0.18 & ~~0.38 & ~~0.03 & ~~0.28 \\
\hline\hline
\end{tabular}
\end{table*}

\subsubsection{Limitations of existing kaon identification algorithms}

It is important to point out that the use of kinematic constraints to achieve kaon identification, without dedicated hardware, has limitations.  By requiring that the recoil proton be reconstructed, we are unable to study charge exchange processes that have a recoil neutron.  In addition, this requirement results in a loss of efficiency of 30\%-50\% for proton recoil topologies and biases the event selection to those that have high momentum transfer, which may make it challenging to conduct studies of the production mechanism. Our studies indicate that it will be difficult to attain very high purity samples with a multivariate analysis alone.  In channels with large cross sections, the \gx{} sensitivity will not be limited by acceptance or efficiency, but by the ability to suppress and parameterize backgrounds in the amplitude analysis; thus, we need high statistics and high purity.  This latter statement is not only true for reactions that contain kaons but also applies to to reactions in which the dominant backgrounds arise from kaons.  Supplemental kaon identification hardware will help suppress these background thereby enhancing the sensitivity.  Finally, it is worth noting that our estimates of the kaon selection efficiency using kinematic constraints depends strongly on our ability to model the performance of the detector.  Although we have constructed a complete simulation, the experience of the collaboration with comparable detector systems indicates that the simulated performance is often better than the actual performance in unforeseen ways.

\subsubsection{Performance with FDIRC detector in \gx{}}
\label{Sec:FDIRCglueX}

As described in Sec.~\ref{Sec:singleTrack}, the single track particle identification of the FDIRC in \gx{} is expected to provide $3\sigma~K/\pi$ separation up to momentum of $\approx4$~GeV/$c$.  This provides vital, independent information to the multivariate analysis that has a very high discrimination power.  The FDIRC information is included in the BDT by converting the measured Cherenkov angle into a probability for each particle mass hypothesis ($\pi, K,$ and $p$).  We define a $\chi^2$ for each particle mass hypothesis as
\begin{equation}
\chi^2_i = \frac{\left(\theta_{C,i}^{exp} - \theta_C^{reco}\right)^2}{\sigma_{\theta_C}^2},
\label{Eqn:chi2}
\end{equation}
\vspace{0.02in}\\
\noindent where $\theta_{C,i}^{exp}$ is the expected Cherenkov angle for mass hypothesis $i$ using the measured track momentum from the drift chambers,  $\theta_C^{reco}$ is the ``reconstructed'' Cherenkov angle, and $\sigma_{\theta_C}$ is the Cherenkov angle resolution.   

As we do not yet have a full FDIRC reconstruction algorithm, we use Eq.~\ref{Eqn:chi2} as a proxy for the FDIRC performance.  We use $\sigma_{\theta_C} = 2.5$~mrad for all tracks (this is an upper bound on the expected resolution; see Sec.~\ref{Sec:singleTrack}).
The track momentum resolution (see Fig.~\ref{Fig:TrackPosMomRes}) is included in $\theta_{C,i}^{exp}$.  The ``reconstructed'' Cherenkov angle is obtained by generating a random number from a Gaussian distribution whose mean is the expected Cherenkov and width is $\sigma_{\theta_C}$.  A confidence level for each particle mass hypothesis ($\pi$, $K$, $p$) is computed from Eqn.~\ref{Eqn:chi2}.  These three values for each track are included in the BDT training, and the performance is evaluated in the same way as the baseline \gx{} detector and shown in Table~\ref{tab:bdt_eff}.
We note that, depending on the choice of readout, the FDIRC may provide an improvement in time-of-flight measurements
for charged particles over our baseline design.  Further study is needed to quantify this improvement; therefore, we neglect 
it in the studies presented below.

At 95\% purity, the signal efficiencies are typically about twice as high including the FDIRC into \gx.   Reaching 99\% purity is not possible for several of these channels without the FDIRC.  It is important to stress here that the purity levels are defined as correctly identified final state candidates divided by all candidates.  In the case that exotic contributions to some channel are at the percent level, extracting such signals will require reaching 99\% purity, which helps ensure that the backgrounds are smaller than the small signal of interest.  Without the FDIRC, this will not be possible for many channels of interest.  Finally, as noted above, the baseline numbers are dependent on the reliability of the simulation. For example, the discrimination power of the kinematic fit confidence level will decrease drastically if the \gx~detector resolution is worse than expected.  The simulation of the FDIRC performance is based only on the Cherenkov-angle resolution.  The value of 2.5~mrad is expected to be achievable; thus, the real-world performance enhancement obtained by adding the FDIRC is likely to be even greater than what is shown in Table~\ref{tab:bdt_eff}.

\subsection{Effects of the FDIRC on other \gx~Systems}

\begin{figure}[]
            \includegraphics[width=0.45\textwidth]{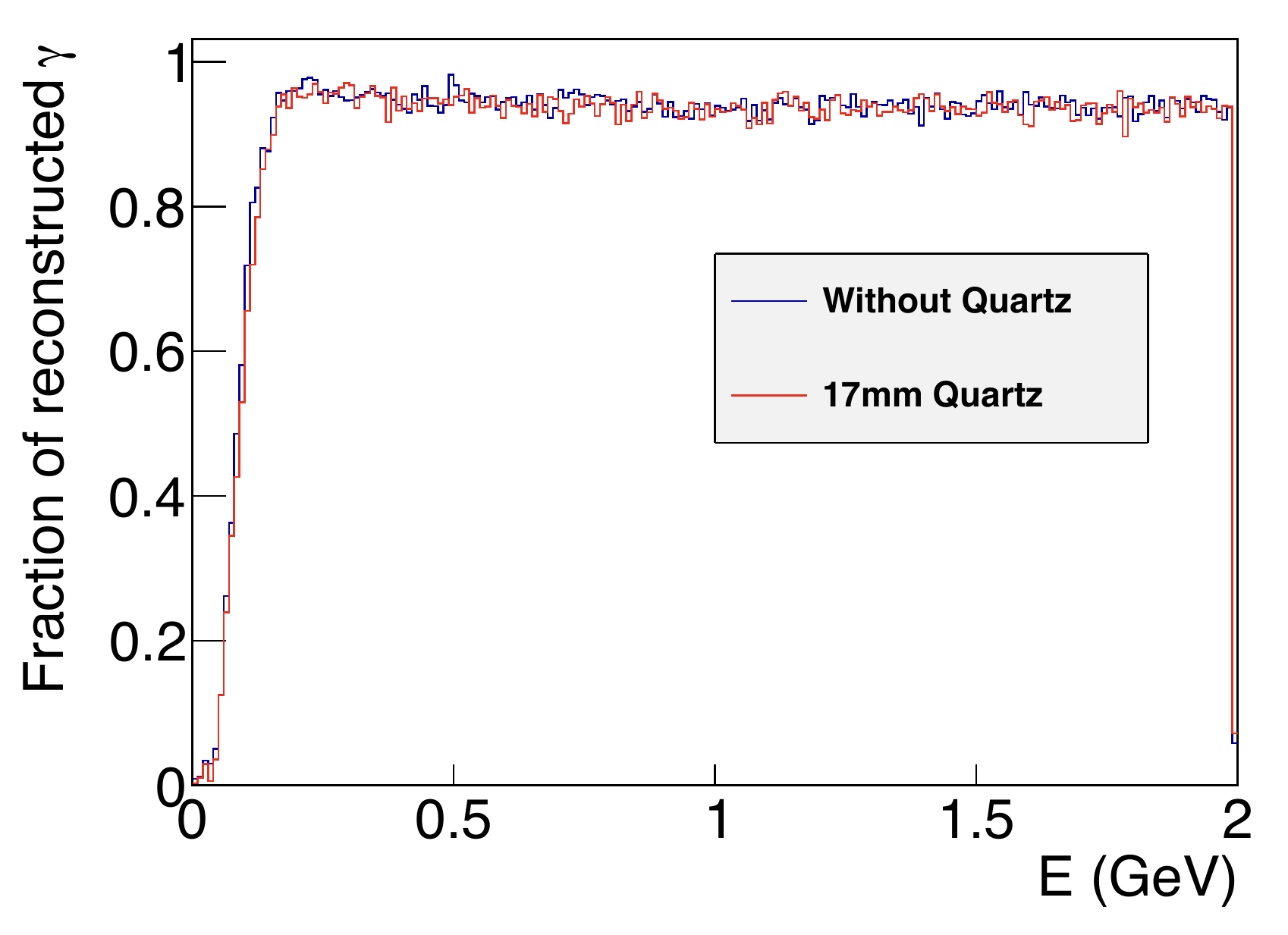}\quad 
\caption{Effects of the FDIRC material on FCAL photon reconstruction.} 
	\label{FCALeff}
\end{figure}

Installing the FDIRC results in a significant increase in material upstream of the FCAL.  We have studied the effects of the FDIRC on the FCAL performance using GEANT and found them to be minimal (see Fig.~\ref{FCALeff}).  The photon energy detection threshold increases from 160~MeV to 180~MeV.  Above 500~MeV the photon reconstruction efficiency is unaffected.  The small electron-positron opening angle from converted photons, along with the small distance between the FDIRC and FCAL, results in a single EM shower; thus, the effect of the FDIRC on photon reconstruction is minimal.  

\section{\gx{} FDIRC construction plan}

In this section we describe our preliminary construction schedule, budget, and provide a discussion of
logistical details concerning construction, specifically transportation of the fragile DIRC components from
SLAC to Jefferson Lab.

\subsection{Preliminary schedule}

As noted earlier, our ultimate goal is to have the \gx{} FDIRC operational for the Phase IV
running that is currently estimated to take place in around 2017-2018.  During 2014 we plan to develop
a technical design for the detector, including a complete cost estimate and detailed
construction schedule.  We expect Jefferson Lab to appoint an external committee to review this
technical design and construction plan during the summer of 2015.  During 2014 we will
continue to work with the LAPPD collaboration to be certain that large area photodetectors
will provide a feasible solution for the FDIRC photon camera.  Additional design efforts in
2014-2015 include optimizing the optics, exploring cost effective alternatives for the fluid in the FOB, 
and conducting materials testing to verify that the vertical orientation of the bar box is structurally sound.
Construction of the support structure and FOB could begin in fiscal year 2016, after a technical
design review.  Integration and installation into the hall could then be complete in time for
operation around 2017-2018.

\subsection{Transportation of DIRC components to Jefferson Lab}

We have had preliminary
discussions with Rock-It Cargo, a world-wide shipper of delicate art and industrial equipment.
We plan to transport the bars from SLAC to Jefferson Lab over road via air ride trailer.  The trailer will be temperature
controlled and equipped with a liquid nitrogen dewar to maintain constant flow of nitrogen through the bar boxes.
There is concern that optical joints between bars may be brittle, therefore physical shock should be avoided
during transport.  Based on discussions with Rock-It Cargo, a crew would
construct a custom shipping crate under our supervision, which would then be transported to SLAC for loading.
The crate would incorporate metal substructure to prohibit flexing of the boxes.  The boxes themselves are relatively
lightweight, which means that foam materials may be used to attenuate transport vibration.  A group from Livermore
Lab studied accelerations of a 12 ton load being transported by air ride trailer~\cite{transport} and found that 
accelerations never exceeded 1.5$g$'s in all three dimensions when the trailer was driven at 40 mph on
typical Oakland, CA roads.  Careful container design will attenuate these shock loads.
If necessary, the mass of the trailer can be increased with an additional dummy load, which 
should further reduce acceleration.  As a conservative limit, we evaluate internal stresses assuming a  3$g$ 
acceleration will be experienced during transit.  

Assuming the bar box and internal bar support buttons remain rigid, acceleration of the box would cause bending of
the bars in between the support buttons.  Any elastic compression of the buttons would mitigate this bending.
We evaluated the bending stress of a quartz bar when subjected to a 3$g$ load applied at a point in the center
between two supports and parallel to the narrow dimension of the bar, which is a worse case scenario.  We found 
the bending stress for a bar supported by buttons with 600 mm spacing, typical near the center of the bar, to be about 
300 psi, which is over an order of magnitude less than typical tensile or rupture strengths for fused silica.  In the vicinity
of a bar-to-bar epoxy joint the button spacing is assumed to be 25 mm and the corresponding bending stress is
about 1 psi.  Reference~\cite{NIM} states a tensile strength of epoxy used in the DIRC that exceeds 1000 psi.
The most sensitive area to such bending appears to be the region between the window and the first bar, a distance
of about 100~mm that is occupied by the wedge and not supported by buttons.  Here a conservative estimate of
the bending stress, assuming the wedge has the same profile as the bar, yields an estimate of 9 psi.  Even with
the consideration that the strength of the epoxy may be degraded due to aging, it seems feasible to transport the 
components over road, provided that appropriate packaging and other considerations are made.

Prior to shipping the actual DIRC bars, we plan to instrument and ship a prototype bar
box at SLAC that is filled with ordinary glass in an appropriate shipping crate.  
This will give us an opportunity to measure $g$-loads that may be experienced in transit.  
In addition small pieces of unpolished fused silica with a cross sectional profile the same
as the actual bars can be procured, glued with the same epoxy as used in construction, 
and tested for strength.

\subsection{SLAC resources}

A significant amount of infrastructure for DIRC construction and testing still remains
at SLAC.  The clean room used for assembling the bar boxes, with its
large granite surface table that is capable of accommodating a full bar box, is still
in usable condition.  Assuming that our final optical design requires us to glue
wedges to the existing bar boxes, it may be optimal to perform these operations in
the SLAC clean room prior to transporting the bar boxes to Jefferson Lab.
Also at SLAC is a cosmic ray test facility that is equipped with a
bar box and muon hodoscope.  This facility provides a unique opportunity to test readout
and electronics with actual Cherenkov signals produced by cosmic ray muons.
Due to the size of the bar box and the precision required of the timing and tracking
system for cosmic rays, such a test setup cannot be easily replicated elsewhere.
Finally, the personnel at SLAC who have experience with the BaBar
DIRC have already provided an enormous amount of beneficial information in
the development of our conceptual design; we hope to draw on this expertise,
if possible, as we continue with the design and construction
of a \gx{} FDIRC.

\subsection{Preliminary budget}

A preliminary material cost estimate for the FDIRC detector at \gx~is shown in Table~\ref{tab:COST}.  
This budget does not include technical or engineering manpower, indirect costs, or project management costs.  
Costs for wedge material and assembly are estimated from vendor quotes for the wedges as
described in this document.  A detailed optimization that balances performance, cost, and construction
feasibility (technical risk) has not yet been performed.  Costs for mechanical structures are
estimated based on experience with building similar structures.  We base our current cost
estimate for the photosensors and readout on our desire to use the LAPPD collaboration
sensors, but recognize this technology is not yet available.  Alternate sensors and readout options using
existing technology are listed in the table for reference.  All readout quotes include low 
voltage, crates, cables, and other necessary infrastructure to integrate with the Jefferson
Lab data acquisition system.

\begin{table*}[!h]
\centering
\caption{Estimated material cost for the \gx{} FDIRC in thousands of dollars.  Costs for alternate photosensor and readout options are shown in brackets but not included in total estimated cost.  These costs do not include manpower, overhead, or project management costs.}
\label{tab:COST}
\begin{tabular}{lr}
  \hline\hline
  Item & Estimated Cost [k\$] \\
  \hline
\underline{Focussing oil box:}  &  \\
New wedge material, machining, and polishing & \$190 \\
Wedge assembly infrastructure & \$15 \\
Oil (CARGILLE) & \$120 \\
Focusing oil box & \$10 \\
Mirrors & \$40 \\
\hline
\underline{Photosensors and readout:}  & \\
LAPPD: 26 tiles $\times$ \$6 & \$156 \\
LAPPD PSEC4 readout: 900 channels $\times$ \$0.20 & \$180 \\
( LAPPD TDC readout: 900 channels $\times$ \$0.30 ) &( \$270 ) \\
( MaPMT: 318 H8500 MaPMTs $\times$  \$2.5 ) & ( \$795 ) \\
( MaPMT MAROC3 readout: 318 $\times$ \$0.83 ) & ( \$262 ) \\
\hline
Detector support structure & \$50 \\
Bar box transport to Jefferson Lab & \$30  \\
Calibration, monitoring, and control systems & \$40 \\
\hline
{\bf Total estimated material cost} & {\bf \$831}   \\
\hline\hline
\end{tabular}
\end{table*}


\section{Conclusion and Acknowledgements}

The ability to reconstruct kaon final states is absolutely critical in the context of attempting
to study mesons and baryons with both explicit and hidden strangeness.  Following
the request of PAC39, we have developed and presented  a conceptual design for an 
FDIRC detector to enhance the particle identification
capabilities of the \gx{} experiment.  The FDIRC utilizes one-third of the quartz bars from the
BaBar DIRC along with the bar boxes that house the bars.  A focussing optical system consisting
of mirrors submerged in oil is proposed.  Our plan is to construct an optical system and readout
around the large area micro-channel plate PMTs under development by the LAPPD Collaboration.
However, alternate options based on multi-anode PMTs, which are more expensive but less
technically risky exist.  

The FDIRC provides enhanced PID capability for the \gx{} experiment 
that will increases the sensitivity and reduce backgrounds for final state topologies that are 
necessary to search for $s\bar{s}$ hybrid mesons and infer their quark flavor content.  We
propose a program consisting of 20 PAC days of beam for commissioning the FDIRC followed
by 200 days of beam at an average intensity of $5\times 10^7~\gamma$/s in order to conduct
a program of studying the spectrum of hadrons that contain valence strange quarks.  The proposed
running is complementary to and could, in principle, be conducted concurrently with that 
approved by PAC40.

We would like to thank J. Va'vra, B. Ratcliff, and B. Wisniewski for their useful discussions and
technical information they provided about the BaBar DIRC.  We thank M. Benettoni and INFN of
Padova for computer models of the BaBar DIRC box.

\section{Appendix}

\begin{figure*}[!h]
            \includegraphics[height=5.2cm]{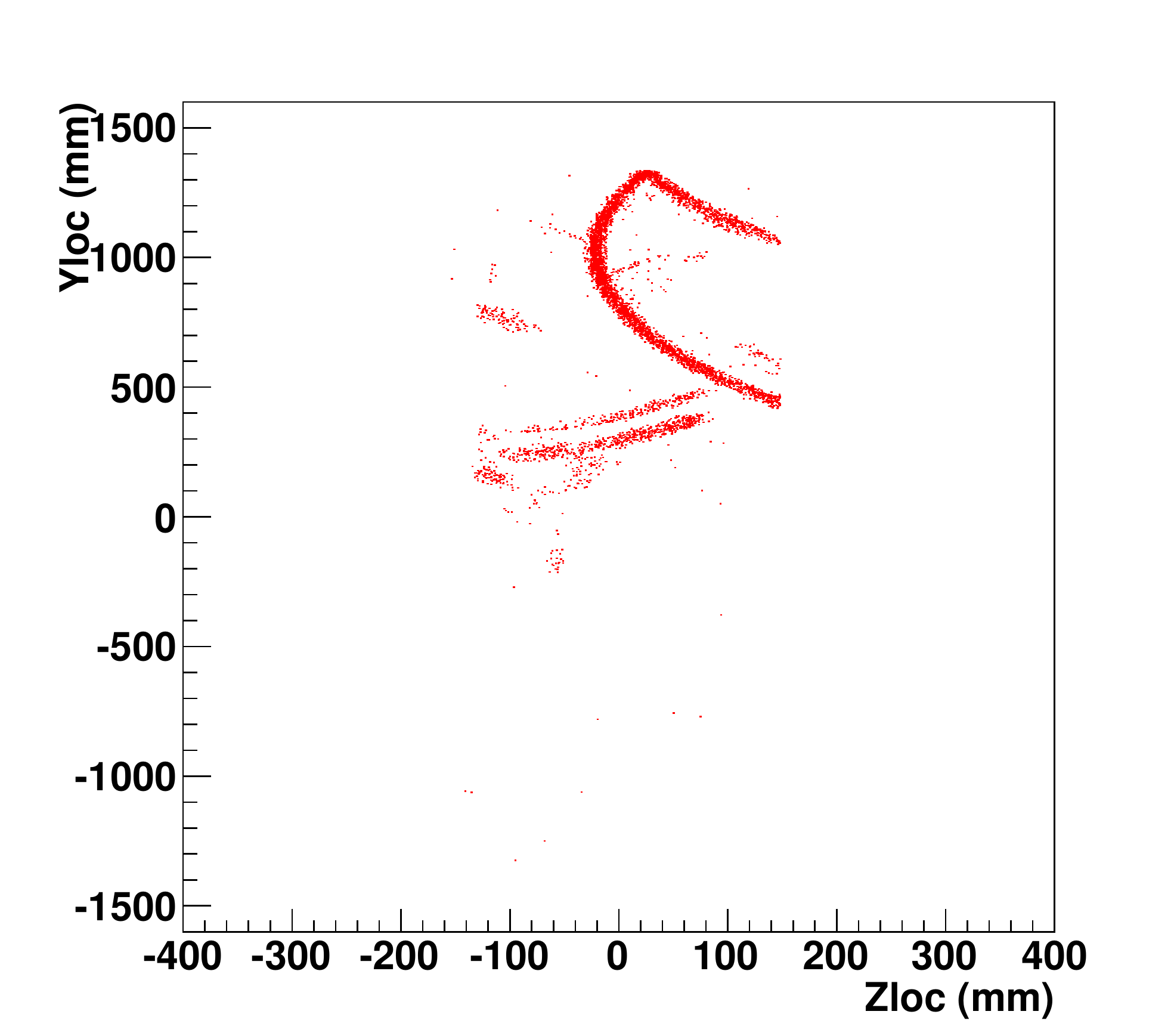}\quad 
            \includegraphics[height=5.2cm]{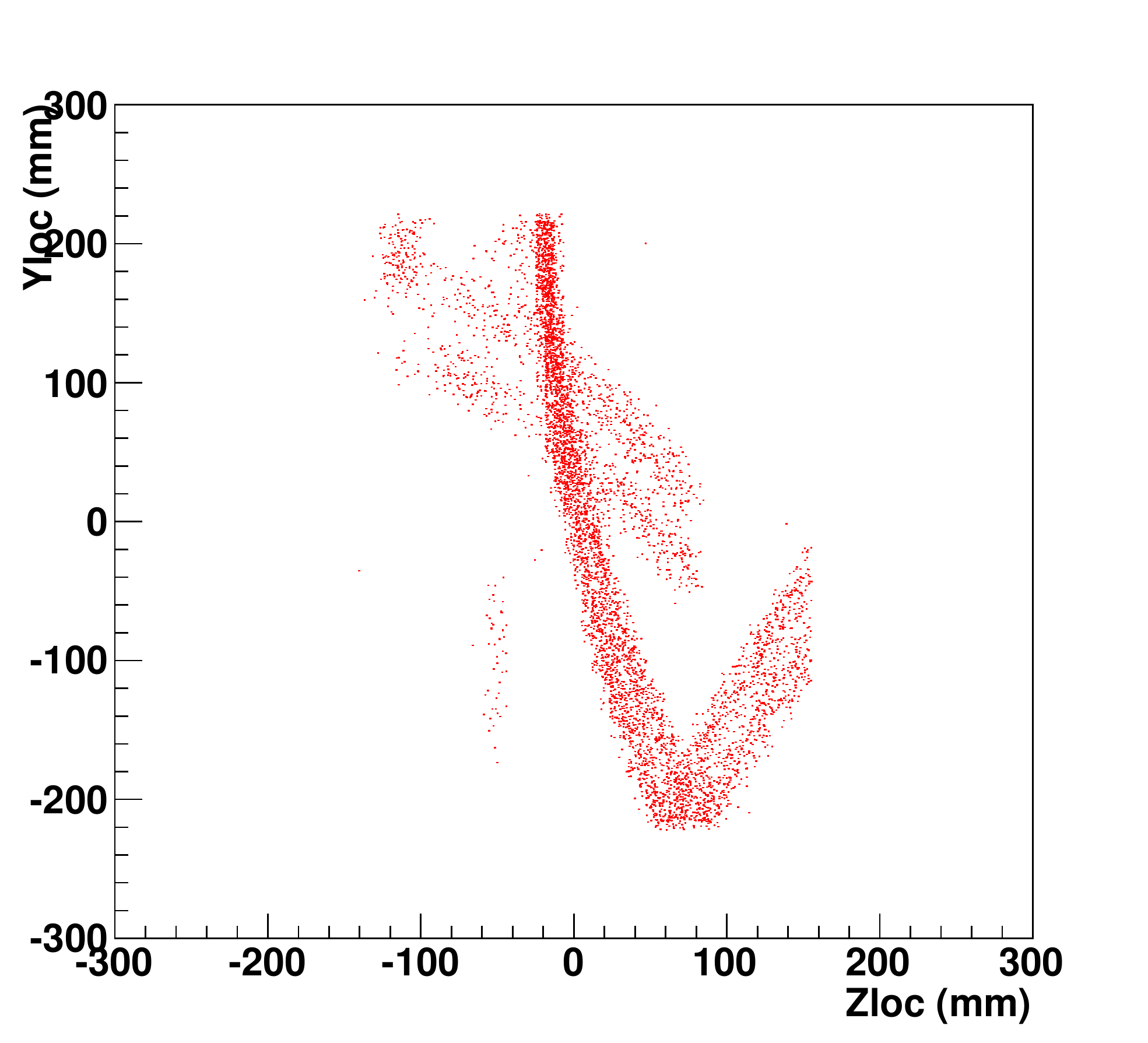}  \\
            \includegraphics[height=5cm]{figures/YZ_barcentrale.pdf}\quad 
            \includegraphics[height=4.8cm]{figures/YZ_MiddleBar.pdf}  \\
            \includegraphics[height=5cm]{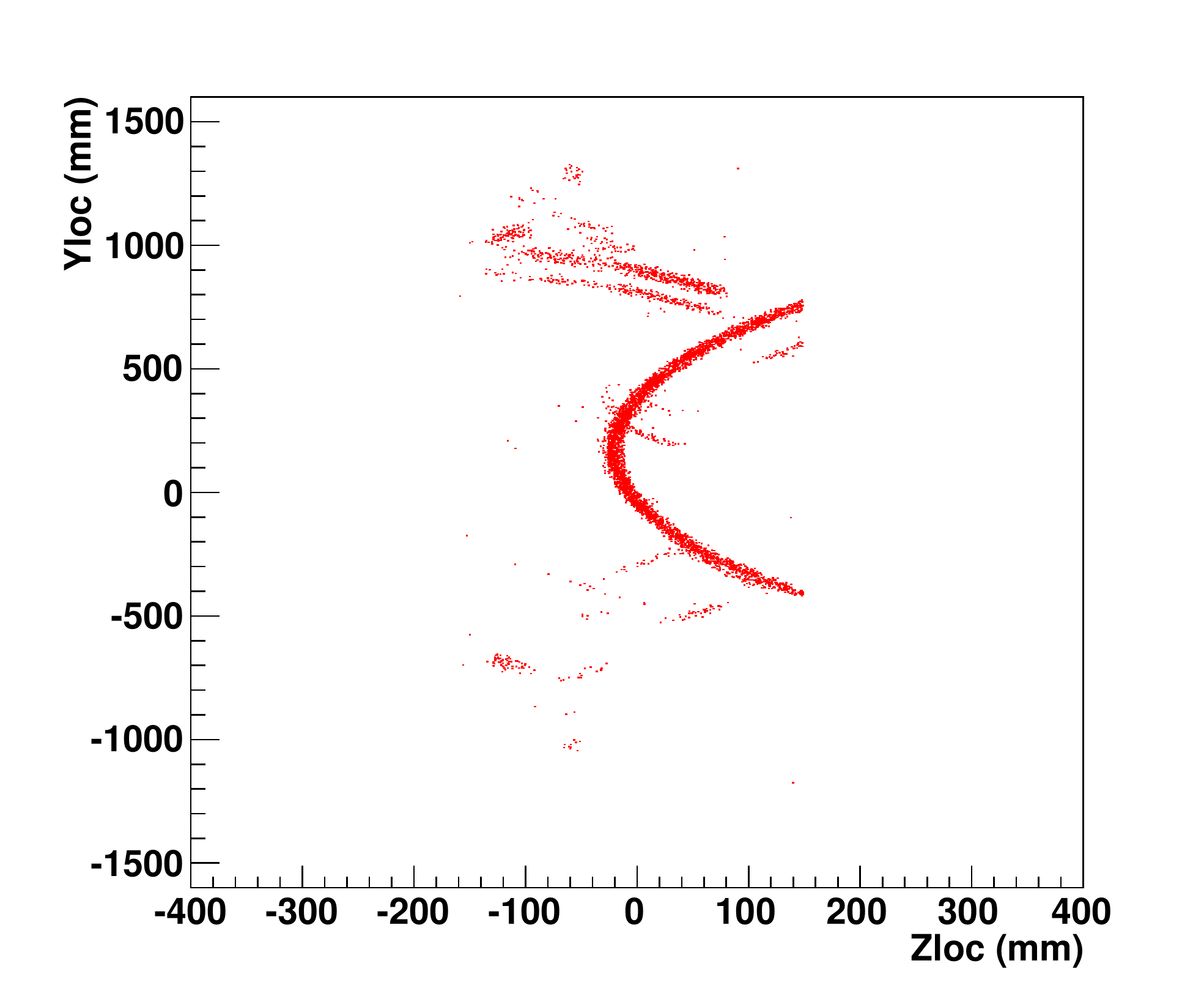}\quad 
            \includegraphics[height=5cm]{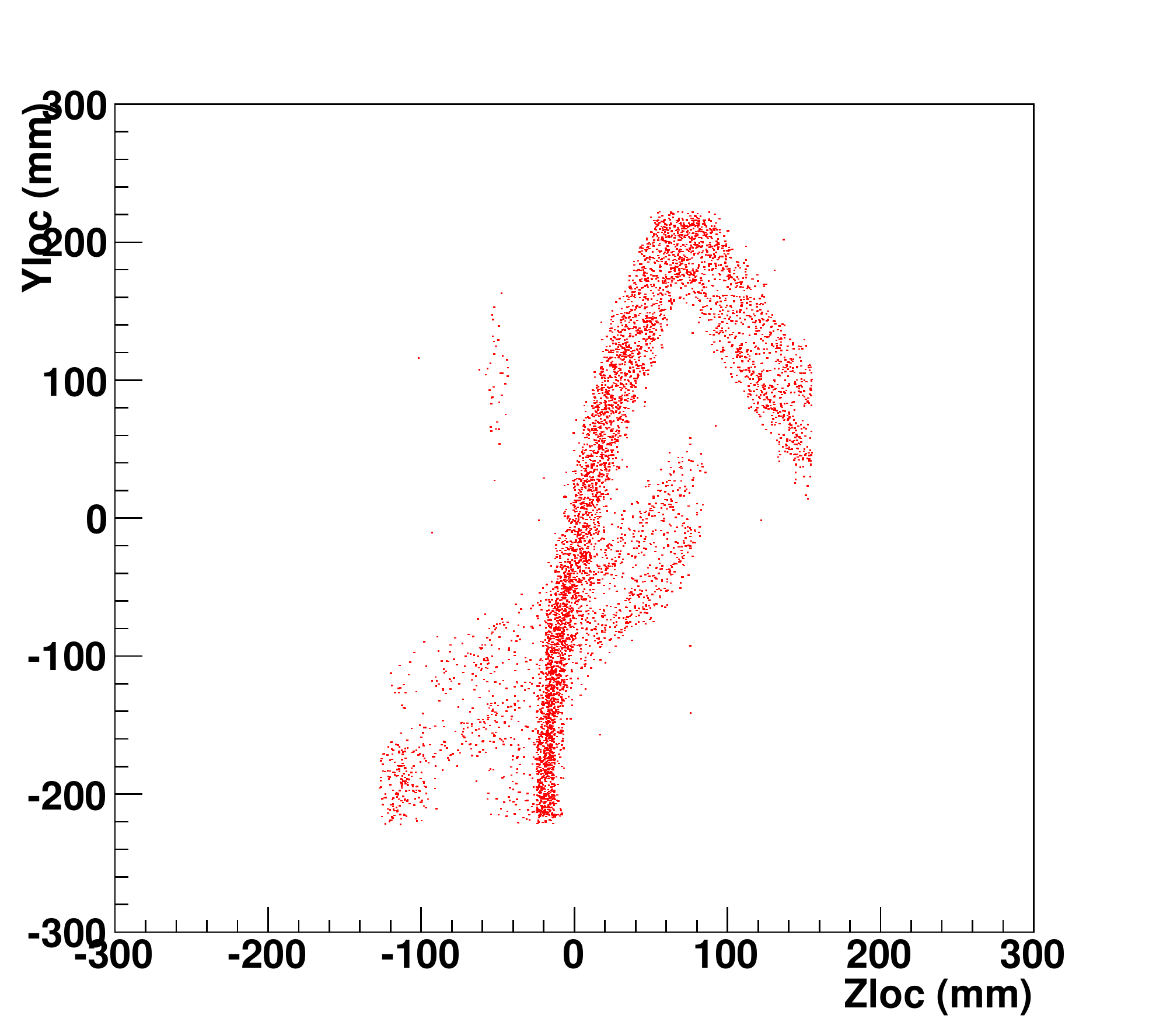}  \\
	\caption{Comparison of the (right) original SLAC design to the (left) adapted version of the camera for \gx. The figures correspond to the following: (top) a hit in the bar the far from the beam line; (middle) a hit in the central part of a box; and (bottom) a hit in the bar closest to the beam line.} 
	\label{YZcomparison}
\end{figure*}

Figure~\ref{YZcomparison} shows the expected distribution of photons on the PMT plane for charged pions intersecting the DIRC at various locations.  The \gx~design greatly reduces the overlap in the patterns.  
There are still side reflections for hits in the bars that are farthest from the beam line.  These reflections could be removed by instrumenting an additional 300~mm along the length of the FOB with PMTs; however, the cost of this extra instrumentation outweighs the benefits as it is unlikely to get particles near the limits of $\pi/K$ separation in this region. 

\begin{thebibliography}{2}

\bibitem{pac39}
  \gx~Collaboration,  ``A study of meson and baryon decays to strange final states with \gx~in Hall D," {\it Presentation to PAC 39}, (2012).  Available at: \href{http://www.gluex.org/docs/pac39_proposal.pdf}{\it http://www.gluex.org/docs/pac39\_proposal.pdf}

  \bibitem{pac40}
  \gx~Collaboration,  ``An initial study of meson and baryon decays to strange final states with \gx~in Hall D," {\it Presentation to PAC 40}, (2013).  Available at: \href{http://www.gluex.org/docs/pac40_proposal.pdf}{\it http://www.gluex.org/docs/pac40\_proposal.pdf}

\bibitem{Dudek:2011bn} 
  J.~J.~Dudek,
  Phys.\ Rev.\ D {\bf 84}, 074023 (2011).
  
  \bibitem{Dudek:2009qf} 
  J.~J.~Dudek, R.~G.~Edwards, M.~J.~Peardon, D.~G.~Richards and C.~E.~Thomas,
  Phys.\ Rev.\ Lett.\  {\bf 103}, 262001 (2009).
  
  \bibitem{Dudek:2010wm} 
  J.~J.~Dudek, R.~G.~Edwards, M.~J.~Peardon, D.~G.~Richards and C.~E.~Thomas,
  Phys.\ Rev.\ D {\bf 82}, 034508 (2010).
  
  \bibitem{Dudek:2011tt} 
  J.~J.~Dudek, R.~G.~Edwards, B.~Joo, M.~J.~Peardon, D.~G.~Richards and C.~E.~Thomas,
  Phys.\ Rev.\ D {\bf 83}, 111502 (2011).
  
  \bibitem{Edwards:2011jj} 
  R.~G.~Edwards, J.~J.~Dudek, D.~G.~Richards and S.~J.~Wallace,
  Phys.\ Rev.\ D {\bf 84}, 074508 (2011).
 
\bibitem{Beringer:1900zz} 
  J.~Beringer {\it et al.}  [Particle Data Group Collaboration],
  Phys.\ Rev.\ D {\bf 86}, 010001 (2012).

 \bibitem{Meyer:2010ku} 
  C.~A.~Meyer and Y.~Van Haarlem,
  Phys.\ Rev.\ C {\bf 82}, 025208 (2010).
  
  \bibitem{Ablikim:2007ab} 
  M.~Ablikim {\it et al.}  [BES Collaboration],
  Phys.\ Rev.\ Lett.\  {\bf 100}, 102003 (2008).
  
  \bibitem{Aubert:2006bu} 
  B.~Aubert {\it et al.}  [BABAR Collaboration],
  Phys.\ Rev.\ D {\bf 74}, 091103 (2006).
  
  
  \bibitem{Shen:2009zze} 
  C.~P.~Shen {\it et al.}  [Belle Collaboration],
  Phys.\ Rev.\ D {\bf 80}, 031101 (2009).

\bibitem{Aubert:2005rm} 
  B.~Aubert {\it et al.}  [BABAR Collaboration],
  Phys.\ Rev.\ Lett.\  {\bf 95}, 142001 (2005).
  
  \bibitem{Coan:2006rv} 
  T.~E.~Coan {\it et al.}  [CLEO Collaboration],
  Phys.\ Rev.\ Lett.\  {\bf 96}, 162003 (2006).
  
  \bibitem{He:2006kg} 
  Q.~He {\it et al.}  [CLEO Collaboration],
  Phys.\ Rev.\ D {\bf 74}, 091104 (2006).
  
  \bibitem{Yuan:2007sj} 
  C.~Z.~Yuan {\it et al.}  [Belle Collaboration],
  Phys.\ Rev.\ Lett.\  {\bf 99}, 182004 (2007).

 \bibitem{Close:2005iz} 
  F.~E.~Close and P.~R.~Page,
  Phys.\ Lett.\ B {\bf 628}, 215 (2005).

  \bibitem{Zhu:2005hp} 
  S.~-L.~Zhu,
  Phys.\ Lett.\ B {\bf 625}, 212 (2005).
  
  \bibitem{Kou:2005gt} 
  E.~Kou and O.~Pene,
  Phys.\ Lett.\ B {\bf 631}, 164 (2005).
    
    \bibitem{Luo:2005zg} 
  X.~-Q.~Luo and Y.~Liu,
  Phys.\ Rev.\ D {\bf 74}, 034502 (2006)
  [Erratum-ibid.\ D {\bf 74}, 039902 (2006)].
      
  \bibitem{Ding:2007pc} 
  G.~-J.~Ding and M.~-L.~Yan,
  Phys.\ Lett.\ B {\bf 657}, 49 (2007).
  
\bibitem{Page:1998gz} 
  P.~R.~Page, E.~S.~Swanson and A.~P.~Szczepaniak,
  Phys.\ Rev.\ D {\bf 59}, 034016 (1999).

\bibitem{Isgur:1985vy} 
  N.~Isgur, R.~Kokoski and J.~E.~Paton,
  Phys.\ Rev.\ Lett.\  {\bf 54}, 869 (1985).


 \bibitem{pac30}
  \gx~Collaboration, ``Mapping the Spectrum of Light Quark Mesons and Gluonic Excitations with Linearly Polarized Protons," {\it Presentation to PAC 30}, (2006).  Available at:  \href{http://www.gluex.org/docs/pac30_proposal.pdf}{\it http://www.gluex.org/docs/pac30\_proposal.pdf}
  
  \bibitem{pac36}
  \gx~Collaboration,  ``The \gx~Experiment in Hall D," {\it Presentation to PAC 36}, (2010).  Available at: \href{http://www.gluex.org/docs/pac36_update.pdf}{\it http://www.gluex.org/docs/pac36\_update.pdf}

  
 \bibitem{NIM}
 I. Adam {\it et al.},  Nucl. Instr. and Meth. {\bf A538}, 281 (2005).

\bibitem{NIM2} C. Field {\it et al.}, Nucl. Instr. and Meth. {\bf A553}, 96 (2005).

\bibitem{NIM3} C. Field {\it et al.}, Nucl. Instr. and Meth. {\bf A518}, 565 (2004).

\bibitem{NIM4} J.F. Benitez {\it et al.}, Development of a focusing DIRC, IEEE Nuclear Science Conference Record, October 29, SLAC-PUB-12236 (2006).

\bibitem{NIM5} J. Va'vra {\it et al.}, The Focusing DIRC- the First RICH Detector to Correct the Chromatic Error by Timing, Presented at Vienna Conference on Instrumentation, February, SLAC-PUB-12803 (2007).

\bibitem{NIM6} J.F. Benitez {\it et al.}, Nucl. Instr. and Meth. {\bf A595}, 104 (2008).

\bibitem{NIM7}
 J. Va'vra {\it et al.},  Nucl. Instr. and Meth. {\bf A718}, 541(2013).



 \bibitem{lappd_proposal}
 LAPPD Collaboration, ``The Development of Large-Area Fast photo-detectors", Proposal to the Department of Energy's Office of High Energy Physics (2009).

 \bibitem{mapmt_superb}
F.~Gargano {\it et al.}, Nucl. Instr. and Meth. {\bf A718}, 563 (2013).

 \bibitem{mapmt_solid}
S.P.~Malace, B.D.~Sawatzky, H.~Gao, JINST {\bf 8}, P09004 (2013).

 \bibitem{mapmt_clas12}
R.A.~Montgomery {\it et al.}, Nucl. Instr. and Meth. {\bf A695}, 326 (2012).

 \bibitem{h8500_spec}
 Hamamatsu Photonics K. K., ``Mutlanode PMT Assembly H8500/H10966 Series", \href{http://www.hamamatsu.com/resources/pdf/etd/H8500_H10966_TPMH1327E02.pdf}{\it http://www.hamamatsu.com/resources/pdf/etd/  H8500\_H10966\_TPMH1327E02.pdf}

 \bibitem{h9500_spec}
Hamamatsu Photonics K. K., ``Mutlanode PMT Assembly H9500, H9500-03", 
 \href{http://www.hamamatsu.com/resources/pdf/etd/H9500_H9500-03_TPMH1309E01.pdf}{\it http://www.hamamatsu.com/resources/pdf/etd/ H9500\_H9500-03\_TPMH1309E01.pdf}

 \bibitem{adarvan}
 Ardavan Ghassemi, Technical Sales Rep for Scientific Projects, Hamamatsu Photonics K. K., Private Communication

 \bibitem{rich_clas12}
 M.~Contalbrigo, E.~cisbani, P.~Rossi, Nucl. Instr. and Meth. {\bf A639}, 302 (2011).

 \bibitem{maroc3}
 S.~Blin {\it et al.}, ``A generic photomultiplier readout chip", JINST {\bf 5}, C12007 (2010).

 \bibitem{ssp}
B. Raydo, ``Sub-System Processor Manual",  
 \href{https://coda.jlab.org/wiki/Downloads/HardwareManual/SSP/SSP_Module_HallD.pdf}{\it https://coda.jlab.org/wiki/Downloads/HardwareManual/ SSP/SSP\_Module\_HallD.pdf}, (2009).

 \bibitem{lappd_tofrich}
M. Wetstein {\it et al.} [LAPPD Collaboration], Nucl. Instr. and Meth. {\bf A639}, 148 (2011).

 \bibitem{lappd_mcp}
O.W.~Siegmund {\it et al.}, Nucl. Instr. and Meth. {\bf A695}, 168 (2012).

 \bibitem{lappd_anode}
 H. Grabas {\it et al.}, Nucl. instr. and Meth. {\bf A711}, 124 (2013).

 \bibitem{lappd_psec4}
E.~Oberia {\it et al.}, submitted to Nucl. instr. and Meth., [arXiv:1309.4397] (2013).

 \bibitem{f1tdc}
Jefferson Lab Data Acquisition Group, ``F1TDC a High-Resolution, Multi-Hit, VME64x Time-to-Digital Converter", 
 \href{https://coda.jlab.org/wiki/Downloads/docs/manuals//f1tdc_v1_2.pdf}{\it https://coda.jlab.org/wiki/Downloads/docs/manuals/ f1tdc\_v1\_2.pdf}, (2005).

 \bibitem{lappd_test}
B.~Adams {\it et al.}, Rev. of Sci. Instr. {\bf 84}, 061301 (2013).

\bibitem{ref:bdt} L. Brieman {\em el al.}, {\em Classification and regression trees}, Wadsworth International Group, Belmont, California (1984).

\bibitem{ref:roe} B.P. Roe {\em et al.}, 
Nucl. Instr. and Meth. {\bf A543}, 577 (2005).

\bibitem{ref:TMVA2007} A.~Hoecker, P.~Speckmayer, J.~Stelzer, J.~Therhaag, E.~von Toerne, and H.~Voss, {\em TMVA: Toolkit for Multivariate Data Analysis}, PoS A CAT 040 (2007). 

\bibitem{transport}
B.~Thomas {\it et al.},  arXiv:1105.5409 [physics.ins-det] (2011).

\end{thebibliography}
\end{document}